\documentclass[a4paper,12pt]{article}
\pdfoutput=1
\def\parn              {  \par\noindent }

\def\parmedskipn        {  \par\medskip\noindent  }
\def\parsmallskipn      {  \par\smallskip\noindent  }
\def\al{\alpha}
\def\be{\beta}
\def\ga{\gamma} \def\Ga{\Gamma}
\def\ep{\epsilon}
\def\lam{\lambda}
\def\Lam{\Lambda}
\def\sig{\sigma}

  \def\calO{{\cal O}}
  
  \def\calU{{\cal U}}

\def\lamtil{\tilde{\lambda}}

\def\psitil{\tilde{\psi}}

\def\adot{{\dot{a}}}
\def\bdot{{\dot{b}}}

\def\aldot{{\dot{\alpha}}}
\def\bedot{{\dot{\beta}}}

\def\onedot{{\dot{1}}}
\def\twodot{{\dot{2}}}

\def\alhat{\hat{\alpha}}

\def\pbar{{\bar{p}}}

\def\wbar{{\bar{w}}}

\def\zbar{{\bar{z}}}

\def\delbar{\bar{\del}}

\def\braket#1#2{\langle #1 | #2 \rangle}

\def\ovsqtwo{{1\over \sqrt{2}}}
\def\del        {  \partial }
\def\half       {  {1\over 2}  }
\def\rootof#1   {  \left( #1 \right)^{1/2}  }

\def\abs#1      {  \vert #1 \vert  }
\def\ie         {{\it i.e.}\,\,}
\def\evalat#1   {  \left\vert_{#1} \right. }

\def\comma          {\, ,}
\def\period         {\, .}
\def\lsim      {\lower .65ex \hbox{\ $\stackrel{<}{\sim}$\ } }
\def\gsim      {\lower .65ex \hbox{\ $\stackrel{>}{\sim}$\ } }

\def\bra#1{{\langle #1 | } }
\def\ket#1{{| #1 \rangle } }
\def\braket#1#2{\langle #1 | #2 \rangle}
\def\com#1#2{{ \left[#1, #2\right] } } 

\def\matel#1#2#3  {{\langle #1 | #2 | #3 \rangle } }

\def\lrvec#1    {\hbox{$\stackrel{\leftrightarrow}{#1}$}}
\def\lvec#1     {\hbox{$\stackrel{\leftarrow}{#1}$}}

\def\vecii#1#2      {  \left(\begin{array}{c}#1\\#2\end{array}\right)  }
\def\veciii#1#2#3   {  \left(\begin{array}{c}#1\\#2\\#3\end{array}
                     \right)  }
\def\veciv#1#2#3#4  {  \left(\begin{array}{c}#1\\#2\\#3\\#4
                                 \end{array}\right)  }
\def\vecfv#1#2#3#4#5 {  \left(\begin{array}{c}#1\\#2\\#3\\#4\\#5
                                 \end{array}\right)  }

\def\matrixii#1#2#3#4            {  \left(\begin{array}{cc}#1&#2\\#3&#4
                                       \end{array}\right) }
\def\matrixiii#1#2#3#4#5#6#7#8#9 {  \left(\begin{array}{ccc}#1&#2&#3\\
                                     #4&#5&#6\\#7&#8&#9\end{array}
                               \right)  }
\def\mativ#1#2#3#4               {  \left(\begin{array}{cccc}
                                       #1\\#2\\#3\\#4\end{array}\right) }
\def\matv#1#2#3#4#5              {  \left(\begin{array}{ccccc}
                                     #1\\#2\\#3\\#4\\#5\end{array}
                              \right)  }
\def\eqabegin         {  \begin{eqnarray}  }
\def\eqaend           {  \end{eqnarray}  }
\def\nn               {  \nonumber  }
\def\bracetwo#1#2     {  \left\{ \begin{array}{l} #1 \\ #2 \end{array}
                         \right.  }
\def\bracetwocases#1#2#3#4  {   \left\{ \begin{array}{ll} #1 &
                                 \qquad #2 \\
                                 #3 & \qquad #4 \end{array} \right.  }
\def\bracebegin#1     {  \left\{ \begin{array}{#1}   }
\def\braceend         {  \end{array}\right.   }
\def\shead#1   { \parmedskipn {\bfall $\Box$\ #1}: \parmedskipn }
\def\head#1    { \parmedskipn {\bfall $\Box$\ #1}: \qquad }
\def\lhead#1   { \parmedskipn {\large\bfall $\Box$\ #1} \parmedskipn }
\def\Lhead#1   { \parmedskipn {\Large\bfall $\Box$\ #1} \parmedskipn }
\def\boxit#1#2      {  \vbox{\hrule\hbox{ \hskip -4.1pt \vrule\kern3pt 
                     \vbox
                    {  \hsize #1 \strut\kern3pt #2 \kern3pt\strut  }
                       \kern3pt  \vrule} \hrule  } }
\def\centerbox#1#2  {  \mbox{  }\par\bigskip  \hfil \boxit{#1}{#2} \hfil
                       \par\bigskip\noindent }
\def\rightbox#1#2   {  \hfill\boxit{#1}{#2}  }
\def\leftbox#1#2    {  \boxit{#1}{#2}  }
\def\fullbox#1      {  \boxit{\textwidth}{#1}  }

\newcommand{\nullify}[1]{}

\def\mpg#1#2{\begin{minipage}[t]{#1} #2  \end{minipage} }

\def\bfall{\boldmath\bf }
\def\nxt{\parsmallskipn}

\def\epsfig#1#2#3{
{\lower #3 \hbox{
 \mpg{#1}{\begin{center} \includegraphics[width=#1,clip]{#2.eps} \\
 Fig. #2\end{center} }}}}

\def\papertitlepage{\baselineskip 3.5ex \thispagestyle{empty}}
\def\Title#1{\baselineskip 1cm \vspace{1.5cm}\begin{center}
 {\Large\bf #1} \end{center} 
\vspace{0.5cm}}
\def\Authors#1{\begin{center} {\it #1} \end{center}}
\def\Abstract{\vspace{1.0cm}\begin{center} {\large\bf Abstract} 
           \end{center} \par\bigskip}
\def\Komabanumber#1#2#3{\hfill \begin{minipage}{4.2cm} UT-Komaba #1
              \parn #2 
              \parn #3 \end{minipage}}
\renewcommand{\thefootnote}{\fnsymbol{footnote}}

\renewenvironment{thebibliography}{\pagebreak[3]\par\vspace{0.6em}
\begin{flushleft}{\large \bf References}\end{flushleft}

\begin{enumerate}\if@twocolumn\baselineskip=0.6em\itemsep -0.2em
\else\itemsep -0.2em\fi\labelsep 0.1em}{\end{enumerate} }

\def\vecX{\vec{X}}

\def\slprod#1#2{\langle #1, #2 \rangle}

\def\bfall{\boldmath\bf}
\usepackage{latexsym}
\usepackage{amsmath}
\usepackage{amssymb}
\usepackage[usenames]{color}
\usepackage{cite}
\definecolor{MyRed}{cmyk}{0,1,1,0.15}
\definecolor{MyBlue}{cmyk}{1,1,0,0.25}

\definecolor{darkblue}{rgb}{0,0,0.7}
\usepackage{ifpdf}
\ifpdf
\usepackage{graphicx} 
\usepackage{epsfig}
\usepackage[setpagesize=false,pagebackref=false, linktocpage, bookmarksopen=true, colorlinks=true, linkcolor=darkblue,citecolor=darkblue,urlcolor=darkblue]{hyperref}
\usepackage{hyperref}
\newcommand{\arXiv}[2]{[\href{http://arxiv.org/abs/#1}{{\tt arXiv:#2}}]}
\def\picture#1#2{\includegraphics[#1]{#2.pdf}}
\else
\usepackage{graphicx}
\newcommand{\arXiv}[2]{[{\tt arXiv:#2}]}
\def\picture#1#2{\includegraphics[#1]{#2_400x400_p1.eps}}
\fi
\def\barz{\bar{z}}
\def\delbar{\bar{\partial}}
\def\fn#1{\footnote{#1}}
\newcommand{\jump}[1]{[\![#1]\! ]}
\def\beq#1{\begin{align}#1\end{align}}
\def\pmatrix#1#2{\left( 
\begin{array}{#1}
#2\end{array} 
\right)}

\def\pp{\rho}
\def\braket#1{\langle #1 \rangle}
\def\alp{auxiliary linear problem } 
\newcommand{\beqa}{\begin{eqnarray}}
\newcommand{\eeqa}{\end{eqnarray}}
\newcommand{\bea}{\begin{array}}
\newcommand{\eea}{\end{array}}
\newcommand{\beqn}{\begin{equation}}
\newcommand{\eeqn}{\end{equation}}

\begin{document}
\papertitlepage
\vspace*{0cm}
\Komabanumber{11-9}{October, 2011}{}
\Title{On holographic three point functions  \\ for GKP  strings  from 
 integrability} 
\Authors{{\sc Yoichi Kazama\footnote[2]{{\tt kazama@hep1.c.u-tokyo.ac.jp}}
 and Shota Komatsu\footnote[3]{{\tt skomatsu@hep1.c.u-tokyo.ac.jp}}
\\ }
\vskip 3ex
 Institute of Physics, University of Tokyo, \\
 Komaba, Meguro-ku, Tokyo 153-8902 Japan \\
  }
\baselineskip .7cm

\numberwithin{equation}{section}
\numberwithin{figure}{section}
\numberwithin{table}{section}
\parskip=0.9ex

\Abstract

Adapting the powerful integrability-based formalism 
invented  previously for the calculation of gluon scattering amplitudes
 at strong coupling,  we develop a method for computing the holographic 
 three point functions for  the  large spin limit of  Gubser-Klebanov-Polyakov (GKP)  strings. Although many of the ideas from 
 the gluon scattering problem can be transplanted with minor modifications, 
 the fact that the information of the external states is  now encoded in 
 the singularities at the vertex insertion points  necessitates several new techniques.  Notably,  we develop a  new generalized  Riemann bilinear identity, which allows one  to express  the  area integral  in terms of appropriate  contour integrals 
 in the presence of such singularities. 
We also give some general discussions on  how  semi-classical vertex operators for heavy string states  should be constructed systematically from the solutions 
 of the Hamilton-Jacobi equation. 
\newpage
\baselineskip 3.4ex
\thispagestyle{empty}
\enlargethispage{2\baselineskip}
\renewcommand{\contentsname}{\hrule {\small \flushleft{Contents}}}
{\footnotesize\tableofcontents}
\nxt\hrule
\newpage
\section{Introduction} 
\renewcommand{\thefootnote}{\arabic{footnote}}
AdS/CFT\cite{AdSCFT:Malda, AdSCFT:GKP, AdSCFT:Witten} is perhaps the most profound and enigmatic paradigm 
of contemporary physics: Profound because it connects  such fundamental yet
disparate theories as gravity  and non-gravitational  gauge theories,  defined 
 in different dimensions and at diametric  coupling strengths. 
It is enigmatic because even 
after more than a decade of intense research producing  a heap of evidence, 
the gist of its  mechanism has not been elucidated. 
Undoubtedly, deeper understanding of  its 
dynamical aspect is needed in order to apply this  powerful duality 
 more reliably to interesting strong coupling  phenomena  in diverse areas of physics 
as well as to further unveil the rich intricate structure of string theory and M theory. 

As in the prototypical example of this duality,  the one between 
the $N=4$ super Yang-Mills (SYM) theory in 4 dimensions and the 
type IIB  superstring theory in $AdS_5 \times S^5$ spacetime, 
 a vast majority of the firm evidence for the AdS/CFT correspondence 
has been on the BPS quantities, which are protected from 
 quantum effects by  high degree of supersymmetry. 
Although research along such a direction will undoubtedly 
 yield further interesting results, in order to reveal  dynamical 
aspects of the duality it is important to  employ 
  methods which do not rely heavily on supersymmetry. 
Perturbative expansion is the simplest such method, which indeed has 
 yielded many results on  the gauge theory side (for a review \cite{Kovacs}).
 On the string side, however, 
 perturbative quantization is quite difficult, even for a single string, 
 due to the curvature of the relevant spacetime.  In any case, the 
 inherent strong/weak nature of the duality calls for methods beyond 
 perturbation theory. 

A powerful and promising framework which in principle can overcome,
 under certain circumstances,  the shortcomings and  difficulties described  above  
 is the use of integrability  and the associated analyticity. 
Of course  fully integrable theories are  scarce but 
 since   there can be various sensible perturbations of such theories 
they are not expected  to be tied to exceptional dynamics. 

Fortunately,  both sides of the celebrated    SYM${}_{\mathcal{N}=4}$/AdS-string 
duality appear to possess integrable properties. Although the conjectured 
full  integrability of the $\mathcal{N}=4$ SYM theory (see \cite{AF, GKV, BT} and for a review \cite{review} and references therein) is still to be confirmed  
and that of the quantum AdS string is  yet to be  formulated\cite{McLoughlin}, 
at least  the integrability of  {\it classical}  strings in  $AdS_5 \times S^5$
\cite{BPR:0305} (and 
in other similar backgrounds) is well-defined and firmly established. 

In this article, we will describe an application of such an integrability to the computation of holographic three point  functions of semi-classical 
 string states, in particular the large spin limit of the Gubser-Klebanov-Polyakov(GKP) folded strings\cite{GKPstring} 
 rotating in $AdS_3$.  Our  study  is  strongly inspired 
 by the  work of Alday and Maldacena\cite{AM:0903, AM:0904} 
 and the related works\cite{GMN:0907, AGM:0911, AMSV:1002}, 
 which developed a beautiful  method of computing the 
 gluon scattering amplitudes at large  't Hooft coupling $\lam$ by 
mapping it to a minimal surface problem for null polygonal Wilson loop 
 in T-dualized $AdS_5$ and 
making clever  use of the integrability of the classical string in $AdS$ space. 
The most  ingenious feature of this method  is that the information 
of the  area can be extracted without  the explicit knowledge of the  form of the 
minimal surface. We will demonstrate that, with certain non-trivial 
modifications,  to be explained shortly,  the basic framework of this 
 powerful method can be transplanted to the important problem of 
 the evaluation of the aforementioned three point functions\footnote{
Recently, a  work in which three point functions 
 of heavy strings in $AdS_2\times S^k$ were studied by similar means 
appeared \cite{JW}. We will make a brief comparison of this work with ours 
 at the end of this section. 
}. 

Needless to say, understanding of the three point functions of the gauge 
invariant composite operators in SYM and of the corresponding dual 
 string states which emanate from and land on the boundary 
 of $AdS$ space is of utmost importance, as they are the basic 
 building blocks encoding the dynamics of conformal theory. 
On the SYM side, the couplings  of  the BPS operators 
were computed  already at the early stage\cite{Lee:1998bxa}, 
while those involving  non-BPS operators have been  studied more 
 recently using  weak coupling expansion and integrability
 structure\cite{Okuyama:2004bd, Roiban:2004va, Alday:2005nd, Escobedo, Georgiou:2011qk, Caetano}. On the string side, the result for the 
 BPS operators was again obtained in \cite{Lee:1998bxa} using type IIB 
supergravity, which agreed precisely with the SYM counterpart. 
As for  non-BPS states,  due to  the lack of perturbative
 quantization scheme, even the computation of two point functions 
 remained  difficult.  Rather recently, the situation 
 was improved for heavy semi-classical states by the development of the 
 saddle point method\cite{Tsuji, JSW, BuchTseytlin},  which makes use of  the known classical configurations. Subsequently, by inserting  a light BPS operator
  (for which the back reaction can be neglected) into the 
 heavy two point functions, various three point 
 functions of heavy-heavy-light type have been computed\cite{Zarembo:1008,Costa:1008,RoibTsey:1008,Hernandez:1011,Ryang:1011,Georgiou:1011,HHL_RusTsey,ParkLee:1012,BuchTsey:1012,BakChenWu:1103,Kristjansen:1103,Arnaudov:1103,Hernandez:1104,Ahn:1105}. Also, when a saddle point can be approximated by point like geodesics, such as in the case of BMN or near BMN strings, calculation of three point functions was performed in \cite{KloseMcL}.  Treatment of the most 
 interesting cases  of three heavy operators has not been possible 
along the same line since  no exact three-pronged solution 
in $AdS \times S$ spacetime is known to date. 

This is precisely the situation where the powerful method developed for 
the gluon scattering problem may be utilized. 
 As we already announced,  we will deal with the case where  each of the three external states 
corresponds to a GKP string  with a large spin propagating
 within $AdS_3$.  Since  we will employ the saddle 
 point approximation, the  main  problem is that of finding 
 the  minimal area surface in $AdS_3$ with prescribed behavior at the 
three punctures  representing  the external states. This is essentially 
 the same type of problem encountered  in the gluon scattering computation.
The only difference is that in that case the information about the external gluons 
 is mapped by T-duality to the  behavior of the surface along  the null polygonal  boundaries at infinity,  instead of at the punctures. 
Thus one expects that the same basic techniques, 
\ie the Pohlmeyer reduction\cite{Pohlmeyer} and  the analysis of the associated auxiliary  linear problem with a spectral parameter $\xi$,  are  applicable. 

To see  more precisely  how and to what extent such techniques can be applied, 
 we must first   describe  the basic structure of the type of  three point function 
of our interest  in the saddle point  approximation.  Schematically, it can be
 written as
\begin{align}
G(x_1, x_2, x_3) &= e^{-S[X_\ast]} \prod_{i=1}^3 
V_i[X_\ast; z_i,  x_i, Q_i ] 
\period \label{corfun}
\end{align}
Here, the arguments $x_i$ of the correlation function $G$ are   the coordinates 
of the boundary of the (Euclideanized) $AdS_3$. 
 $S[X_\ast]$ is  the value of the action at the saddle point configuration $X_\ast$ 
and is of the order $\sqrt{\lam}$, where  $\lam$ is the large  't Hooft coupling 
 constant. 
$V_i$ denotes the  vertex operator  inserted at the point $z_i$ 
 on  the worldsheet\footnote{The dependence on $z_i$ 
 eventually disappears due to $SL(2,C)$ invariance.}.  It  carries  some  large charge $Q_i$ so that, evaluated at the saddle configuration $X_\ast$,  $\ln V_i$ is also of the order  $\sqrt{\lam}$.

 In the case of the gluon scattering, due to the T-duality
 transformation,  the effect of the vertex operators is included  in the behavior of 
the area  $S[X_\ast]$ at the null polygonal boundary at infinity so that 
one need not deal with $V_i$ part explicitly. The area  $S[X_\ast]$ is infrared 
divergent and one can apply  the integrability technique to evaluate the 
 regularized area $S[X_\ast]_{reg}$ after separating the divergence. 

For a holographic three point function, $S[X_\ast]$ is 
 also divergent, this time  due to the presence of the pole singularities 
 on the worldsheet at $z_i$.  One can regularize and separate such a divergence 
 and split the area as $S[X_\ast] = S[X_\ast]_{reg} + S_[X_\ast]_\ep$, 
 where $S[X_\ast]_{reg}$ is finite and $\ep$ stands for the   regularization 
parameter.  At the same time, the contribution of the vertex operator 
is also divergent for the same reason and it must be regularized in a similar way.
This will be denoted by $V_i[X_\ast]_\ep$.  
In this way the correlation function above can be reorganized as 
\begin{align}
G(x_1, x_2, x_3) &= e^{-S[X_\ast]_{reg} }
 \left[ e^{-S[X_\ast]_\ep }  \prod_{i=1}^3 
V_i[X_\ast; x_i, Q_i ]_\ep \right] \period \label{corfundecomp}
\end{align}
We will make an argument  in Sec.4 that the contributions put together in the square brackets becomes finite and give an explicit example. There we also discuss 
 the systematic  procedure to construct the semi-classical vertex operator 
from the Hamilton-Jacobi analysis. In the case of  the GKP string, however,
 we have not yet been able to construct $V_i$ by such a procedure.  

What we have been able to  compute reliably by the integrability technique 
in this work 
is the ``regularized area" $S[X_\ast]_{reg}$. This computation, 
 however, required certain non-trivial modifications  of the method used in 
 the gluon scattering. 

One difference concerns the method of extraction of  the correlation  of the solutions in different regions of the worldsheet. For the present problem, 
 different regions mean the vicinities of different  singularities and the important 
 information is obtained from the analysis of  the local and the global 
monodromies around these singularities of a  set of  differential equations. 
In the gluon scattering case,  the corresponding information is deduced from 
a system of functional equations (called T-system  or Y-system) which 
relate the behavior of solutions in the so-called different Stokes sectors 
produced by the boundary conditions along the null polygon. 

Another modification, which is more involved, is the necessity to derive 
a generalization of the Riemann bilinear identity, which is the key to 
 express the area integral in terms of appropriate contour integrals 
on a double cover of the worldsheet.  As we shall describe in detail 
 in  Sec.~3, the presence of double pole singularities will introduce 
 additional logarithmic cuts in the relevant Riemann surface,  a 
 new feature  not encountered in the gluon scattering problem. 
We have developed an appropriate generalization of the Riemann 
bilinear identity to systematically deal with this situation. 

With these and other minor modifications, we can compute  $S[X_\ast]_{reg}$
 as a function of the parameters, which we call $\kappa_i$, in terms of which 
 the AdS energy $E_i$ and the spin $S_i$ of the $i$-th GKP string
 are  expressed. 

The organization of the rest of this article is as follows. 
In section 2,  we give a brief review of  the  formalism  of  the 
Pohlmeyer reduction and the associated auxiliary linear problem and 
  describe the large spin limit of the GKP string solution in that framework. 
Section 3, which constitutes the main part of this paper, describes the 
method to compute the essential  portion of the three point function 
for the GKP strings, namely the regularized area,  defined 
 in subsection 3.1.  After making a comparison 
 with the gluon scattering problem in subsection 3.2, we develop a generalized Riemann  bilinear identity in subsection 3.3. We then analyze  the monodromy matrices and  the $SL(2,C)$-invariant products of the eigenfunctions for the auxiliary linear problem in subsection 3.4  and explicitly evaluate them in terms
 of the data extracted from the WKB solutions of the linear problem in 
 subsection 3.5.  Finally in subsection 3.6 we combine the result of subsequent subsections to obtain the formula for  the regularized area in terms
 of the parameters which  characterize the GKP strings. The structure of the 
remaining part  of the three point function involving 
the vertex operators is discussed  in section 4.  Although the concrete result 
 for the GKP strings is left for the future, we will describe the systematic procedure
 to construct the semi-classical vertex operators from the Hamilton-Jacobi 
analysis and give an instructive example. We conclude and indicate future 
 directions in section 5. An appendix is provided to give some
technical details. This article became somewhat long as we tried to explain the intricate 
but essential analyses in as clear a fashion as possible.  We hope that 
 such scrupulous description will be beneficial to the reader.

Before ending the introduction, it is appropriate to make the following remark. 
As we were preparing the  manuscript, the work of R.~Janik and A.~Wereszczy\'nski\cite{JW} appeared, which computed the contribution of  the AdS part 
to the  correlation functions of three heavy strings  in $AdS_2 \times S^k$
 by a similar method. 
 Since the basic idea of adapting the 
powerful method  developed for the gluon scattering 
problem is the same, our work has a sizable  overlap with  \cite{JW}.  There  are,  however,  a number of  notable differences. (i) We deal with  the GKP strings carrying large spin which propagate  entirely in  $AdS_3$ and hence with vanishing energy-momentum tensor.  On the other hand,   the work of  \cite{JW} focuses on the $AdS_2$ part of the string  carrying no spin but non-vanishing energy-momentum tensor   that  balances the contribution  from  the  (unspecified motion in the)   $S^k$ part. It is interesting to note, however, that, despite the difference of the systems considered, our result for the regularized area is very similar in structure to the $AdS$ part of the OPE coefficient obtained in \cite{JW}, if the anomalous dimension $\Delta$ in \cite{JW} 
 is identified with our $\kappa  \propto E -S$, the difference of the $AdS$ energy and the  spin of  the GKP string\footnote{We thank R.~Janik for pointing out  this intriguing similarity.}. 
 (ii) In order to properly deal with the Riemann surfaces with  logarithmic as well as square root cuts,  we developed 
and systematically applied a new generalized Riemann bilinear identity. 
In contrast, the work of \cite{JW} did not need to consider logarithmic cuts.
We believe that our method will be useful in further investigations, such as in the study of higher point  functions. (iii) We make a strong  use of the analyticity in the spectral parameter $\xi$ so that the $Z_2$ symmetry is not explicitly needed in the computations. (iv) We thoroughly investigate the cases 
 with general configuration of parameters, which require non-trivial  study of the behavior of the relevant WKB curves.  These features make our work complementary  to the preceding investigation\cite{JW}. 

\section{{\bfall Pohlmeyer reduction for strings in $AdS_3$}}
\subsection{Basic formulation:\ A brief review }
Just as in the computation of the gluon scattering amplitudes discussed in 
 \cite{AM:0903, AM:0904},  we will make full use of the framework of Pohlmeyer reduction to treat  the strings in the $AdS_3$ space.  So we begin by briefly reviewing this  formalism. 
Except for a few  places, we closely follow the notations of  \cite{AM:0903, AM:0904}. 

A string in $AdS_3$ is described by the action on the Euclidean 
worldsheet, proportional to the area $A$, of the form 
\begin{align}
S &= \frac{T}{4}A={T \over 2} \int d^2z   \left( \del  X^\mu \delbar  X_\mu 
 + \Lam (X^\mu X_\mu +1) \right) \comma  
\end{align}
where $T$ is the string tension and 
$X_\mu =\vec{X} = (X_{-1}, X_0, X_1, X_2)$ are the embedding 
coordinates  with the metric  defined by $A^\mu B_\mu  =\vec{ A} \cdot
\vec{ B} = -A_{-1}B_{-1} -A_0 B_0 + A_1 B_1+A_2B_2$.   $\Lam$ is the Lagrange multiplier enforcing 
 the $AdS_3$ constraint $X^\mu X_\mu =-1$. 
The complex worldsheet (plane)  coordinates are denoted by $z, \zbar$ and 
 the derivatives are written as  $\del = \del/\del z, \delbar =\del/\del \zbar$.  
After eliminating $\Lam$, the equation of motion for $X_\mu$ is given by 
\begin{align}
\del \delbar X_\mu = (\del X^\nu \delbar X_\nu) X_\mu \period
\end{align}
In addition we have the following  Virasoro constraints:
\begin{align}
\del X^\mu \del X_\mu  &= \delbar  X^\mu \delbar X_\mu =0 \label{Vira}\period
\end{align}
The system is obviously invariant under the global isometry group $SO(2,2)$ acting
 on $X_\mu$ linearly. 

The basic idea of the Pohlmeyer reduction is to describe the dynamics 
 in terms of a convenient  moving frame consisting of four basis vectors 
 $\{q_1, q_2, q_3, q_4\}$ defined by 
\begin{align}
q_1 &= \vec{X}\comma \quad q_2= e^{-\al} \delbar \vec{X} \comma \quad  q_3= e^{-\al} \del \vec{X} \comma \\
 q_4&= \vec{N} \equiv  \half e^{-2\al} \ep_{\mu \nu \rho\sig} X^\nu \del X^\rho\delbar X^\sig\comma 
\end{align}
where the $SO(2,2)$ invariant function $\al$ is defined by 
\begin{align}
e^{2\al} &\equiv \half \del \vecX \cdot \delbar \vecX \period
\label{etwoal}
\end{align}
These basis vectors  are normalized in the following way:
\begin{align}
q_1^2&=-1\comma \quad  q_2 \cdot q_3= 2\comma 
\quad q_4^2 = 1\comma 
\quad \mbox{other}\ \    q_i\cdot q_j=0\period
\end{align}
Besides the function $\al$,  the following two  $SO(2,2)$ invariant quantities  will be of fundamental  importance\footnote{Because various equations in section 3 
 will become simpler, the signs of $p$ and $\pbar$ are taken to be 
 opposite to those of \cite{AM:0903, AM:0904}.}:
\begin{align}
p &\equiv +\half \vec{N} \cdot \del^2 \vecX \comma \qquad 
\pbar \equiv -\half \vec{N} \cdot  \delbar^2 \vecX \period 
\label{defppbar}
\end{align}

Now by using the equations of motion and the Virasoro conditions, 
 one confirms  that the $\del$ and $\delbar$ derivatives of $q_i$ 
 can all be  expressed in terms of linear combinations of $q_i$'s again. 
To express  the result, it is convenient to introduce  the  W matrix 
consisting of  $q_i$  in the following way:
\begin{align}
W_{\al\aldot, a\adot} &\equiv \half \matrixii{q_1+q_4}{q_3}{q_2}{q_1-q_4} _{\al\aldot} \period
\end{align}
Here $\al$ and $\aldot$ denote the row and the column of the matrix, while 
 the indices $a$ and $\adot$ are associated with the bispinor representation 
 of the each entry of $W$, which is a null  $SO(2,2)$ vector.  For example, an explicit representation for $q_1=\vecX $ can be taken to be 
\begin{align}
(q_1)_{a\adot} &= X_{a\adot} = X_\mu \sig^\mu_{a\adot} =  \matrixii{X_{-1} + iX_0}{X_1 +iX_2}{X_1 -iX_2}{X_{-1} - iX_0} \comma \label{Xaadot} \\
\sig^\mu &=  (1, i\sig_3, \sig_1, -\sig_2) \period
\end{align}
Then the  derivatives of $W$ can be related back to  $W$ itself as 
\begin{align}
& \del W + B^L_z W + W (B^R_z)^T =0 \comma \\
& \delbar W + B^L_\zbar W + W (B^R_\zbar)^T =0 \comma 
\end{align}
which we call $W$-equations. 
 The $2\times 2$ matrices $B^L_z, B^L_\zbar, B^R_z, B^R_\zbar$ appearing here 
are given by 
\begin{align}
B^L_z &= \matrixii{\half \del \al}{-e^\al}{-p e^{-\al}}{-\half \del \al} \comma \qquad 
B^L_\zbar =  \matrixii{-\half \delbar \al}{-\pbar e^{-\al}}{-e^\al}{\half \delbar \al} \comma \\
B^R_z &= \matrixii{-\half \del \al}{pe^{-\al}}{-e^{\al}}{\half \del \al}
\comma \qquad 
B^R_\zbar = \matrixii{\half \delbar \al}{- e^{\al}}{\pbar e^{-\al}}{-\half \delbar \al} \period 
\end{align}
$B$'s are traceless and hence are called $SL(2)$ connections. 
The $W$-equations contain the information of the equations of motion and the 
Virasoro conditions for the invariants $\al, p$ and $\pbar$. They read 
\begin{align}
&\del \delbar \al -e^{2\al} +p \pbar e^{-2\al}   =0 \comma \label{mSGeq} \\
& \del \pbar = \delbar p =0 \period \label{ppbareq}
\end{align}
The first line is identified with a 
 modified sinh-Gordon equation and the second line 
 dictates that $p$ and $\pbar$ are holomorphic and anti-holomorphic
 respectively. 
Furthermore from the $W$-equations one can deduce that  the  $SL(2)$ connections
are flat, namely
\begin{align}
\com{\del + B^L_z}{\delbar +B^L_\zbar} &= 0 \comma  \label{flatcondL}\\
\com{\del + B^R_z}{\delbar +B^R_\zbar} &= 0 \period \label{flatcondR}
\end{align}
These conditions reproduce the equations  (\ref{mSGeq}) and (\ref{ppbareq}). 
What is of crucial importance is that one can generalize these flat
 connections to include a complex spectral parameter $\xi$ 
 in the following way:
\begin{align}
B_z(\xi) &= {1\over \xi} \Phi_z +A_z \comma \qquad 
B_\zbar(\xi) = \xi \Phi_\zbar + A_\zbar \comma \label{Bxicon1}\\
A_z &\equiv   \matrixii{\half \del\al}{0}{0}{-\half \del\al}
\comma \qquad A_\zbar \equiv  \matrixii{-\half\delbar\al}{0}{0}{\half\delbar \al}
\comma \label{Bxicon2}\\
\Phi_z & \equiv  \matrixii{0}{-e^\al}{-p e^{-\al}}{0} \comma\qquad 
\Phi_\zbar  \equiv \matrixii{0}{-\pbar e^{-\al}}{-e^\al}{0} \period\label{Bxicon3}
\end{align}
One can easily check that the flatness is preserved, namely 
\begin{align}
\left[\del + B_z(\xi),  \delbar  + B_\zbar(\xi)\right] =0 \period
\end{align}
This is the reflection of the fact that the system is integrable. 
These connections obey the following $Z_2$ symmetry relation:
\beq{
\sigma _3 A \sigma _3 = A\comma \qquad \sigma _3 \Phi \sigma _3 = -\Phi \period\label{Z2}
} 
As discussed in \cite{AM:0904}, this symmetry originates in the Virasoro constraints (\ref{Vira}).

It is easy to see that these generalized connections are related to 
the original connections $B^L, B^R$ at special values of $\xi$. 
Explicitly, 
\begin{align}
B^L_z &= B_z(\xi=1) \comma \qquad B^L_\zbar = B_\zbar(\xi =1) 
\comma \\
B^R_z &= \calU^\dagger B_z(\xi=i) \calU 
\comma \qquad B^R_\zbar = \calU^\dagger B_\zbar(\xi =i) \calU 
\comma \\
\calU &= e^{i\pi/4}\matrixii{0}{1}{i}{0}  \period
\end{align}

Now the flatness or the commutativity properties,  (\ref{flatcondL}) and (\ref{flatcondR}), mean that the following ``left" and ``right" auxiliary linear problems 
can be  consistently set up:
\begin{align}
&\del \psi^L_\al + (B^L_z)_\al{}^\be \psi^L_\be =0  \comma 
\qquad 
 \delbar \psi^L_\al + (B^L_\zbar)_\al{}^\be \psi^L_\be =0  
\comma \label{leftalp}
 \\
&\del \psi^R_\aldot + (B^R_z)_\aldot{}^\bedot \psi^R_\bedot =0 \comma 
\qquad 
 \delbar \psi^R_\aldot + (B^R_\zbar)_\aldot{}^\bedot \psi^R_\bedot =0 
\period \label{rightalp}
\end{align}
There are two independent solutions for each of $\psi_\al^L$ and $\psi_\aldot^R$, which will be  denoted by $\psi^L_{\al,a}$ and $\psi^R_{\aldot, \adot}$, where 
$a=1,2, \adot=\onedot, \twodot$. 
 Then in matrix notation, the equations above read
\begin{align}
(\del + B^L_z ) \psi^L_a =0 \comma \qquad (\delbar + B^L_\zbar) \psi^L_a =0 \comma \label{psiLeq}\\
(\del + B^R_z ) \psi^R_\adot =0 \comma \qquad (\delbar + B^R_\zbar)
 \psi^R_\adot =0  \period \label{psiReq}
\end{align}
If we use the $\xi$-dependent flat connections,  one can consider instead
\begin{align}
(\del + B_z(\xi)) \psi (\xi) &= 0 \comma \qquad 
(\del + B_\zbar(\xi))\psi(\xi) =0  \period \label{eqalpxi}
\end{align}
The $Z_2$ symmetry relation, (\ref{Z2}), is reflected in the following property of $\psi (\xi)$.
  \beq{
  (\del + B_z (\xi))\sigma _3 \psi (e^{i\pi} \xi )=0 \comma \qquad 
(\del + B_\zbar(\xi))\sigma _3\psi (e^{i\pi}\xi) =0  \label{Z2-2}\period
  }
Once we find $\psi(\xi)$, then $\psi^L$ and $\psi^R$ can be 
 easily deduced  from the relations between $B^L, B^R$ and $B(\xi)$
 shown above, with the result
\begin{align}
\psi^L&= \psi(\xi=1) \comma \qquad \psi^R = \calU^\dagger \psi(\xi=i)
\period \label{psiLR}
\end{align}

The remaining important  ingredient of the formalism is the 
 reconstruction formula, which expresses the components of $W$, 
  including  in particular the original embedding coordinates $q_1=\vecX$, 
 in terms of the solutions $\psi^L, \psi^R$ of the auxiliary linear problems. 
For this purpose, one needs to specify the normalization of  $\psi^L$ and $\psi^R$.  Define the $SL(2)$-invariant product 
 for any pair of $SL(2)$ spinors $\psi$ and $\chi$ by 
\begin{align}
\slprod{\psi}{\chi} &\equiv \ep^{\al \be} \psi_\al \chi_\be \comma 
\qquad (\ep^{\al\be} = -\ep^{\be\al}\comma 
\quad \ep^{12} \equiv 1) \period  \label{sl2prod}
\end{align}
Then, by using the equations of the auxiliary linear problems, (\ref{leftalp}) and 
 (\ref{rightalp}), one can easily show that  for the solutions 
 of the auxiliary linear problems the products   $\slprod{\psi_a^L}{\psi_b^L}$
 and $\slprod{\psi_a^R}{\psi_b^R}$ are constant and anti-symmetric 
 in $(a,b)$. Therefore one can normalize the solutions by the 
 conditions 
\begin{align}
\slprod{\psi^L_a }{ \psi^L_b} &=  \ep_{ab} \comma \qquad 
\slprod{\psi^R_\adot}{\psi^R_\bdot} =  \ep_{\adot\bdot}
\comma 
\label{normpsisol}
\end{align}
where $\ep_{ab}$ is the  anti-symmetric tensor with $\ep_{12} \equiv 1$. 

With such normalized solutions, the entries of the $W$ matrix can be
 reconstructed  succinctly as 
\begin{align}
W_{\al\aldot, a\adot} &= \psi^L_{\al,a} \psi^R_{\aldot,\adot}
\period \label{RecW}
\end{align}
To prove it, first note that each entry of $W$ is a $SO(2,2)$ null vector, 
which has the bispinor representation in terms of the  product of twistors
 as $p_{a\adot} = \lam_a \lamtil_\adot$. Therefore the components of $W$ 
 can be written as $W_{\al\aldot, a\adot} = \lam_{a, \al\aldot} \lamtil_{\adot, \al\aldot}$. Substituting this form into the $W$-equations, one finds 
 that $\lam_{a, \al\aldot}$ and $\lam_{\adot, \al\aldot}$ 
must be proportional to  the solutions $\psi^L_{\al,a}$ and $\psi^R_{\aldot, \adot}$ of the linear problem\footnote{Actually $\lam_{a,\al\aldot}$ does not depend on $\aldot$ and $\lamtil_{\adot, \al\aldot}$
 does not depend on $\al$.}. Taking into account the normalization 
 conditions (\ref{normpsisol}) above, one obtains  the formula (\ref{RecW}). 

The form of the formula (\ref{RecW}) directly reflects the 
global symmetry of the target space. Clearly, the left and the right 
 auxiliary linear problem separately admits linear transformation of the 
 solutions. Taking into account the normalization conditions, this corresponds
 to the action of  $SL(2,C)_L \times SL(2,C)_R \simeq  SO(4,C)$ 
 on the indices $a$ and $\adot$ of the $W$ matrix. 
 If we do note require the reality  condition, this is precisely 
the action of the global symmetry  on  the embedding coordinates of the target space.

Before we finish  this subsection, some remarks are in order. 
\begin{itemize}
	\item The method of the Pohlmeyer reduction does not by itself 
allow  one to solve the non-linear problem. It  merely transforms the 
 equations of motion and the Virasoro constraints into  reduced 
equations for smaller degrees of freedom, in this case 
the modified sinh-Gordon equation and the holomorphicity conditions 
for the invariant functions $\al, p$ and $\pbar$. If one can solve 
 such reduced set of equations by other means,  the solutions for the original
 variables can be readily reconstructed. 

\item However, as was demonstrated clearly in its application to 
 the minimal surface problem relevant to gluon scattering, 
 introduction of the complex spectral parameter $\xi$  into the auxiliary linear 
 problem enormously strengthens the power of the formalism. 
The dependence on $\xi$, including the analyticity, encodes 
 essential  information so that  if  judiciously extracted one can compute 
 important quantities without  solving for the solutions explicitly. 
This feature is what we are going to utilize in the present problem, 
 the computation of the three point function for the strings 
with large quantum numbers. 
\end{itemize}
\subsection{Case of  GKP string in the large spin limit}
Although the basic idea can be applied  in principle to various semi-classical 
string  states, in this work we will deal almost exclusively with the 
large spin limit of the GKP string  propagating in $AdS_3$,
 which we shall call LSGKP. In this subsection, 
 in order to make use of the integrability-based  techniques in later sections,  
we will describe the LSGKP solution in the framework of  Pohlmeyer reduction. 
\subsubsection{LSGKP solution in the embedding  coordinate }
\begin{figure}[htbp]
\begin{center}
\picture{clip,height=4cm}{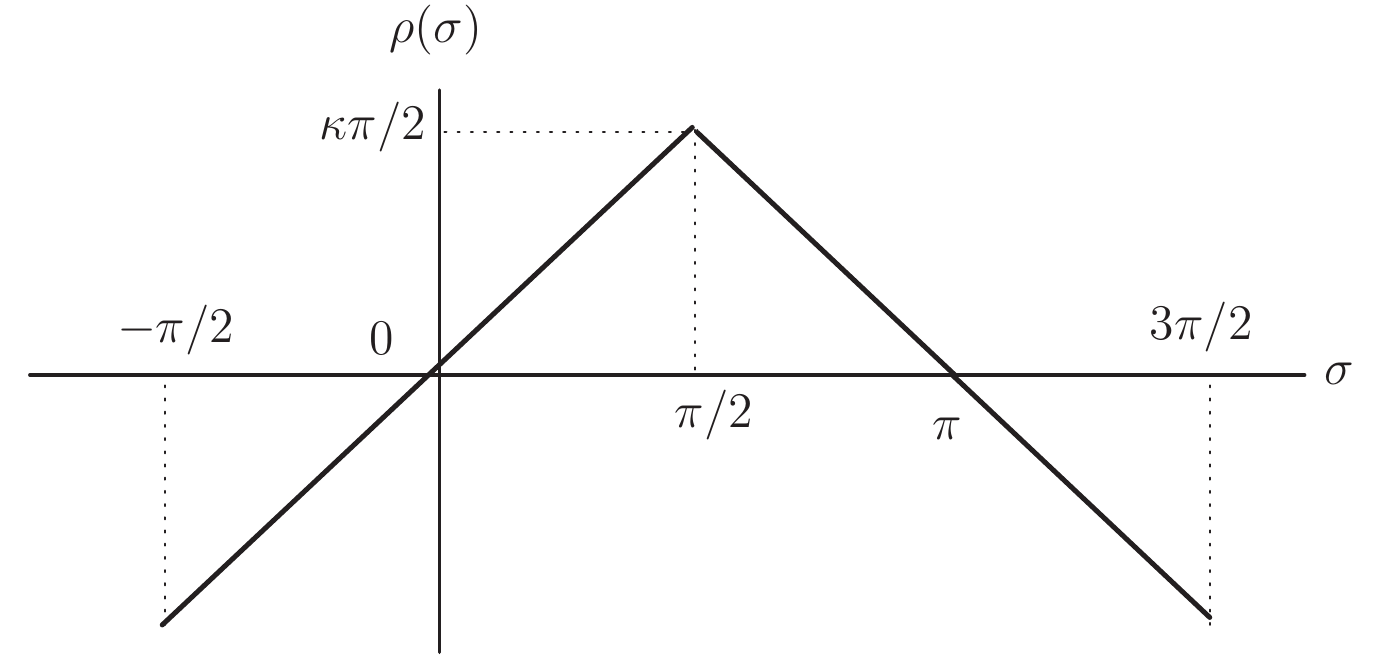}
\end{center}
\caption{Profile of the function $\rho(\sig)$ which appears in the 
large spin limit of the GKP solution.} \label{figgkp}
\end{figure}

The LSGKP solution in the embedding  coordinate is given using  the Euclidean 
cylinder worldsheet variables  $(\tau, \sig)$ by 
\begin{align}
X_{-1} \pm i X_0 &= e^{\pm \kappa \tau} \cosh \rho(\sig) 
\comma \label{GKPone}\\
X_1 \pm i X_2 &= e^{\mp \kappa \tau} \sinh \rho(\sig)
\comma  \label{GKPtwo}
\end{align}
where the function $\rho(\sig)$,   periodic with period $2\pi$, has the profile 
depicted in the figure \ref{figgkp}. 
It is obtained from a smooth function expressible in terms of 
the Jacobi elliptic function by taking the limit of large $\kappa$. 
It  is  given in the interval $[ -{\pi \over 2}, {3\pi \over 2}]$ by 
\begin{align}
\rho(\sig) =\bracetwo{ \kappa \sig  \comma  \hspace{1.7cm} \left( -{\pi \over 2}  \le \sig \le {\pi \over 2}\right)}{\kappa \left( \pi  -\sig \right) 
\comma 
\quad \left({\pi\over 2} \le \sig \le {3\pi \over 2}\right)} 
 \period
\end{align}
In this limit we must treat the two segments separately and glue them 
 together. 
There are two conserved quantities in this system, namely the AdS energy
 $E$ and the angular momentum (spin)  $S$ in the $1$-$2$ plane, which 
 can be expressed in terms of the parameter $\kappa$  in the following way:
\begin{align}
E &= T\kappa \int_0^{2\pi} d\sig\,\cosh^2 \rho = T (\kappa \pi 
 + \sinh \kappa \pi) \comma  \label{AdSE} \\
S &=  T\kappa \int_0^{2\pi} d\sig\,\sinh^2 \rho = T (-\kappa \pi 
 + \sinh \kappa \pi) \period \label{Spin}
\end{align}

Now, introducing the complex coordinates 
$w=\tau+i\sig, \wbar=\tau -i\sig$, it is straightforward to 
evaluate the basic invariant quantities $p(w), \pbar(\wbar)$ and $\al(w,\wbar)$
from their definitions (\ref{etwoal}) and (\ref{defppbar}) for this 
solution. The result is 
\begin{align}
p(w) &= -{\kappa^2 \over 4} \comma \qquad \pbar(\wbar) = -{\kappa^2 \over 4} \comma \qquad 
e^{2\al(w,\wbar)} = {\kappa^2 \over 4} \period 
\end{align}
Going to the plane coordinate $z, \zbar$ defined by
\begin{align}
z =  e^w \comma \quad \zbar = e^\wbar \comma 
\end{align}
$p, \pbar$ and $e^{2\al}$, which are $(2,0), (0,2)$ and $(1,1)$ forms respectively,  become  
\begin{align}
p(z) &= -{\kappa^2 \over 4z^2} \comma \qquad \pbar(\zbar) =
-{\kappa^2 \over 4\zbar^2}\comma  \label{ppbarforgkp}\\
e^{2\al(z,\zbar)} &= 
\sqrt{p\pbar}  \period \label{alforgkp}
\end{align}
Recall that in the case of the gluon scattering problem,  $p(z)$ is a polynomial 
in $z$ of degree  $n-2$ for the scattering of $2n$ gluons. Here it behaves 
 as $\propto 1/z^2$, with a double pole at the origin, expressing the emergence of a string at $z=0$  (or $\tau = -\infty$) from an appropriate vertex operator. 
Actually there is another pole at $z=\infty$, which can be seen by 
going to the inverse coordinate $z^{\prime} = 1/z$. In this coordinate 
 $p(z^{\prime})$ becomes $p(z^{\prime}) = -\kappa^2/4z^{\prime^2}$, showing
the same structure at infinity.  The relation (\ref{alforgkp}) is characteristic 
 of the large spin limit of the GKP solution and will play important roles. 

We now study  how the solutions of the linear auxiliary problem 
behave  for LSGKP and check  the reconstruction formula explicitly. 
These data will be important when we discuss the three point functions 
 for LSGKP. 

First it is convenient to simplify the linear problem by 
 making the  following gauge transformation:
\begin{align}
\psi &= \mathcal{A} \tilde{\psi}\comma\qquad \mathcal{A}=\pmatrix{cc}{p^{-1/4}e^{\alpha/2}&0\\0&p^{1/4}e^{-\alpha/2}} 
\period\label{gaugetrA}
\end{align}
Then, the equations $0=(\del + B_z(\xi))\psi\comma$ $(\delbar + B_\zbar(\xi))\psi=0$ drastically simplify in terms of the cylinder coordinates $w\comma \bar{w}$:
\begin{align}
0&=  \left[ \del _w
 - {i\kappa \over 2\xi} \matrixii{0}{1}{1}{0}  \right] \tilde{\psi}
\comma \label{delpsieq}\\
0&=\left[ \delbar _{w}+  {i\kappa\xi\over 2} \matrixii{0}{1}{1}{0} \right] \tilde{\psi} \period \label{delbarpsieq}
\end{align}
It is easy to solve these two equations, (\ref{delpsieq}) and (\ref{delbarpsieq}), and obtain two linearly independent solutions,
\begin{align}
&\tilde{\psi}=\exp \left( \pm \frac{\kappa i}{2}\left(\xi ^{-1}w-\xi \bar{w}\right)\right)\pmatrix{c}{1\\ \pm1}\period \label{psitwoeq}
\end{align}
Note that as we go around  the origin once, this expression 
 gets multiplied by a  factor 
 $e^{\mp \kappa \pi  \left({1\over \xi} +\xi \right)} $. 
This is an important quantity characterizing the LSGKP string  and will appear 
in the next section  as the ``local monodromy" datum. Applying the relation 
 between $\psi(\xi=1\comma i)$ and $\psi^L, \psi^R$ given in (\ref{psiLR}), we obtain 
\begin{align}
\psi^L_1& = \ovsqtwo e^{-\kappa i(w-\bar{w})/2} \vecii{-ie^{(w-\bar{w})/8}}{e^{-(w-\bar{w})/8}}  \comma \qquad 
\psi^L_2 = \ovsqtwo e^{\kappa i(w-\bar{w})/2} \vecii{-e^{(w-\bar{w})/8}}{ie^{-(w-\bar{w})/8}} \comma \\
\psi^R_\onedot &=-{ie^{-i\pi/4} \over \sqrt{2}}  e^{\kappa (w+\bar{w})/2} \vecii{e^{-(w-\bar{w})/8}}{e^{(w-\bar{w})/8}} \comma \qquad 
\psi^R_\twodot =-{e^{-i\pi/4} \over \sqrt{2}}  e^{-\kappa (w+\bar{w})/2} \vecii{-e^{-(w-\bar{w})/8}}{e^{(w-\bar{w})/8}} \period  
\end{align} 
It is easy to check that these solutions are  properly normalized, namely 
$\slprod{\psi^L_1}{\psi^L_2} =\slprod{\psi^R_\onedot}{\psi^R_\twodot}=1 $. 

With the solutions of the auxiliary linear problem at hand, we can now
 reconstruct the embedding coordinates of the LSGKP solutions. From the structure 
 of $W$ matrix and the reconstruction formula (\ref{RecW}), $(q_1)_{a\adot}$ 
 is given by 
\begin{align}
 (q_1)_{a\adot} &=
 \psi^L_{1a} \psi^R_{\onedot \adot}
+\psi^L_{2a} \psi^R_{\twodot \adot} \period 
\end{align}
Then we obtain the simple result
\begin{align}
q_1 &= -\ovsqtwo \matrixii{e^{\kappa \tau} e^{\kappa \sig}}{e^{-\kappa \tau} e^{\kappa \sig}}{-e^{\kappa \tau} e^{-\kappa \sig}}{e^{-\kappa \tau} e^{-\kappa \sig}} \period
\end{align}
By making an appropriate  $SL(2,C)_L \times SL(2,C)_R$ global transformation, this can be 
put into the familiar form (\ref{Xaadot}), with $X_{-1} \pm iX_0$ and $X_1 \pm iX_2$ given precisely by the standard LSGKP solution  (\ref{GKPone}) and (\ref{GKPtwo}). 
Before closing this section, we would like to make some important remarks:
\begin{itemize}
\item If we regard $\tau$ as being real, the embedding coordinates of
 the solution  described by (\ref{GKPone}) and (\ref{GKPtwo}) are  not all real.  The appearance of complex trajectories  of this kind is often encountered  in the saddle point calculation of  correlation functions. 
In some cases,  Wick rotation of the target space ``time"  coordinate, $X_{0}\rightarrow iX_{0,e}$,  can be performed so that such solutions can be regarded as real solutions in the {\it Euclidean} $AdS$. 
However, this prescription does not apply universally. In fact in the present case, 
 $X_2$ in (\ref{GKPone}) and (\ref{GKPtwo}) is also complex-valued and 
 it is unnatural to make it real by another ``Wick rotation".  
The  viewpoint  we adopt in this work is  that the solution described by  (\ref{GKPone}) and (\ref{GKPtwo})   is  simply a complex saddle point of the path integral  of the string theory in {\it Lorentzian} $AdS_5$. This should be the correct interpretation  since what $AdS/CFT$ correspondence predicts   is a duality between $\mathcal{N}=4$ super Yang-Mills theory in the Lorentzian signature and the superstring theory in the Lorentzian $AdS_5\times S^5$.

\item The embedding coordinates  $(X_{-1},X_{0},X_{1},X_{2}, X_3, X_4)$
 of the $AdS_5$ space are related to the  Poincar\'e coordinates $(x^{0},x^{1},x^{2},x^{3},z)$ as
\beq{
X_{-1}+X_{4}=\frac{1}{z}\comma \qquad X_{-1}-X_{4}=z+\frac{x^{\mu}x_{\mu}}{z}\comma \qquad X_{\mu}=\frac{x_{\mu}}{z}\label{poincare}
\period}
Since we are considering the $AdS_3$ subspace spanned by $(X_{-1},X_{0},X_{1},X_{2})$, the coordinates  $X_3$ and $X_4$  in (\ref{poincare}) 
vanish in our case. This imposes a constraint of the form  $x^0x_0 + x^1x_1 + x^2x_2 +z^2=1$ on the Poincar\'e coordinates and hence the positions of the vertex operators on the boundary at $z=0$
are restricted to  be in the region  $x^0x_0 + x^1x_1 + x^2x_2 =1$.  
If one wishes to avoid this constraint, one can instead consider solutions in $(X_{-1},X_{1},X_{2},X_{4})$ from the beginning by replacing   $iX_{0}$ in (\ref{GKPone}) and (\ref{GKPtwo}) with   $X_{4}$. 
 With this choice of embedding coordinates, the equations (\ref{GKPone}) and (\ref{GKPtwo}) describe  a  folded spinning string which is emitted from the boundary of $AdS$ and gets absorbed at  the horizon. 
We can then perform an appropariate  $SL(2,C)_L \times SL(2,C)_R$ global transformation to make it into an appropriate saddle point configuration for a two point function on the boundary. 
\end{itemize}

\section{Regularized area from integrability}

In this section, we shall describe in detail how one can compute one of the 
two building blocks  of the semi-classical three point function, 
namely the contribution  of the form   $\exp(-S[X_\ast])$, where $S[X_\ast]$ is the action (or the area) 
 evaluated at the saddle point solution $X_\ast$ in the presence of  vertex operators. As briefly mentioned in the introduction, $S[X_\ast]$ itself 
 contains divergences coming from the vicinity of the points where 
 the vertex operators are inserted.  We shall discuss later in  Sec.~4 
that  such divergences will in the end be canceled by the contribution 
 of the vertex operators (or the wave functions),  which is also divergent, 
  and  the correlation function as a whole will be finite. 
In what follows, we shall  first clarify how we should compute the 
action in the case of LSGKP folded strings and then discuss the 
 isolation of  the divergent part  of the area by using the knowledge of the behavior of the GKP strings  near the insertion points. 
Subsequently, the remaining finite part of the area will  be calculated 
by using a generalization of the Riemann bilinear identity and the WKB analysis of the auxiliary linear problems. 
\subsection{Definition of the regularized area}
We begin by clarifying precisely what area  we need to compute and how 
we will actually evaluate it. 

\begin{figure}[htb]
\begin{center}
\picture{clip, width=11cm}{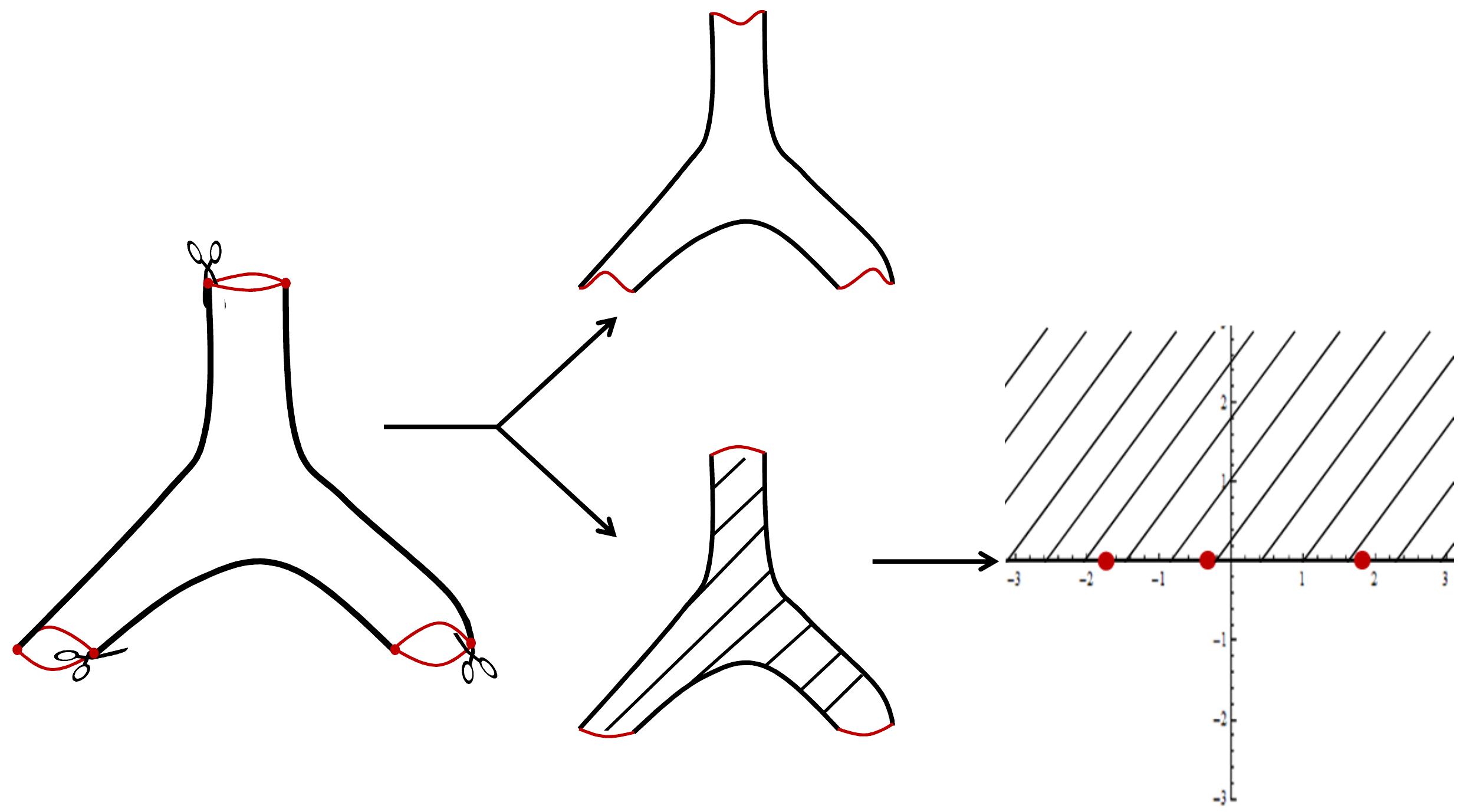}
\end{center}
\caption{Splitting the worldsheet into two identical halves. Each sheet can be 
 mapped to the upper half plane. }
\label{cut-and-map}
\end{figure}

As was reviewed in the previous subsection, in the large spin limit,  the 
 GKP string propagates in  completely folded fashion. As long as we stick to 
 this limit this feature should 
 continue to hold for the 3-pronged saddle configuration which is relevant 
 for the three point function. Therefore the area swept out  by one half of 
 the folded  string is the same as that  swept out by the other half and we only need to compute one of them and multiply by a factor of two. This is 
 depicted in figure \ref{cut-and-map}. The worldsheet for such ``half string"  is 
 like that of an open string and can be represented by 
  the upper half plane, as shown in figure \ref{cut-and-map}.  Now a very convenient way 
 to compute the area for this portion of the worldsheet is to first extend it 
smoothly to the whole complex plane,  compute the area  and then  divide 
 the result by two.  But as we have to double this result at the end, 
 the net result  we want  coincides with  the contribution from the whole
  plane itself. 

To justify this method, we should clarify a possibly confusing point. 
An important point is that the method above is valid in so far as we use it 
for the computation of the area, which is not affected by the non-trivial behavior of the actual string solution in the  vicinity of  the folding points.  
If one wishes to reconstruct the entire 
LSGKP solution  including the acute folding behavior automatically using the 
integrability-based method of the previous section,  one must modify the behavior of $\al(z, \zbar)$ 
 at the folding points and this in turn changes the monodromy of the 
solutions of the auxiliary linear problem, as exemplified in the work \cite{Dorey}. 
Therefore, if we are interested in the saddle point configuration 
 itself for the three point function, we must either start from  the 
exact form of the GKP string\cite{GKPstring}  and properly take into account the subtle behavior 
 of $\al(z, \zbar)$ etc.  in the large spin limit,  or find the solution for the 
 half string from the analysis in the upper half plane and then paste its copy 
on top  of it to construct the closed string solution. 
In either way we cannot use the smooth extension into the lower half plane 
 as in the above method. 

 For the computation of the area, we can circumvent such  complicated procedures by the simpler method described  above.  As is clear from figure \ref{cut-and-map}, 
after the smooth extension, the worldsheet is symmetric under 
the reflection about the real axis and hence the solution $\al(z,\zbar)$ 
of the modified sinh-Gordon equation  must enjoy  the same symmetry. 
Since the area is expressed in terms of $\al(z,\zbar)$, the contribution from 
 the lower half plane must equal that from the upper half plane. Moreover, since the area is conformally invariant, we can make an arbitrary 
conformal transformation to move the positions of the singularities 
(\ie the positions of the vertex operators) from along  the real axis to 
 wherever are more convenient on the complex plane. We will make use 
 of  such a  freedom in developing the method of computation later. 
The conclusion is that as far as the computation 
 of the area is concerned we may effectively forget about the effect of folding.

 The area can be expressed as 
\beq{
&A = 2\int d^2z \, \del \vec{X} \cdot \delbar\vec{X} = 4\int d^{2}z \,e^{2\alpha}\period
}
As discussed in the previous section, near the insertion points $z_i$ the quantity
 $e^{2\al}$ behaves like 
\begin{align}
e^{2\al} \simeq  \sqrt{p \pbar} \comma 
\label{asympt} 
\end{align}
which in the case of GKP string goes like $ 1/|z-z_i|^2$, giving rise to 
log divergences for the area.  Thus one can separate the finite  and the 
 divergent  part of the area as 
\begin{align}
A &= A_{fin} + A_{div}\comma \\
A_{fin} &= 4\int d^2 z \,\left( e^{2\alpha}-\sqrt{p\pbar}\right) \comma 
\qquad A_{div} = 4\int d^2z \,\sqrt{p\pbar} \period
\end{align}
Now by using the modified sinh-Gordon equation (\ref{mSGeq}) 
\beq{
&\del\delbar \alpha -e^{2\alpha}+p\pbar\, e^{-2\alpha}=0\comma\label{mSGeq2}
}
one can rewrite  $A_{fin}$ into the form 
\beq{
A_{fin} &= 2\int d^2z \, \left(e^{2\alpha}+p\pbar \,e^{-2\alpha}-2\sqrt{p\pbar}+\del\delbar\alpha\right)\period
}
%
The contribution from the total derivative term  $\del\delbar \alpha$
comes solely from the ``boundaries", which in our case consist of 
 infinitesimal circles around  the vertex insertion points and a  large  circle 
 at infinity.  For the contributions from the singularities, we may use 
  (\ref{asympt}) to  replace $\del\delbar \al$ by $\frac{1}{4} \del \delbar \ln (p \pbar)$. 
Furthermore, from the behavior of $p \pbar$ at the singularity 
and the familiar formula $\del \delbar \ln z\zbar =\pi \delta^2(z)$, 
the integral $\frac{1}{4}\int d^2 z\,\del\delbar \ln p\pbar$ simply yields $\pi/2$ 
for each of the singularity.  On the other hand, for large $z$, as will be discussed in Sec.~3.3.2, $p(z)$ goes as $\sim 1/z^4$, which again leads to the behavior  $\al \simeq (1/4)\ln p \pbar$ from (\ref{mSGeq2}). Using the Stokes theorem the contribution of $\frac{1}{4}\int d^2 z\,\del\delbar \ln p\pbar$ from this region yields  $-\pi$.  Combining,  for $N$-point correlation functions  $A_{fin}$ can be written as 
\beq{
A_{fin}&= 2 A_{reg} + \pi (N-2) \comma \label{area0}\\
A_{reg} &\equiv \int d^2z \,\left(e^{2\alpha}+p\pbar \,e^{-2\alpha} -2\sqrt{p\pbar}\, \right) \comma \label{area}}
where we defined $A_{reg}$, to  be called the ``regularized area". 
This  will be the quantity to be evaluated in the remainder of this section. 
The divergent part $A_{div}$ will be discussed briefly in Sec.4,  where we argue that 
when combined with the contribution of the vertex operators it 
will give a finite result.     
\subsection{Comparison with the gluon scattering problem}
Before proceeding to the actual evaluation of the regularized area, 
it should be helpful to quickly review  the salient features of the 
calculation  of the corresponding quantity in the case of 
 the gluon scattering amplitudes  given in  
\cite{AM:0904} and describe the similarity to and the difference 
 from the computation that we are going to perform.

In \cite{AM:0904},  the momenta of the massless gluons  were  mapped by T-duality to the segments of a null polygonal  Wilson loop and the scattering amplitude of such gluons is given by the area of the  minimal surface  bordered by that loop.  For a simple configuration of gluon momenta, the loop can be taken to 
lie in a $R^{1,1}$ subspace of $R^{1,3}$ and the relevant minimal surface 
 is contained  in  $AdS_3$. Although  such an area is divergent  reflecting 
 the infrared divergence of the massless on-shell amplitude,  one can extract 
  the important  finite part, called the  remainder function, by regularizing the area as in (\ref{area}). 
As  it is  difficult to construct  the explicit solution of  the minimal surface for
  case with  more than four gluons,  they utilized the framework of Pohlmeyer reduction and performed  integrability-based analysis of the associated auxiliary linear problem  to calculate the  regularized area without the knowledge of such 
solutions. 

In that analysis, the holomorphic function $p(z)$ which provides the input data of the gluons for the auxiliary linear problem behaves, in the case of $2n$ gluons, as 
a polynomial in $z$ of degree $n-2$, viz, 
\beq{
p(z)=z^{n-2}+\cdots \comma 
}  
which has a singularity at infinity. Due to this singularity,
 the solutions  to the auxiliary linear problems exhibit the so-called Stokes
 phenomena. Namely, the solutions change their asymptotic 
 forms at infinity discontinuously  as  they cross  a set of lines called Stokes lines. 
These lines partition the whole complex plane into several Stokes sectors. 
In each Stokes sector, labeled by $i$, there are two solutions
$b_i$ and $s_i$, which are called ``big" and ``small"  according as they increase or decrease asymptotically at infinity. It is extremely important to identify the small solution since it is uniquely defined contrary to  the big solution which can contain a multiple of small solution  and hence is ambiguous.  It was shown  that the essential  information 
 is carried by the $SL(2,C)$-invariant products  $\braket{s_j, s_k}$
of these small solutions  and  they are related to each other through  
 a system of functional equations called T-system, which  depends  on  the spectral 
 parameter $\xi$:
\beq{
T_j (e^{+i\pi /2}\xi )T_j (e^{-i\pi /2}\xi )=T_{j-1}T_{j+1}+1\comma \hspace{11pt} T_j \equiv \braket{s_0,s_{j+1}}(e^{-i(j+1)\pi /2}\xi )\period
}
On the other hand, with the aid of the Riemann bilinear identity, the regularized area 
of interest can be expressed in terms of certain boundary contour integrals on a bordered double cover of the worldsheet, where the important function defined by 
\beq{
\Lambda (z) \equiv \int^z \sqrt{p(z')} dz'  
}
is single-valued. The crucial fact is that  these  contour integrals 
appear in the WKB expansions\footnote{Following \cite{GMN:0907}
 we shall refer to the expansion in powers of $\xi$ or $1/\xi$ as WKB 
 expansions.}
 of  the $SL(2,C)$-invariant products  $\braket{s_j, s_k}$ discussed above.  Therefore, by solving the $T$-system equations, one can compute  the value of the regularized area. 

As we shall explain in detail in the subsequent subsections, 
 the procedure for the computation of the regularized area relevant 
 to the three point function will be quite similar to what we described above. 
There are, however, some differences which require non-trivial 
modifications. Below let us preview them by  comparing with the gluon 
scattering computation sketched above. 

For the gluon scattering, the worldsheet is of the disk topology as it is an open string
 process.   The boundary of the disk, however,  is not directly visible in 
 the actual calculation as it is mapped to a point at infinity. Its effect is 
encoded  in the singularity of $p(z)$ at infinity.  In contrast,  in the case of three point functions  of closed strings, the worldsheet is basically a sphere with  the  vertex operators inserted at three points $z_i$ and their effect is  reflected in the  singularities of $p(z)$ of the form\fn{For LSGKP string, $\delta_i=i\kappa_i /2$, as shown in (\ref{ppbarforgkp}).} 
\beq{
p(z) \overset{z\to z_i}{\sim} \frac{\delta _i^2}{(z-z_i)^2} \period
\label{pzatzi}
}
These double pole singularities produce two main modifications in the calculation of the regularized area.

One  significant modification is that one needs to generalize the Riemann bilinear identity in order to express the area in terms of contour integrals.  In contrast to the case of the gluon scattering, due to the double pole singularities 
 of $p(z)$   the function  $\Lambda(z)$ acquires logarithmic branch cuts, 
 on top of  the usual square root cuts which are also present for the gluon 
 scattering. The existence  of such  logarithmic branch cuts invalidates the application of the standard Riemann bilinear identity and one needs to derive an appropriate  generalization to deal  with this new  situation. 

The other modification stems from the difference of the way the information 
 of the external states is encoded. In place of the  behavior of the 
 solution in different Stokes sectors in the case of the gluon scattering problem, 
such information is carried by the so called monodromy data for the 
solutions of the \alp. As will be explained more fully in subsection 3.4, a 
 basis of solutions $(\eta_1, \eta_2)$ of the \alp gets transformed into 
 a different  basis  $(\eta_1', \eta_2')$ by a monodromy matrix $M$ 
when analytically continued around each pole:
\beq{
\pmatrix{c}{\eta _1^{\prime}\\ \eta _2^{\prime}}=M\, \pmatrix{c}{\eta _1\\ \eta _2}\period
}
It will be shown that the contour integrals which appear in the generalized 
 Riemann bilinear identity can be expressed  in terms of the 
$SL(2,C)$ invariant products between the  eigenvectors of the monodromy matrices, which in turn can be expressed in terms of the parameters $\delta_i$ 
appearing in (\ref{pzatzi}) from the properties of the monodromy matrices. 
This process actually corresponds to the solution of the T-system.  
To actually evaluate the $SL(2,C)$ invariant products we will need to make use 
 of the notion of the small and big solutions and their WKB expansions, as in the 
 case of the gluon scattering problem. In particular  it will be important to find 
 the precise relation between the $SL(2,C)$ products between the 
eigenfunctions of the monodromy matrices and those between the small solutions. 

Having explained the relation to the gluon scattering problem, we now proceed to  the actual computation of  the regularized area
 relevant to  the three point functions.  
\subsection{Generalization of Riemann bilinear identity}
As in the case of the gluon scattering problem, 
we wish to reexpress the regularized area in terms of certain contour integrals.  
For the gluon scattering,  this  was made possible  with the aid of the Riemann bilinear identity, which is essentially the familiar  Stokes theorem applied to 
a Riemann surface with  finite genus.  Unfortunately, in the  present case 
 the usual  Riemann bilinear identity is not directly applicable due to 
double pole singularities of $p(z)$. The presence of these singularities leads to logarithmic branch cuts  in addition to  the square root cuts, which must be 
 properly treated. 

In this subsection, we will derive an appropriate  generalization of the Riemann bilinear identity, which is applicable  in the presence of any number of singularities. 
Although we mainly consider, in this paper,  three point functions of the large spin limit of the GKP strings, the identity we derive can be applied to $N$ point functions of general operators.  
\subsubsection{From surface integrals to boundary contour integrals}
As the first step, we rewrite $A_{reg}$ 
defined in (\ref{area}) as an integral over the product of two functions 
 of the form 
\begin{align}
A_{reg} &= \int d^2z \lam \,  u \comma \label{surface}
\end{align}
where $\lambda$ and $u$ are given by
\beq{
&\lambda = \sqrt{p} \comma\hspace{11pt} u = 2\sqrt{\bar{p}}(\cosh 2 \hat{\alpha} -1)\comma\\
&\alhat \equiv \al -{1\over 4}  \ln p \pbar \period  \label{defalhat}
}
It is easy to show that one can construct  a closed 1-form of the form 
 $ud\zbar + vdz$ with the help of a function $v$ given by 
\beq{
v=\frac{1}{\sqrt{p}}(\del \hat{\alpha}) ^2 \period
} 
Since $\lambda$ is holomorphic and $ud\barz + vdz$ is closed, (\ref{surface})  can be reexpressed as 
\beq{
\int d^2z \lambda \,  u = \frac{i}{2}\int \lambda dz \wedge (ud\barz + vdz) \period
}

Now due to the presence of the square root branch cuts,  it will be 
 convenient to consider a double cover of the worldsheet, to be 
 denoted by $D$, with appropriate boundary $\del D$, which will be specified  more explicitly later. To distiguish the points and contours on the first and the second sheet
 clearly,  we will put  overhats on  quantities on the second sheet. 
 $D$ and $\del D$ are defined so that the function
\beq{
\Lambda (z) = \int _{z_0}^{z} \lambda (z^{\prime})dz^{\prime}
\label{defLam}
}
is singled-valued and the functions $\lambda,\, u$ and $v$ have no singularities on $D$. The end point $z_0$ of the integral can be chosen arbitrarily on $D$ 
but in the following discussion we shall take $z_0$ to be on the first sheet. 
The corresponding point on the second sheet is denoted by $\widehat{z}_0$. 

With this setting, we can convert the area integral
 (\ref{surface})  into a boundary contour integral  in the following way\footnote{Compared to the previous version, 
the sign of the last term has been reversed, in order to be consistent with  the directions of the contours depicted in the figures. This resulted in obvious overall sign changes for various equations in this subsection.}.
\beq{
A_{reg} &=\frac{i}{4}\int _{D} d\Lam  \wedge (ud\barz + vdz)=\frac{i}{4}\int _D d(\Lambda (ud\barz + vdz)) 
=-\frac{i}{4}\int _{\del D} \Lambda (ud\barz + vdz) \period 
\label{Aregint}
}
To proceed further we now  specify $D$ and $\del D$ explicitly
 for the worldsheet relevant to $N$ point functions for various $N$. 
\subsubsection{Double cover of the worldsheet for N
 point functions}
What we need to know is the  structure of the singularities and the zeros of 
 $p(z)$.  In the case of $N$ point function, there are $N$ double poles at $z_i$ 
for $i=1,2, \ldots, N$,   where  $p(z)$ behaves like 
\beq{
p(z) \overset{z\to z_i}{\sim} \frac{\delta _i^2}{(z-z_i)^2} \period
}
This means that $p(z)$ is of the form 
\begin{align}
p(z) &= {f(z) \over (z-z_1)^2\cdots (z-z_N)^2} \comma 
\label{pzfz}
\end{align}
where  $f(z)$ is a polynomial with the property $f(z_i) = \delta_i^2 \prod^N_{j=1( \ne i)} 
(z_i-z_j)^2$.  An important requirement is that there is no singularity at infinity. 
Since $p(z)$ is a $(2,0)$ form,  this means that, by making 
 a conformal transformation $z \rightarrow 1/z$,   $(1/z^4)  p(1/z)$ should be 
 non-singular at $z=0$.  This  in turn is equivalent to the requirement 
 $p(z) \sim \calO(1/z^4)$  for large $z$.  From (\ref{pzfz}) we see that  $f(z)$ is  
 a polynomial  in $z$ of order  $2N-4$ and hence $p(z)$ generically 
has $2N-4$ zeros.  We have thus found that the function $\Lam(z)$ 
 defined in (\ref{defLam}) has $N$ logarithmic branch cuts running from 
 the singularities and $N-2$ square root cuts connecting the $2N-4$ zeros 
 of $p(z)$. 

With this knowledge, let us describe the appropriate domain $D$ and its 
boundary $\del D$ for the cases of two, three  and multi point functions. 

\begin{figure}[htbp]
\begin{center}
\picture{height=2.5cm,clip}{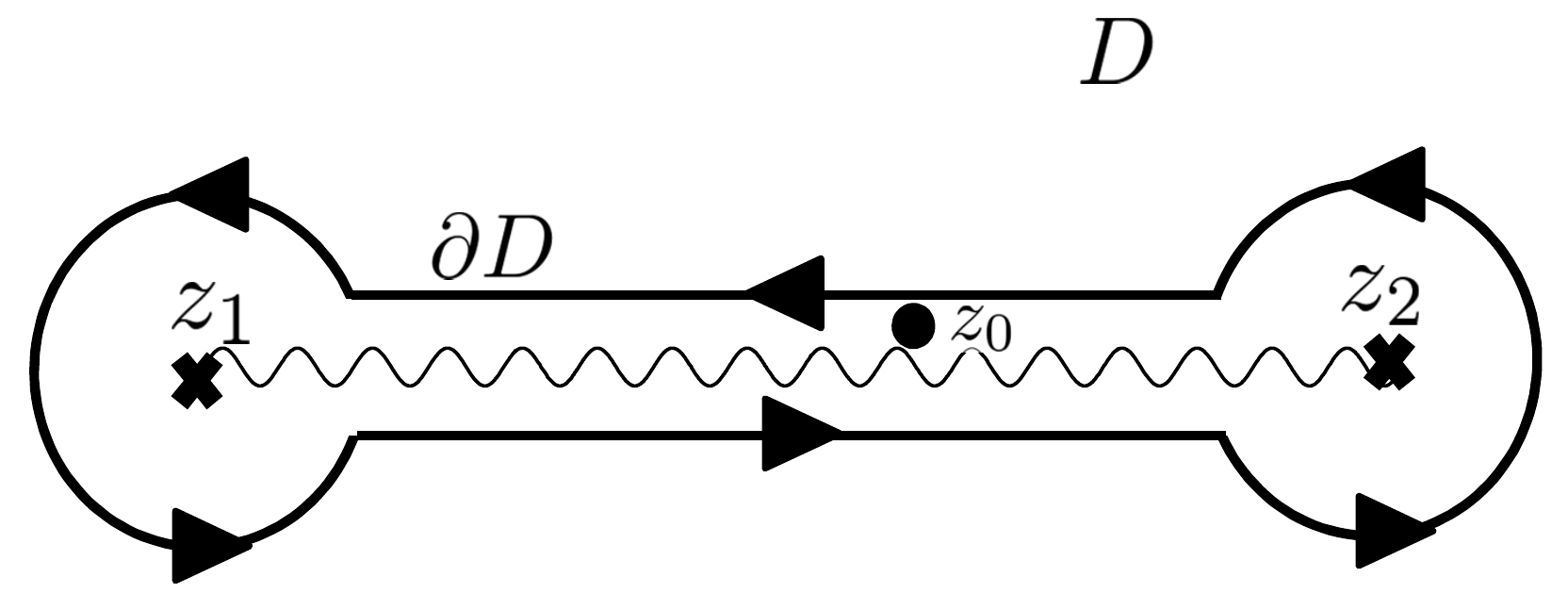}
\end{center}
\caption{$D$ and $\del D$ for  a  two point function. The wavy line represents 
 the logarithmic cut for the function $\Lam$.}
\label{contour2pt}
\end{figure}

First, in the case of a two point function, $\Lam$ has one logarithmic branch cut 
which connects  the two singularities and no square root cut. Therefore, 
we should choose $\del D$ as a contour encircling two singularities and the log 
 branch cut and take  $D$ to be the domain outside, as in figure \ref{contour2pt}. 

\begin{figure}[htbp]
 \begin{minipage}{0.5\hsize}
  \begin{center}
   \picture{height=4.5cm,clip}{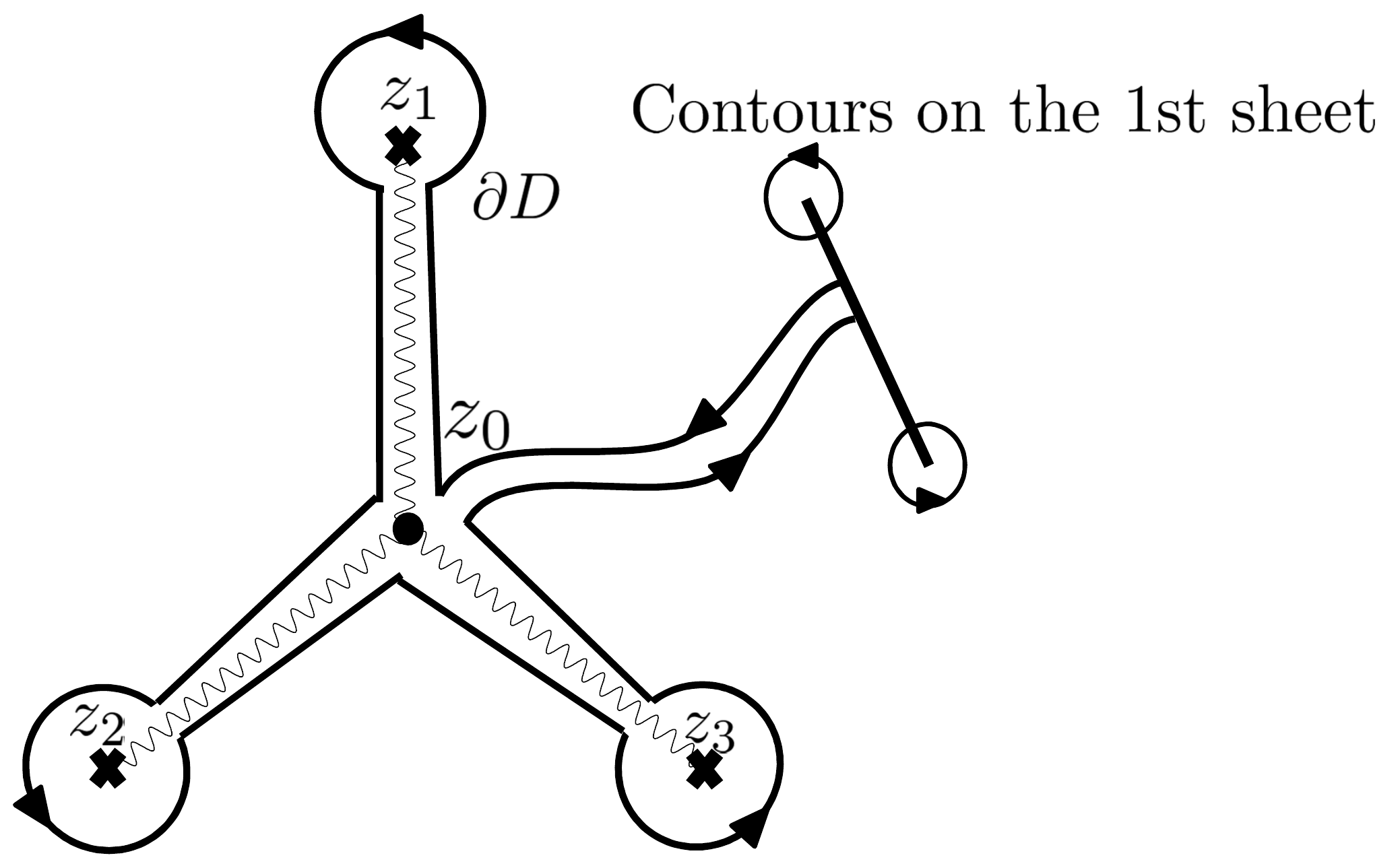}
  \end{center}
 \end{minipage}
\hspace{1cm}
\begin{minipage}{0.5\hsize}
  \begin{center}
   \picture{height=4.5cm,clip}{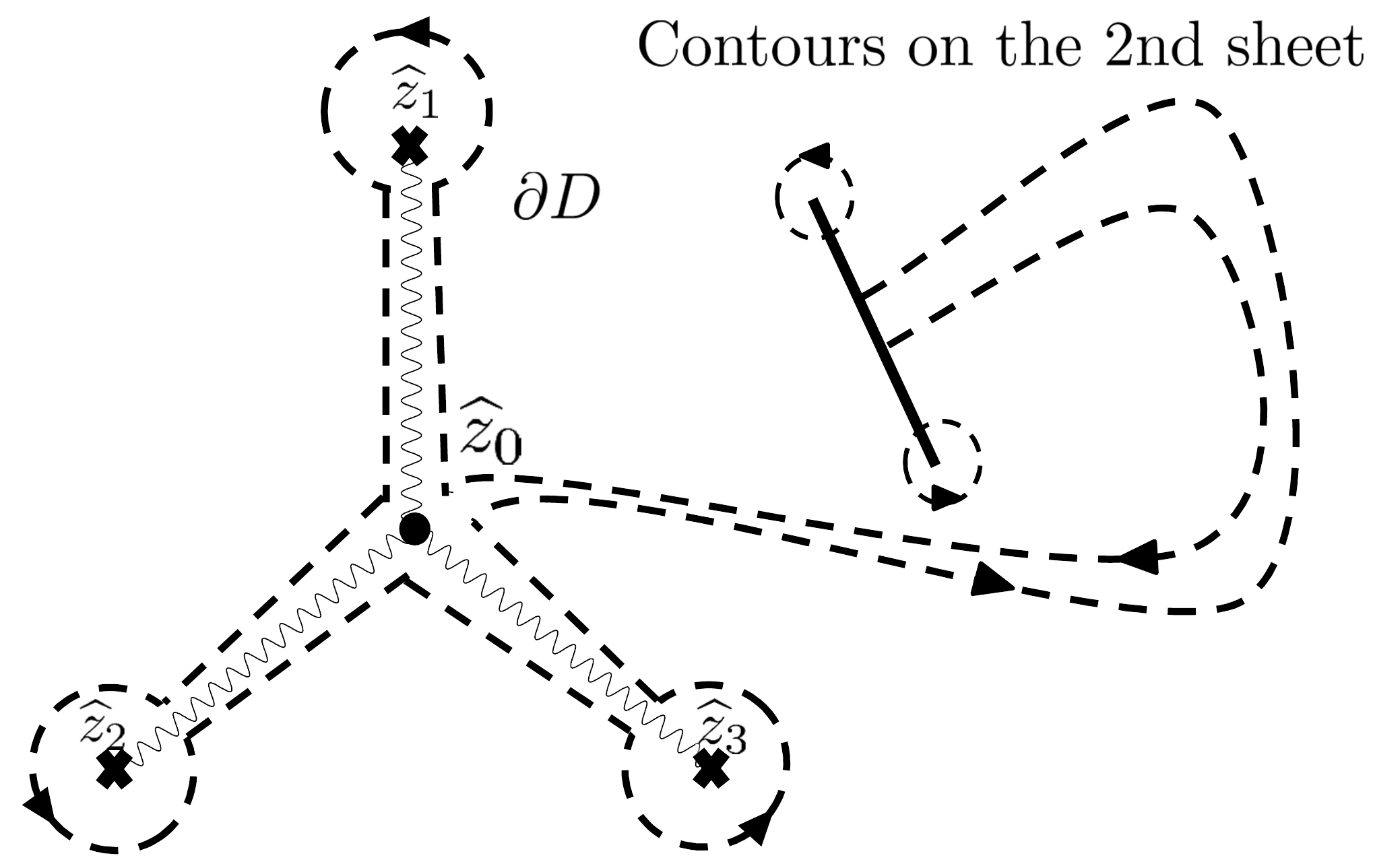}
  \end{center}
 \end{minipage}
\caption{$D$ and $\del D$ for a three point function. Contours on the first sheet 
 are drawn as  solid lines on the left, while the contours on the second sheet
 are shown as dashed lines on the right. } 
\label{contour3pt}
\end{figure}

Next consider the case of a three point function. For later purpose, let us display 
 the explicit form of $p(z)$, which is completely fixed  in terms of $\delta_i$ 
alone:
\begin{align}
p(z) &=  \left( {\delta_1^2 z_{12} z_{13} \over z-z_1}
 +  {\delta_2^2 z_{21} z_{23} \over z-z_2} + {\delta_3^2 z_{31} z_{32} \over z-z_3}\right) {1\over (z-z_1)(z-z_2)(z-z_3)} \comma \label{pz3pt}\\
z_{ij} &\equiv z_i-z_j \period \nn
\end{align}
In this case, $\Lam$ has three
 logarithmic branch cuts running out from the singularities and 
one square root branch cut.  
 We choose the positions of logarithmic branch cuts so that they end 
 at a single point on each sheet.  For later convenience, on the first sheet 
we choose this point to be  $z_0$, which is the initial point of the integral 
in  the definition of $\Lambda$ (\ref{defLam}), while on the second sheet
 the point is chosen to be  $\widehat{z}_0$, right below  $z_0$. 
Then, $D$ and $\del D$ can be chosen as depicted in figure \ref{contour3pt}.

\begin{figure}[htbp]
 \begin{minipage}{0.5\hsize}
  \begin{center}
   \picture{height=4cm,clip}{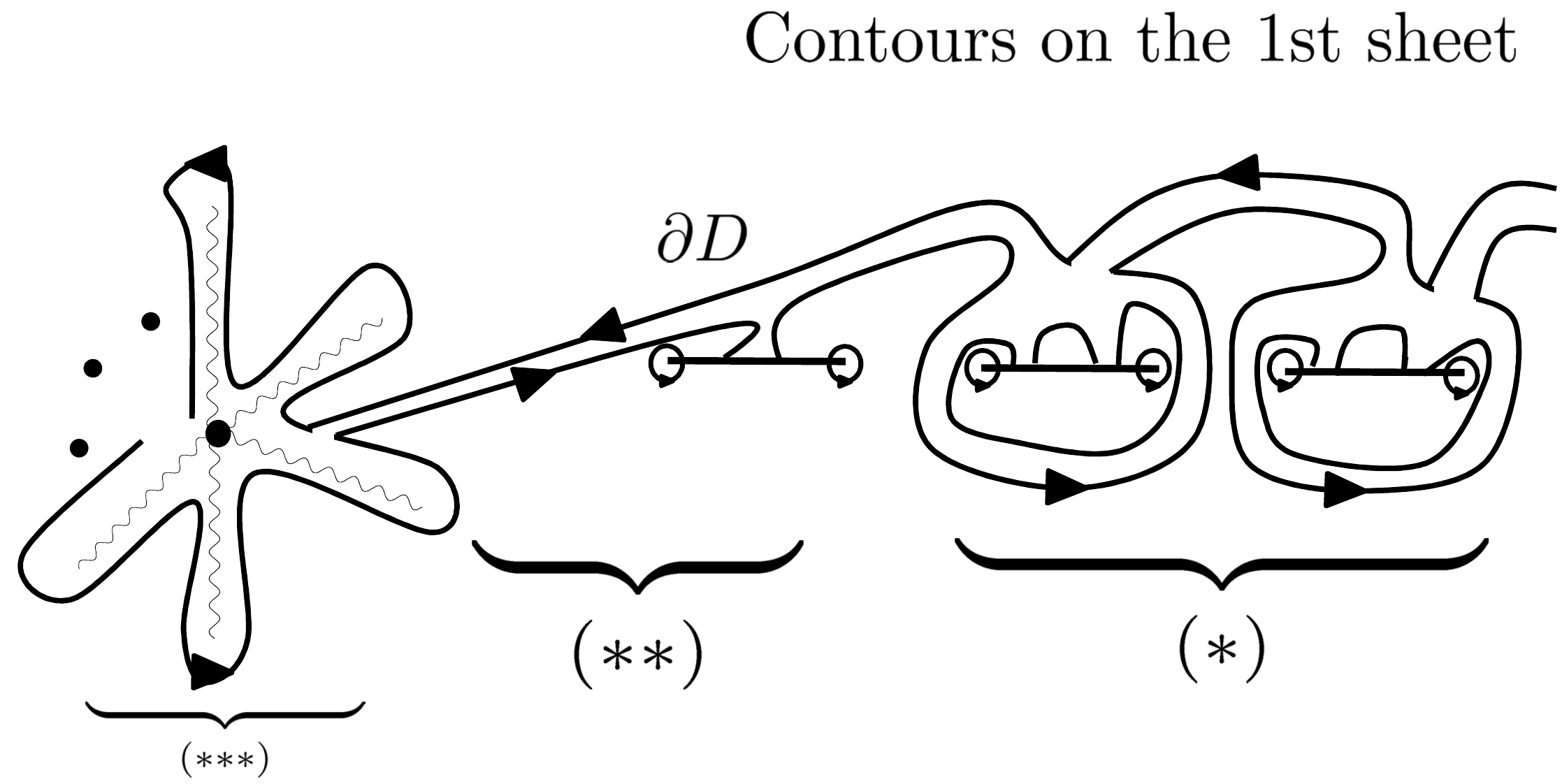}
  \end{center}
 \end{minipage}
 \begin{minipage}{0.4\hsize}
  \begin{center}
   \picture{height=4cm,clip}{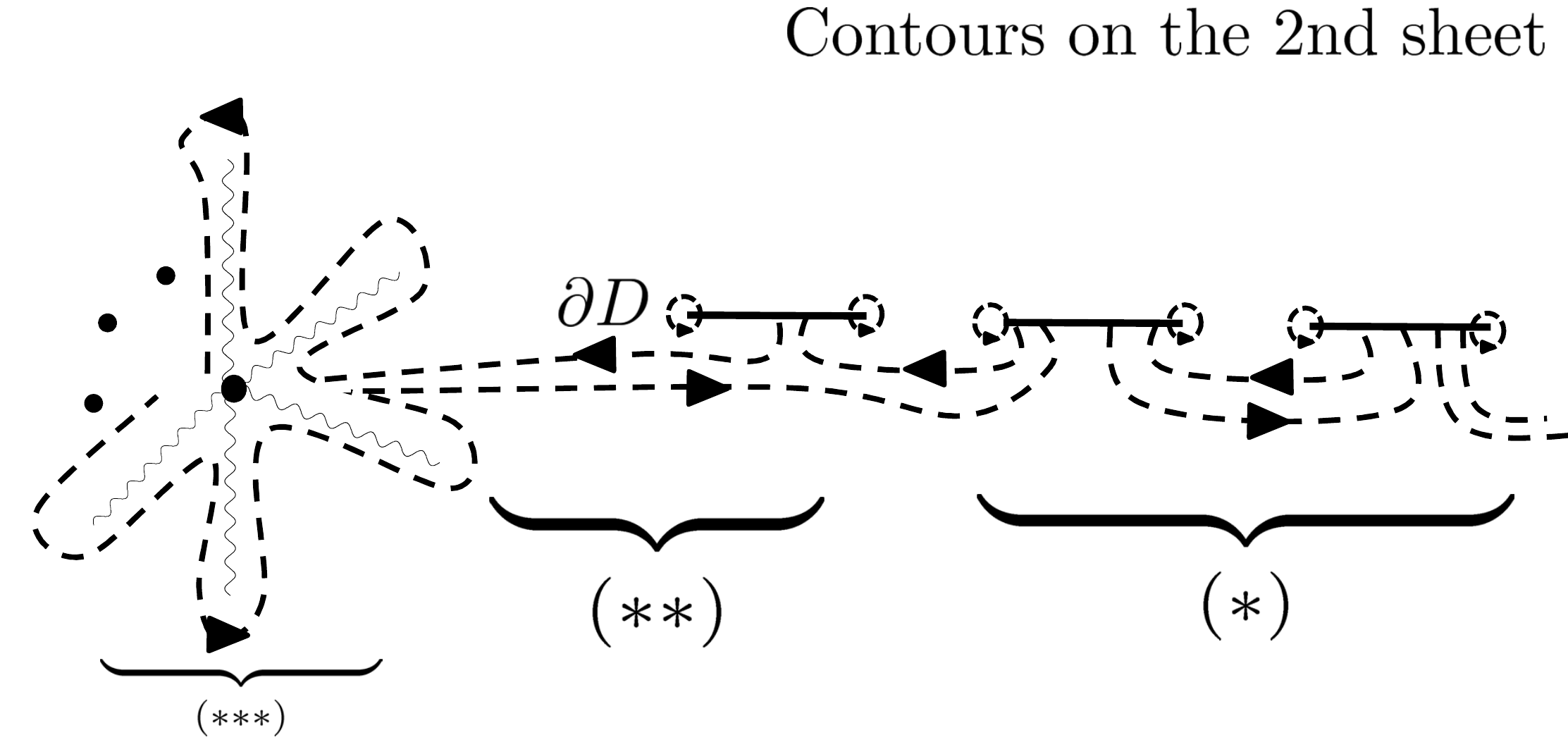}
  \end{center}
 \end{minipage}
\caption{$D$ and $\del D$ for a multi-point function} 
\label{contourNpt}
\end{figure}

Finally, for the case of $N$ point function with $N>3$, there are more 
 logarithmic as well as square root cuts. The basic idea,  
 however,  is the same. As in the case of the three point function, 
we choose the positions of logarithmic cuts so that they all end at $z_0$. 
Then $D$ and $\del D$ are  taken as shown  in figure \ref{contourNpt}.

\subsubsection{Derivation of the generalized Riemann bilinear identity}
With the preparation above, we now derive a generalization of the  Riemann bilinear identity. Basically, what we need to do is to evaluate the contributions 
 from all the portions of the boundary $\del D$, both in the first and the 
 second sheet, shown in figure \ref{contourNpt}. 
We will do this by dividing $\del D$ into several portions. 

\begin{figure}[htbp]
\begin{center}
\picture{height=2.5cm,clip}{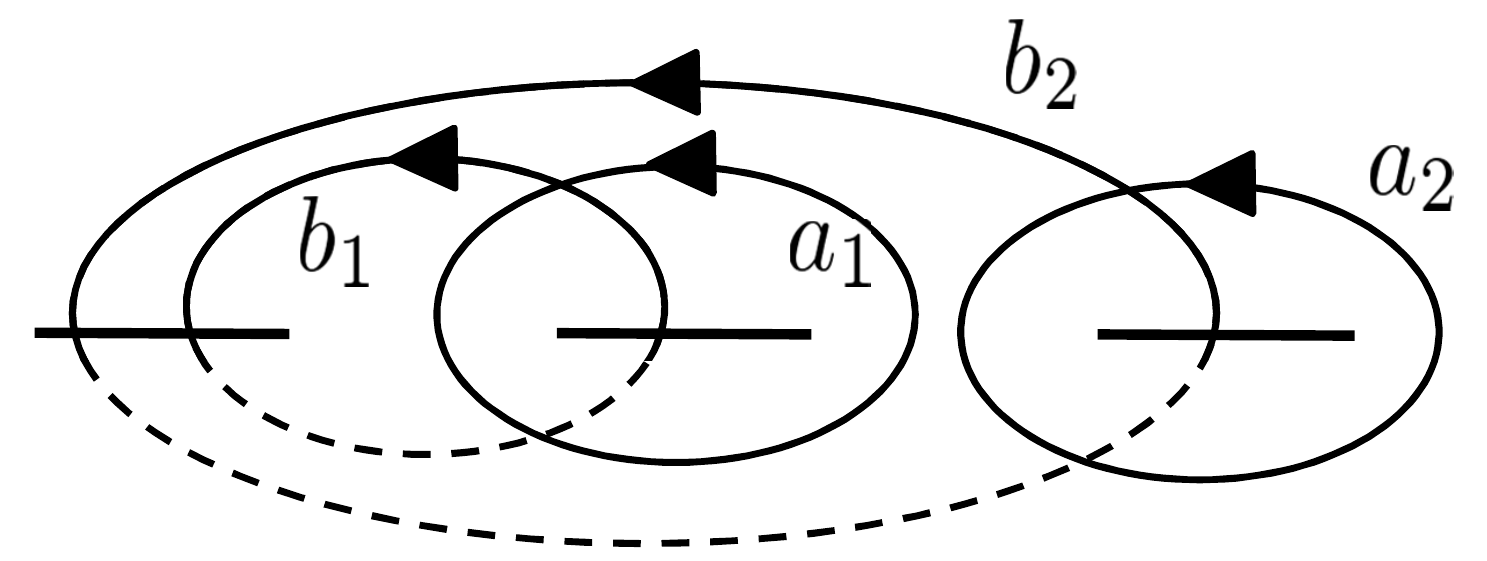}
\end{center}
\caption{Type of contours,  consisting of cycles $(a_i, b_i)$, for which 
the standard Riemann bilinear identity is applicable. }
\label{contourab}
\end{figure}

First we focus on the part consisting of contours which 
 connect or encircle the square root branch cuts, forming the right half on each sheet
 of the figure \ref{contourNpt}, $(\ast)$, and redrawn in a slightly different fashion in figure
\ref{contourab}.  As one can recognize immediately,  they are precisely the 
 contours which appear in the standard Riemann bilinear identity and 
 hence the  contribution can be written as 
\beq{
\frac{i}{4}\left( \sum_{i}\oint_{a_i} \lambda dz \oint_{b_i} (ud\barz +vdz)-\oint_{b_i} \lambda dz \oint_{a_i} (ud\barz +vdz)\right)\comma\label{cont1}}
where the cycles $(a_i,b_i)$ are defined as  in figure \ref{contourab}.
This type of contribution, however, is absent for three point functions since 
 there is only one square root cut. 

\begin{figure}[h]
 \begin{minipage}{0.5\hsize}
  \begin{center}
   \picture{height=4cm,clip}{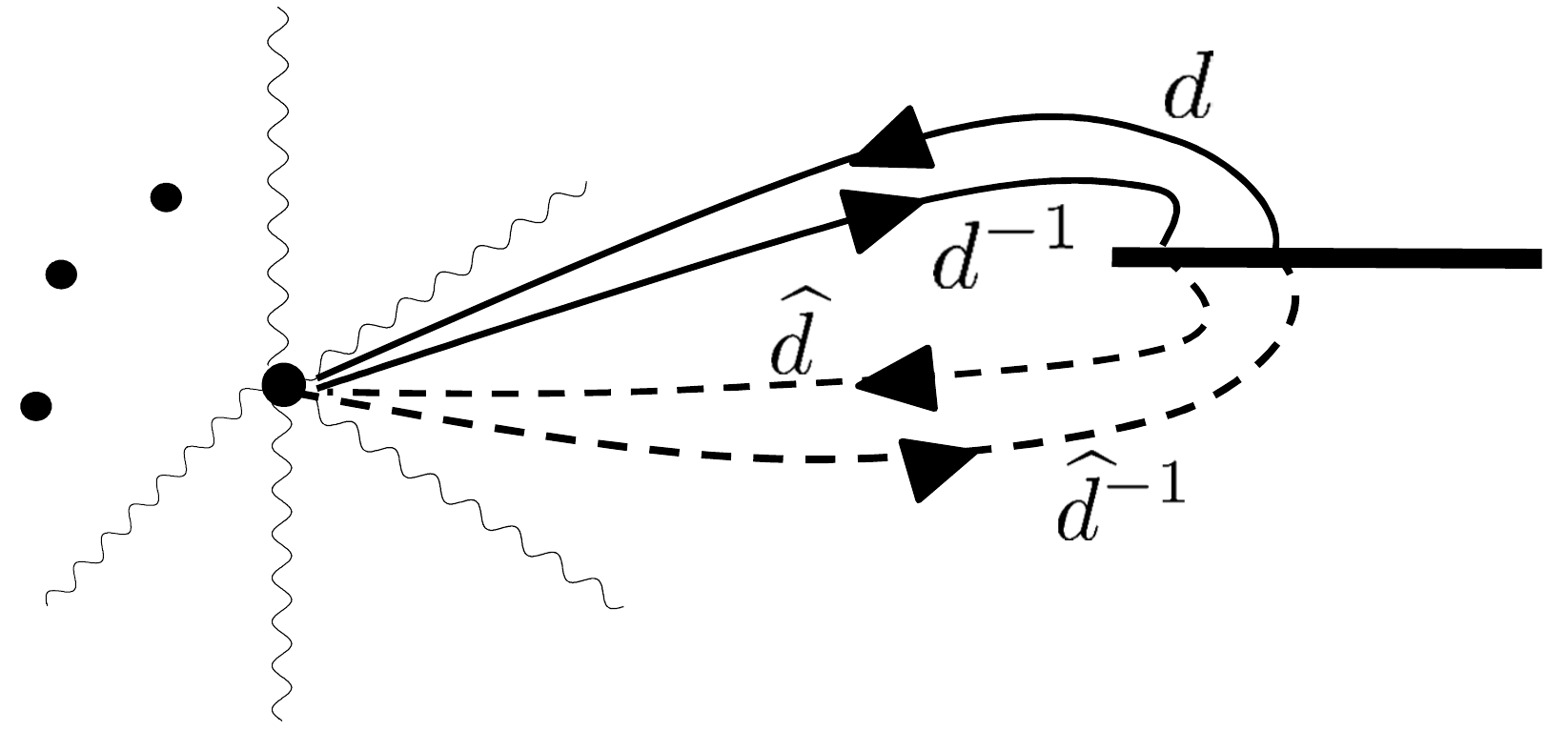}\\
$(a)$  
\end{center}
 \end{minipage} 
\begin{minipage}{0.5\hsize}
  \begin{center}
   \picture{height=4cm,clip}{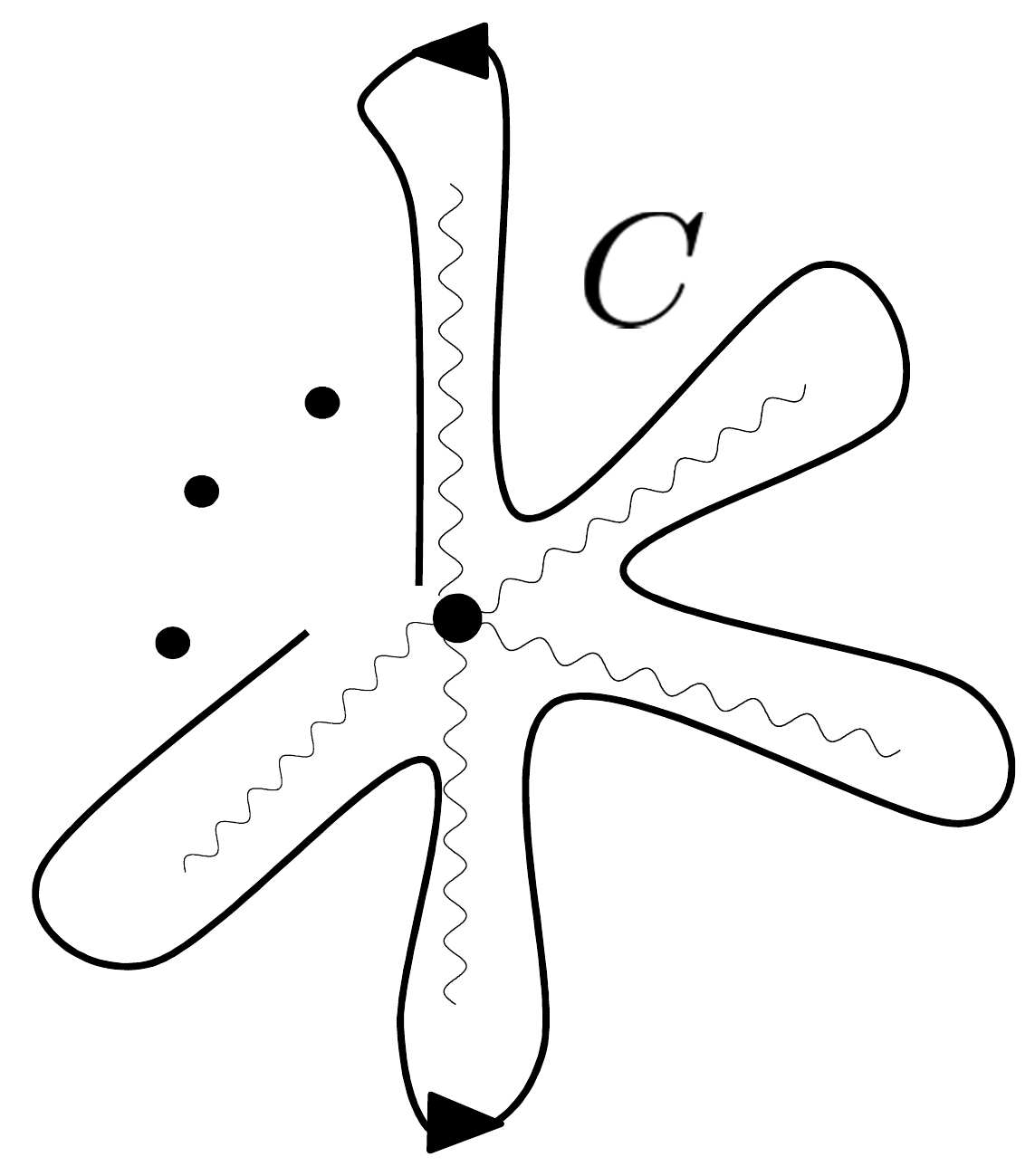}\\
$(b)$  
\end{center}
 \end{minipage}
\caption{Definitions of the contours $d, d^{-1},\hat{d}, \hat{d}^{-1}$ 
and $C$. $d^{-1}$ denotes a path parallel to  $d$ but running in opposite direction. Hatted paths run on the second sheet.} 
\label{contourdC}
\end{figure}

Next, we consider the portion which  connects  the leftmost square root  cut shown in figure \ref{contourNpt} and the point $z_0$ located 
at the junction of  the logarithmic cuts ($(\ast\ast)$ in figure \ref{contourNpt}).  Along this line the quantity 
 $ud\zbar + vdz$ is single-valued but the function $\Lam$ has different 
 values along the two opposite sides (depicted by double lines 
 in figure \ref{contourNpt}). The difference of these values is given by the integral of $\Lam$
 along the contour which connects these two sides in $D$,  namely the one 
 going around the logarithmic cuts. Thus, the contribution can be written as 
\beq{
\frac{i}{4}\oint _{C} \lambda dz\oint _{d} (ud\barz +vdz)\comma  \label{cont2.0}
}
where the contour $d$ and $C$ are shown in figure \ref{contourdC}. The contribution from $\hat{d}$ on the second sheet can be obtained by substituting $d$ and $C$ in (\ref{cont2.0}) with $\widehat{d}$ and $\hat{C}$. This turns out to be the same as to (\ref{cont2.0}) since the only difference between two sheets is the sign of $\lambda\,dz$ and $ud\barz +vdz$. Therefore, the net contribution is
\beq{
\frac{i}{2}\oint _{C} \lambda dz\oint _{d} (ud\barz +vdz)\period  \label{cont2}
} 

\begin{figure}[h]
\begin{minipage}{0.5\hsize}
\begin{center}
\picture{height=4.5cm,clip}{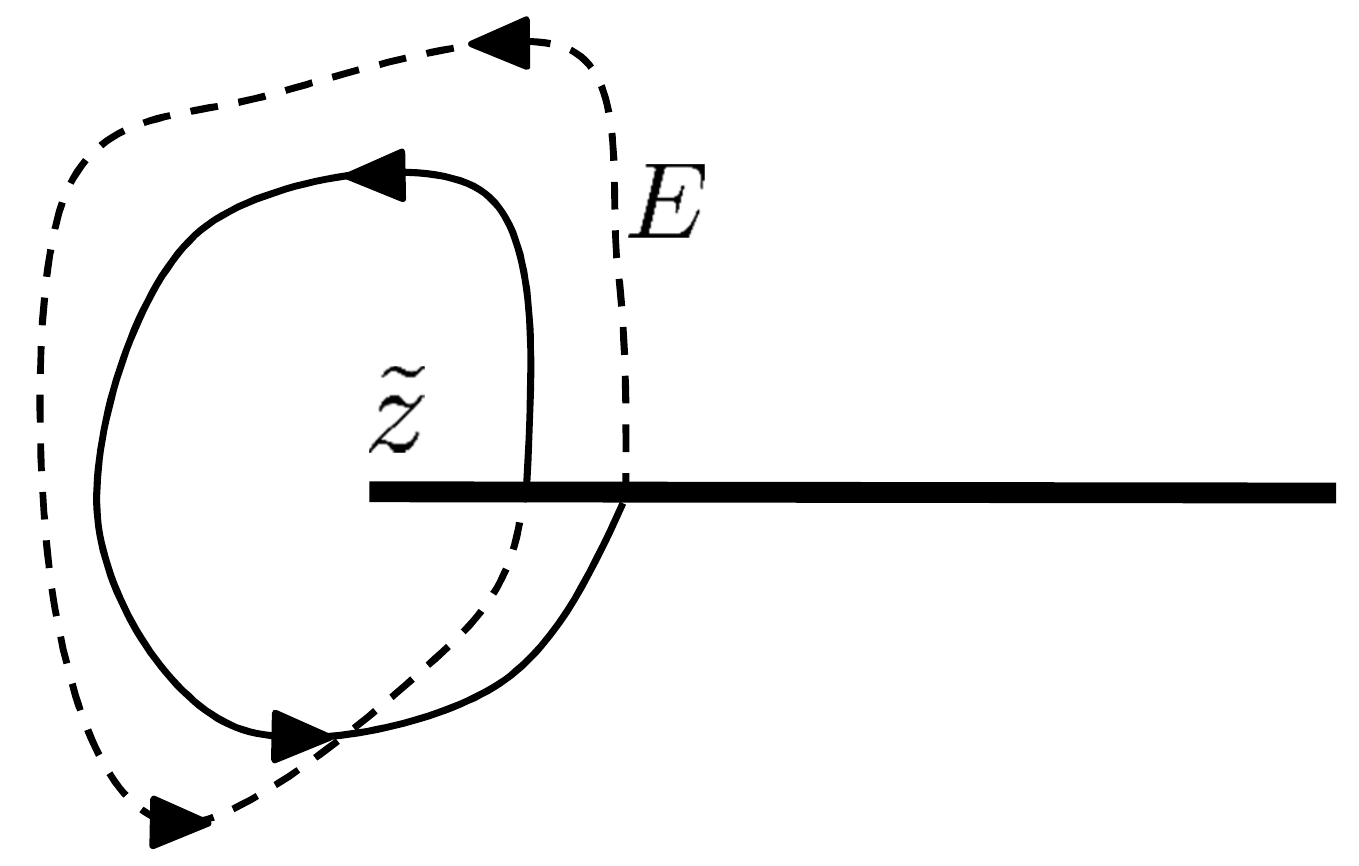}
\end{center}
\caption{Contours encircling a zero of $p(z)$}
\label{contourE}
\end{minipage}
\hspace{11pt}
\begin{minipage}{0.5\hsize}
\begin{center}
\picture{height=4.5cm, clip}{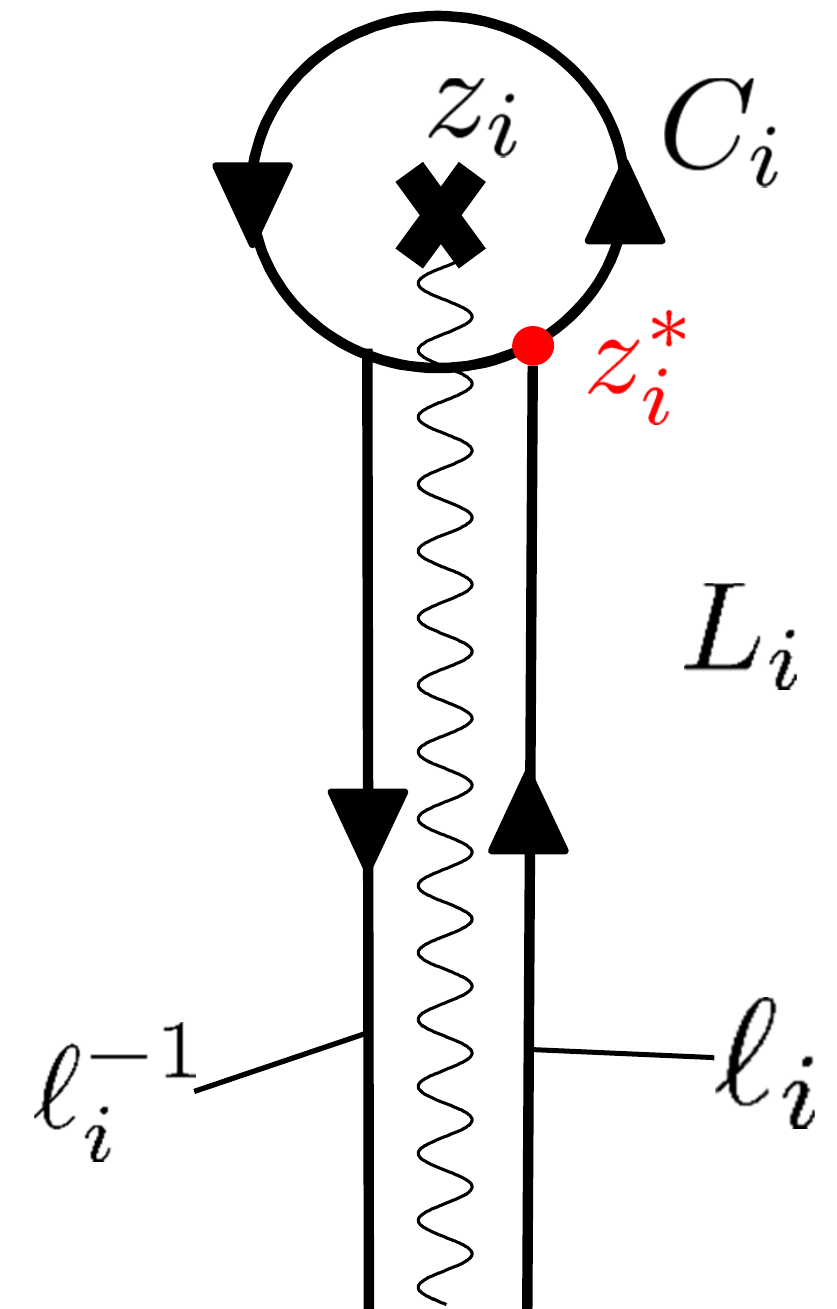}
\end{center}
\caption{Contours $C_i$, $L_i$ and the point $z_i^\ast$. }
\label{contourliCi}
\end{minipage}
\end{figure}

Thirdly, we consider the contribution from contours which encircle the zeros of $p(z)$ as shown in figure \ref{contourE}. This contribution comes from singularities of $v= \frac{1}{\sqrt{p}}(\del\hat{\alpha})^2$ at zeros of $p(z)$. In the vicinity of a zero of $p(z)$, $v$ can be approximated by
\beq{
v= \frac{1}{\sqrt{p}}(\del\hat{\alpha})^2 \sim \frac{1}{\tilde{c}\sqrt{z-\tilde{z}}}\left( \del \alpha -\frac{1}{4(z-\tilde{z})}\right)^2 \comma
} 
where $\tilde{z}$ is the position of the zero and $\tilde{c}$ is the constant which we do not need to specify here.
Since $\alpha$ is not singular at zeros, the leading singularity around the zero is given by
\beq{
\frac{1}{16\tilde{c}(z-\tilde{z})^{5/2}}\period
}
It turns out that this leading part gives a finite contribution as we show below,  
while  the contributions from the sub-leading terms vanish.
The integral on the contour  $E$  encircling $\tilde{z}$ (see figure \ref{contourE}) can be evaluated as
\beq{
\frac{i}{4}\int _{E} \Lambda (ud\barz +vdz) &=\frac{i\Lambda (\tilde{z})}{4}\oint _{E}dz\, \frac{1}{16\tilde{c}(z-\tilde{z})^{5/2} }+  \frac{i}{4}\oint _{E} dz\,\frac{1}{16(z-\tilde{z})^{5/2}}\int _{\tilde{z}}^{z}dz^{\prime}\, (z^{\prime}-\tilde{z})^{1/2} \comma \nn\\
&= -\frac{\pi}{24} \comma 
}
where the first term, which contains $\tilde{c}$, vanishes upon performing 
 the integration. 
Since all $2N-4$ zeros give the same result, the net contribution is a constant which depends on $N$:
\beq{
\frac{\pi}{12}(N-2) \period\label{cont7}
}

The remaining contribution is  from the contour 
encircling the logarithmic cuts ($(\ast\ast\ast)$ in figure \ref{contourNpt}). As a  contour this is the same as $C$ used in (\ref{cont1}). There it appeared effectively to express  the difference of the values 
 of $\Lam$ along $d$ and $d^{-1}$. Here, on the other hand, it is a  genuine portion 
 of $\del D$ along which to compute the integral (\ref{Aregint}). 

To evaluate this contribution, it is convenient to decompose $C$ into 
 three portions $C_1+L_1, C_2+L_2, C_3+L_3$, where  $C_i$ denotes  a 
 small  circle  around $z_i$ and $L_i$ consists of a line $l_i$ starting from  $z_0$
 and ending on  $z_i$ and the return path $l_i^{-1}$ from $z_i$ to $z_0$ 
in  opposite direction. (See figure \ref{contourliCi}.)  In what follows, we will describe the contributions from 
the above contours on the first  sheet. The corresponding contributions  from the second sheet can be obtained  similarly, as will be briefly indicated later. 
Consider first the contribution from $L_i$.  In exactly the same way as 
 in the previous analysis of the contour $d$, the difference of the value of $\Lam$
 along $\ell_i$ and $\ell^{-1}_i$ is represented by the integral $\oint_{C_i}\lam dz$
 and hence the net contribution from $L_i$ can be evaluated as 
\beq{
\frac{i}{4}\oint _{C_i}\lambda dz\int_{\ell_i} (ud\barz +vdz)\period \label{cont3}}  

Next, to evaluate the contribution from the circle $C_i$,  it is convenient 
 to  re-express $\Lambda(z)$ around $z_i$ as 
\beq{
\Lambda (z) = \Lambda (z^{\ast}_i)+\int^z_{z_i ^{\ast}} \lambda  dz' \comma
} 
where $z_i^{\ast}$ is the initial point of the contour $C_i$ as shown in figure \ref{contourliCi}. 
Then, the contribution from $C_i$ can be evaluated as
\beq{
&-\frac{i}{4}\oint _{C_i}\Lambda \, (ud\barz +vdz) 
=-\frac{i}{4}\Lambda (z_i^{\ast})\oint _{C_i} (ud\barz +vdz)-\frac{i}{4}\oint _{C_i}(ud\barz +vdz)\int^{z}_{z_{i}^{\ast}}\lambda dz^{\prime} \period\label{cont4}
}
Now note that $\Lambda(z_{i}^{\ast})$  can be written in the 
 following way, starting from its value at $z_0$ and integrating along 
the contour from $z_0$ to $z_i^\ast$:
\beq{
\Lambda (z_i^{\ast})=\Lambda (z_0)+\int _{\ell_i}\lambda dz+\sum _{j<i}\oint_{C_j}\lambda dz\period 
} 
The last term signifies the contributions from around $C_j$'s  which are encountered 
 on the way from $z_0$ to $z_i^\ast$. Using this expression in (\ref{cont4}) and 
summing over the contributions from all $C_i$'s, we 
 arrive at   the following simpler expression:
 \beq{
 &-\frac{i}{4} \Lambda (z_0) \oint_{\sum_i C_i(=C)}(ud\barz +vdz)-\frac{i}{4}\sum _{i}\int_{\ell_i}\lambda dz\oint_{C_i}(ud\barz +vdz) \nn\\
&-\frac{i}{8}\sum_{i\neq j} \oint_{C_{i}} \lambda dz \oint _{C_{j}} (ud\barz +vdz)-\frac{i}{4}\sum_{i}\oint _{C_i}(ud\barz +vdz)\int^{z}_{z_{i}^{\ast}}\lambda dz^{\prime}
\period\label{cont5}
 }

So far, we have only considered the contribution from the first sheet. However, the  expression we obtained (\ref{cont3}), (\ref{cont5}) can be easily generalized to the contours on the second sheet, since the only difference  is the sign of $\lambda dz$ and $ud \barz+vdz$. Thus, while (\ref{cont3})  will not be changed on the second sheet, the result   (\ref{cont5})  will be modified  as
\beq{
&\frac{i}{4} \Lambda (\widehat{z}_0) \oint_{C}(ud\barz +vdz) -\frac{i}{4}\sum _{i}\int_{\ell_i}\lambda dz\oint_{C_i}(ud\barz +vdz)\nn\\
&-\frac{i}{8}\sum_{i\neq j} \oint_{C_{i}} \lambda dz \oint _{C_{j}} (ud\barz +vdz)-\frac{i}{4}\sum_{i}\oint _{C_i}(ud\barz +vdz)\int^{z}_{z_{i}^{\ast}}\lambda dz^{\prime}\period \label{cont6}
}

This completes the evaluation of contributions coming from  various 
 parts of the contour along $\del D$. Gathering all the results obtained above, namely 
 (\ref{cont1}), (\ref{cont2}),(\ref{cont7}), (\ref{cont3}), (\ref{cont5}) and (\ref{cont6}), 
we can now write down the  generalized Riemann bilinear identity, 
which expresses  the integral  (\ref{surface}) over the worldsheet in terms of 
line integrals.  To exhibit it compactly,  let us introduce the following notion:
\beq{
\jump{ A,B}\equiv \int_{A} \lambda dz \int_{B} (ud\barz +vdz)\period
}
Then, the generalized identity takes the form\footnote{
The equation (3.38) in  the previous version of this paper contained an 
 error due to  an incorrect choice of the contour. This is fixed 
 in the equation below. See Erratum \cite{Erratum} for further explanation of the correction.} 
\begin{align}
\int d^2z \lambda \,  u 
&=\frac{\pi}{12}(N-2)+\frac{i}{4}\left( 2(\jump{ C,d}- \jump{d,C} ) +\sum_{i}\jump{ C_i,C_i}+\sum_i ( \jump{ a_i,b_i}-\jump{b_i,a_i}) \right.\nn\\
&\left.+
2\sum _{j} (\jump{ C_j,\ell _j}- \jump{ \ell _j ,C_j } )\right)
-\frac{i}{2}\sum_{i}\oint _{C_i}(ud\barz +vdz)\int^{z}_{z_{i}^{\ast}}\lambda dz^{\prime} \period  \label{riemann}
\end{align}
Note the following feature: Except for the  terms in the last sum, 
each term on the 
right hand side is of the form of a product of two contour integrals, just like  the standard bilinear identity.  Unfortunately, the last sum  consists of 
 undecoupled double integrals which  apparently have  no simple expressions  in terms of some contour  integrals of $\lam dz$  and/or  $ud\barz+vdz$. 
\subsubsection{Simplification for the LSGKP string}
\begin{figure}
\begin{minipage}{0.3\hsize}
  \begin{center}
   \picture{clip,height=3.7cm}{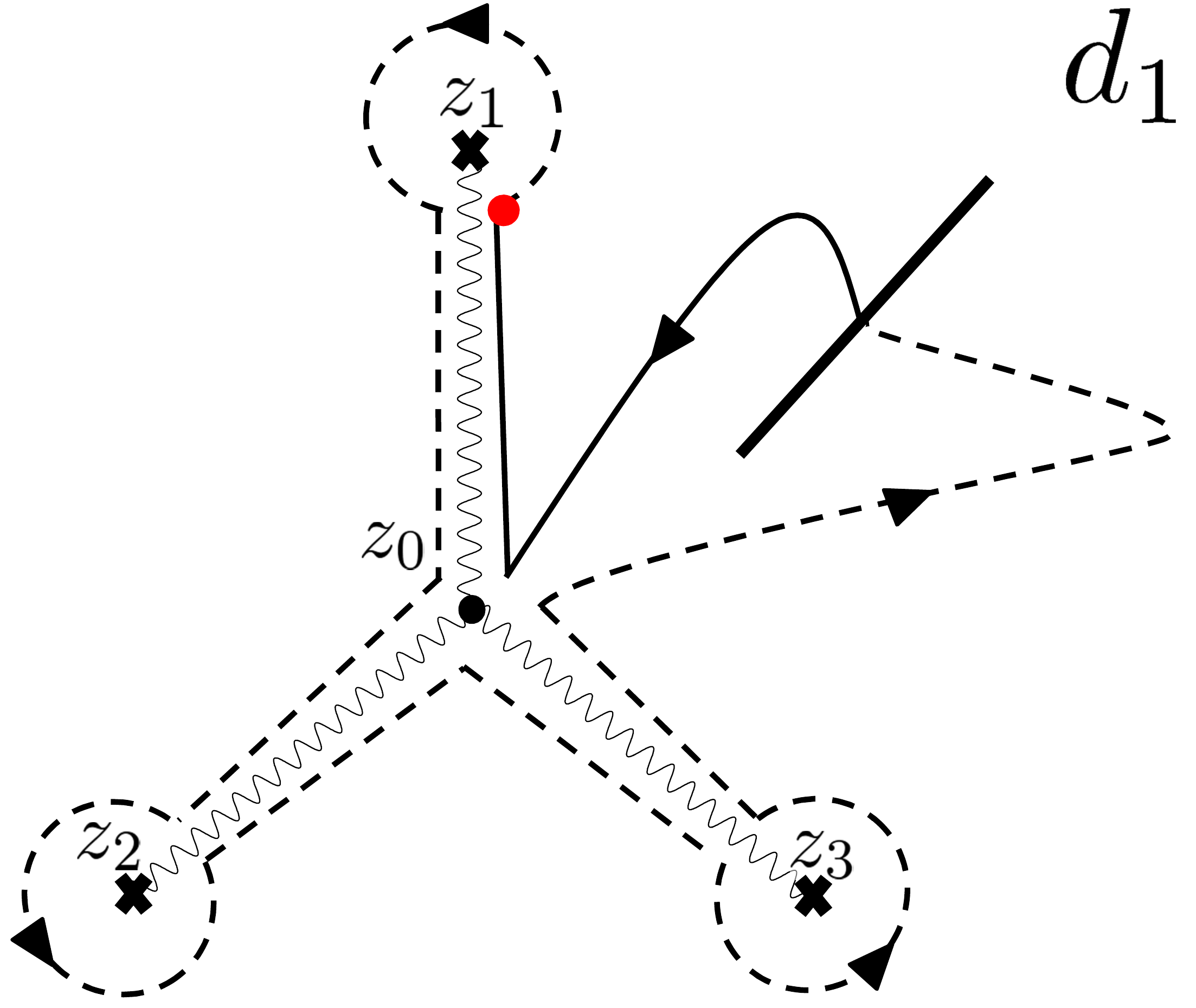}
  \end{center}
 \end{minipage}
 \begin{minipage}{0.3\hsize}
  \begin{center}
   \picture{height=3.7cm,clip}{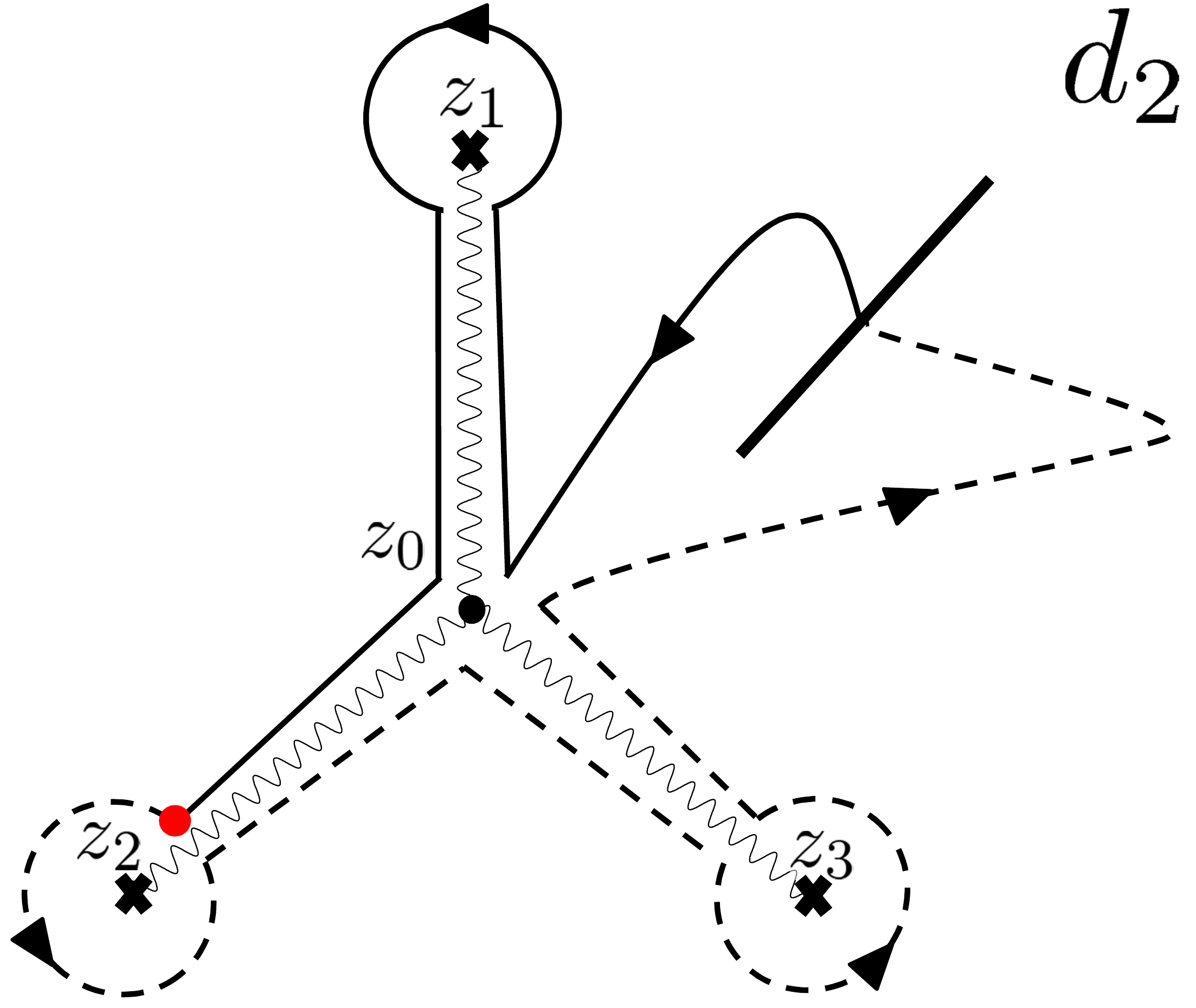}
  \end{center}
 \end{minipage}
\begin{minipage}{0.3\hsize}
  \begin{center}
   \picture{height=3.7cm,clip}{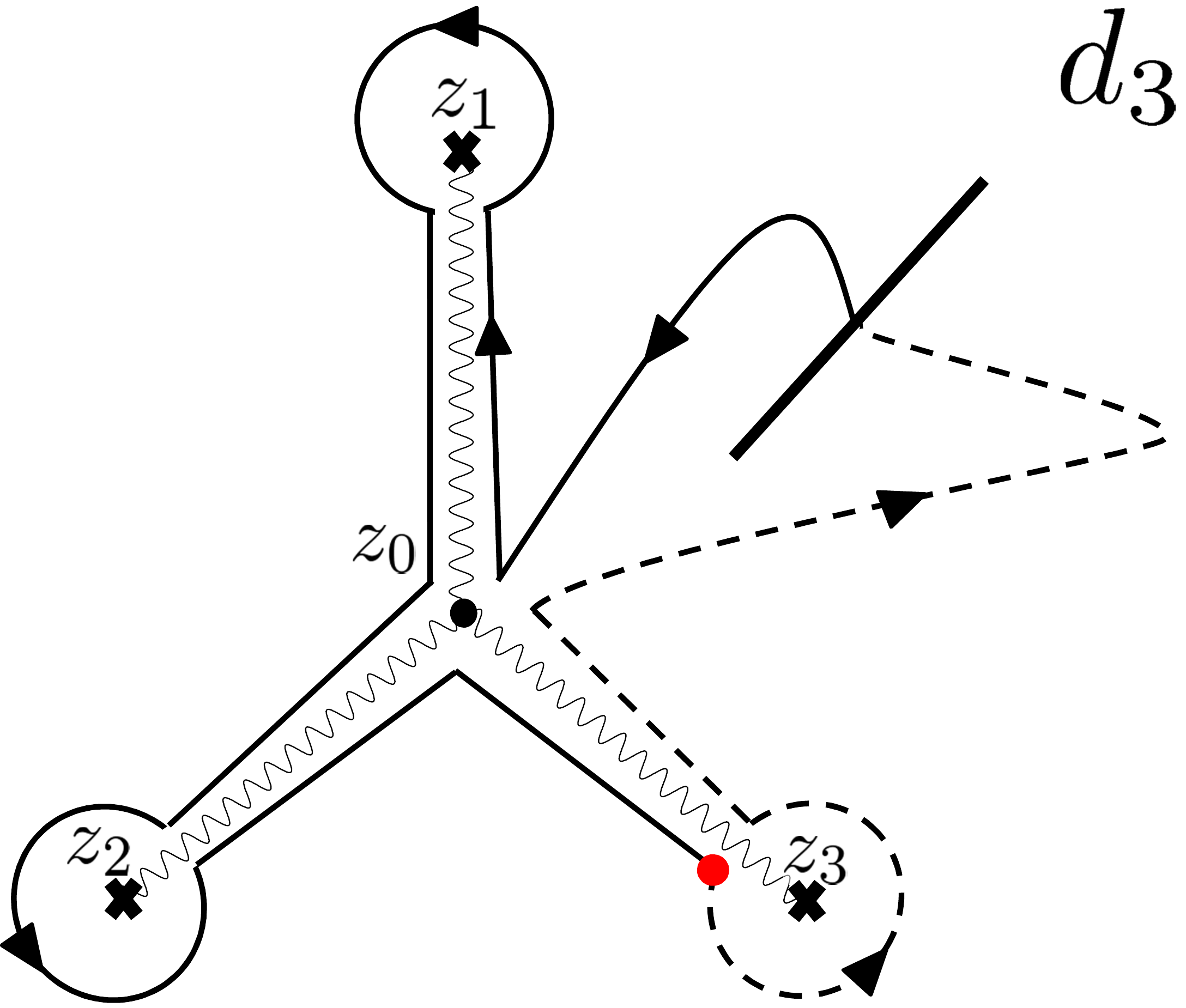}
  \end{center}
 \end{minipage}
\caption{ Definitions of the contours $d_1, d_2$ and $d_3$. } 
\label{contourdj}
\end{figure}

Because the contributions from different types of  branch cuts are properly taken 
 into account, the generalized Riemann bilinear identity derived above appears to be 
somewhat  complicated in form.  However, when applied to 
 the case of LSGKP strings, the identity simplifies significantly. The main reason 
 is that the function $\alhat$, which appears in the definitions of $u$ and $v$, 
vanishes together with its derivative $\del \alhat$  near each singularity $z_i$. 
Therefore  $ud\zbar + vdz$ vanishes at these points and hence 
all the terms which contain the factor  $\oint _{C_i}(ud\barz+vdz)$
  vanish.  The last term of (\ref{riemann}), which is still a double  integral, 
  also vanishes since the integral over $z$, which involves the factor 
$ud\zbar + vdz$, is around  $C_i$.  Thus the identity (\ref{riemann}) reduces, for 
 LSGKP strings, to 
\beq{
\int d^2z \lambda \,  u 
&=\frac{\pi}{12}(N-2)+\frac{i}{4}\left( 2\jump{C,d}+\sum_i ( \jump{a_i,b_i}-\jump{b_i,a_i} ) +
2\sum _{j} \jump{ C_j,\ell _j}\right) \period \label{riemann2}}
Moreover, in the case of three point functions, further simplification of 
 the formula occurs as there is only one square root cut in the double cover 
 of the worldsheet. This means that the terms involving the contour 
 integrals over the cycles $a_i$ and $b_i$ are absent and the 
 identity becomes 
\beq{
\int d^2z \lambda \, u=\frac{\pi}{12}+\frac{i}{2}\left( \jump{C,d}+
\sum _{j} \jump{ C_j,\ell _j}\right)\period \label{riemann3}
}
For later convenience we reexpress (\ref{riemann3}) as
\beq{
\int d^2z \lambda \, u=\frac{\pi}{12}+\frac{i}{4}\left( \sum _j \jump{C_j,d_j}\right)\comma\label{riemann4}
}
where $d_j$ is a contour which starts from $\widehat{z}_j$, goes around the logarithmic cuts counterclockwise and reaches $z_j$ as shown in figure \ref{contourdj}. To see the equality of   (3.40)  and (3.41), 
simply express   $C$ and $d_j$'s in terms of ``components"
 such as $C_i, \hat{C}_i, l_i, \hat{l}_i$  and use the fact that 
$\int_{C_i} (u d\zbar + vdz)$ vanishes in the case of  GKP strings. 

\subsection{Monodromy matrices and eigenfunctions for auxiliary linear problems}
In order to evaluate the regularized area using the generalized Riemann 
 bilinear identity just derived, we need to know the behavior of the solutions 
 of the \alp in which  the function $p(z)$ has three double pole singularities.
In this subsection, we will extract some important information 
 from the so-called  monodromy properties. 
   As the \alp is a system of first order  
linear  differential equations for a 2-component vector function, there are two 
 independent solutions.  Let  $(\eta_1, \eta_2)$ be a particular basis of 
 such solutions. Then, as one analytically continues  them once 
around a singularity, 
 they  get transformed  into a linear combination  in the manner 
\beq{
\pmatrix{c}{\eta _1^{\prime}\\ \eta _2^{\prime}}=M \pmatrix{c}{\eta _1\\ \eta _2}\comma
}
where the $2\times 2$ matrix $M$ is called  the  monodromy matrix
 around that singularity. In what follows we will determine 
 the three monodromy matrices $M_i$ for the three singularities,  find 
 their eigenvectors   and compute the $SL(2,C)$ invariant products between 
 them, which will play crucial roles in the evaluation of the contour 
integrals  appearing  in (\ref{riemann4}). 
\subsubsection{Eigenvalues of monodromy matrices and ``plus" and ``minus" eigenvectors}
Since the   monodromy matrices $M_i$ belong to $SL(2,C)$, each 
 of them can be diagonalized  in the form\footnote{It is generically  not possible to 
diagonalize all the monodromy matrices simultaneously.}
\beq{
\pmatrix{cc}{e^{i\pp_i(\xi )}&0\\0&e^{-i\pp_i(\xi )}} \period 
}
The (logarithm of the)  eigenvalues  $\pp_i(\xi)$  carry  the basis-independent 
 information and will be referred to as the local monodromy data\footnote{
The quantity $\rho_i$ is called  ``pseudo momentum"  and often 
 denoted by $p_i$. In this paper, we use a different letter to distinguish
 it from the function $p(z)$.}. 
The dependence of $\pp_i(\xi)$ on the spectral parameter encodes the 
information of the (infinite number of ) conserved charges of the integrable 
 system, which includes that of the global charges such as the AdS energy $E$ 
 and the spin $S$  of  the GKP solution. Therefore the information of the 
 external string states is contained in $\pp_i(\xi)$ and the regularized area
 should be expressible  in terms of these local monodromy data.

Let us now look at the structure of the eigenvectors of the monodromy 
 matrices closely.  Near each of the singularity $z_i$, the eigenvectors 
 must behave like the external string state of interest. Since the 
LSGKP strings of our interest satisfy  the relation $e^{2\al} = \sqrt{p\pbar}$, 
 the eigenvectors near $z_i$  take the form (see (\ref{psitwoeq}))%
\beq{
\psi \sim \exp\left[ \pm\left(\frac{1}{\xi}\int \sqrt{p(z)}dz+\xi\int \sqrt{\pbar(\barz)}d\barz\right)\right] \period \label{asymsol1}
}
Here and in what follows, we define $\sqrt{p(z)}$ and $\sqrt{\pbar(\barz)}$ as the  values on the first sheet. 
Then, we can  distinguish the two independent eigenfunctions of $M_i$
 by the behavior at the point $z_i$. More precisely, 
 we call the eigenvector  $i_\pm $  (``plus" and ``minus" solution ),
 if it behaves asymptotically like 
\beq{
& \sim \exp\left[ \pm \left(\frac{1}{\xi}\int \sqrt{p(z)}dz+\xi \int \sqrt{\pbar (\barz)}d\barz\right) \right] \label{asymsol2} }
at $z_i$. 
For later convenience, we impose the normalization condition
\beq{
\braket{i_+,i_-}=1\period
}
Note that there still remains a freedom of rescaling $i_+$ and $i_-$ oppositely, 
 namely,   $i_+\to A \, i_+,\hspace{3pt}i_- \to A^{-1}   i_-$. In this description 
 the eigenvalue  of the monodromy $\pp_i(\xi)$ can be defined 
 by the phase factors acquired by $i_+$ and $i_-$  
\beq{
i_+^{\prime} = e^{i\pp _i(\xi )}i_+\comma \hspace{11pt} i_-^{\prime}=e^{-i\pp_i (\xi )}i_-\comma
}   
where $i^{\prime}_+\comma i^{\prime}_-$ are the analytic continuation 
 of  $i_+\comma i_-$  along a path which encircles the singularity counterclockwise.

We should emphasize that the notion of   ``plus" and  ``minus"  solutions
 introduced above is conceptually quite distinct from  that of  ``small" and ``big" 
 solutions that appeared in the gluon scattering problem.  ``Small" and ``big" 
refer to the corresponding behavior of the solution as one approaches 
 the singularity (which in the gluon scattering problem corresponds to 
 the segment of the polygon at infinity). Only the small solution is unambiguously 
 defined since the big solution can contain a multiple of small solution. 
On the other hand, the plus  and minus solutions are simply the convenient and 
 unambiguous characterization of the two independent eigenvectors 
 of the monodromy matrix.  Later when we perform the WKB analysis 
 of the auxiliary linear problem,  we will need the notion of small and big solutions as well. 
The precise relation between the plus-minus  solutions and the small solution
will be of utmost importance when  we  compute the contour integrals 
 in terms of  the $SL(2,C)$ invariant products in section 3.6. 
\subsubsection{Calculation of explicit forms of monodromy matrices}
Let us now determine the explicit form of the  monodromy matrices $M_i$,  up to conjugacy classes.  Since we can diagonalize any one of $M_i$, we choose 
a  basis $\{1_+, 1_-\}$, normalized as $\braket{1_+, 1_-} =1$,  such that $M_1$ is diagonal:
\beq{
M_1=\pmatrix{cc}{e^{i\pp_1(\xi)}&0\\0&e^{-i\pp_1(\xi)}}\period 
}
We then write $M_2$ in the form 
\beq{
M_2 = \pmatrix{cc}{a&b\\c&d}\comma}
where,  from the $SL(2,C)$ and the trace properties, the components $a$ to $d$
 satisfy the following relations
\beq{
ad-bc &=1\comma \label{monod0}  \\
 a+d &= e^{i\pp_2(\xi)}+e^{-i\pp_2(\xi)}\period\label{monod1}
}
Since the monodromy around all the singularities must be trivial,  we 
 must have the relation 
\beq{
M_1  M_2  M_3 =1\period
}
This determines $M_3$  as $M_3=(M_1M_2)^{-1}$ and taking the trace of 
 this relation we get
\beq{
e^{i\pp_1(\xi)} a + e^{-i\pp_1(\xi)} d = e^{i\pp_3(\xi)}+e^{-i\pp_3(\xi)}\period\label{monod2}
}
The relations (\ref{monod0}), (\ref{monod1}) and (\ref{monod2})
 provide  three  equations for four variables $a,b,c,d$. However, as 
 $b$ and $c$ appear in (\ref{monod0}) only in the product $bc$, 
we can completely solve for $a, d$ and $bc$ in terms of $\rho_i$. 
The result is 
\beq{
&a=\frac{\cos \pp_3 -e^{-i\pp_1}\cos \pp_2}{i \sin \pp_1}\comma\hspace{11pt} d=\frac{e^{i\pp_1}\cos \pp_2-\cos \pp_3}{i \sin \pp_1} \comma\label{ad1}\\
&bc=\frac{\cos ^2 \pp_1+\cos ^2 \pp_2+\cos ^2 \pp_3-2\cos \pp_1\cos \pp_2\cos \pp_3-1}{\sin ^2 \pp_1}\period \label{ad2}
}
\subsubsection{Eigenvectors of monodromy matrices}
Having found the form of the non-trivial monodromy matrices $M_2, M_3$,  
we now compute their eigenvectors by diagonalizing them. This amounts to 
 expressing  the eigenvectors $\{2_+\comma 2_-\}$ and $\{3_+\comma 3_-\}$ in terms of $\{1_+\comma 1_-\}$. It will allow us to compute various  $SL(2,C)$ invariant products 
 between  them, which will play  important roles later in the calculation of the 
 regularized area.

First  consider the eigenvectors $\{2_+\comma 2_-\}$ of $M_2$. 
Using the expressions of $a,d$ and $bc$ obtained in 
(\ref{ad1}) and (\ref{ad2}),  they can be computed straightforwardly 
as 
\beq{
&2_+\propto\pmatrix{c}{\frac{2i}{\sin \pp _1} \left( \cos (\pp _1-\pp _2)-\cos \pp _3\right)\\ 2c} \comma \\
&2_-\propto \pmatrix{c}{\frac{2i}{\sin \pp _1} \left( \cos (\pp _1+\pp _2)-\cos \pp _3\right)\\ 2c} \period 
}
Note that these expressions contain $c$,  the undetermined component 
 of $M_2$. This reflects the rescaling ambiguity still left in $\{1_+, 1_-\}$. 
If we further impose the normalization condition $\braket{2_+,2_-}=1$, 
the eigenvectors become 
\beq{
&2_+= A_2\pmatrix{c}{\frac{i}{\sin \pp_1} \left( \cos (\pp _1-\pp _2)-\cos \pp _3\right)\\ c}\comma\\
&2_-=\frac{1}{2A_2 \sin \pp _2} \pmatrix{c}{\frac{1}{c\sin \pp _1} \left( \cos (\pp _1+\pp _2)-\cos \pp _3\right)\\ -i}\comma 
}
where $A_2$ indicates   the remaining scaling ambiguity for $\{2_+\comma 2_-\}$. 

Next, consider  the  eigenvectors  $\{3_+\comma 3_-\}$ of the matrix
 $M_3$,  which is given by 
\beq{
&M_3=M_2^{-1}  M_1^{-1} \nn\\
&= \pmatrix{cc}{\frac{\cos \pp_2-e^{-i \pp_1}\cos \pp_3}{i \sin \pp_1}&-\frac{\cos ^2 \pp_1+\cos ^2 \pp_2+\cos ^2 \pp_3-2\cos \pp_1\cos \pp_2\cos \pp_3-1}{c\sin ^2 \pp_1}e^{-i \pp_1}\\-ce^{i\pp_1}&\frac{e^{i\pp_1}\cos \pp_3 -\cos \pp_2}{i \sin \pp_1}}\period 
} 
With the imposition of the condition $\braket{3_+,3_-}=1$ the 
 eigenvectors are obtained as
\beq{
&3_+=A_3\pmatrix{c}{\frac{ie^{-i\pp _1}}{\sin \pp _1}\left( \cos \pp _2 -\cos (\pp _1-\pp _3)\right)\\ c } \comma \\
&3_-=\frac{1}{2A_3 \sin \pp _3}\pmatrix{c}{\frac{-1}{c \sin \pp _1}\left( \cos \pp _2 -\cos (\pp _1+\pp _3)\right)\\  ie^{i\pp _1}}\comma 
}
where again $A_3$  denotes  an undetermined constant. 
\subsubsection{SL(2,C) invariant products between different eigenvectors}
With  the expressions of  the eigenvectors obtained above, we can easily compute 
 the $SL(2,C)$ invariant products between them. 
Those  between the ``plus" eigenvectors are evaluated as 
\beq{
&\braket{1_+,2_+} = c A_2 \comma \\
&\braket{2_+,3_+} = c A_2 A_3\left( e^{i(\pp _3-\pp _1)}-e^{-i\pp _2}\right)
\comma \\
&\braket{3_+,1_+}=-c A_3  \period 
}
On the other hand, the ones  between the ``minus" eigenvectors are given by 
\beq{
&\braket{1_-,2_-} = -\left( 2A_2 c \sin \pp _1\sin \pp _2\right)^{-1}\left( \cos (\pp _1+\pp _2)-\cos \pp _3\right)\comma\\
&\braket{2_-,3_-} = -\left( 4c A_2 A_3\sin \pp _2\sin \pp _3\right) ^{-1}\left(  e^{i (\pp _1+\pp _2)}-e^{-i\pp_3}\right)\comma\\
&\braket{3_-,1_-}=-\left( 2A_3 c \sin \pp _1\sin \pp _3\right)^{-1}\left( \cos \pp _2-\cos (\pp _1+\pp _3)\right) \period
}
Lastly, the products  between the``plus" and the``minus" eigenvectors 
are obtained as 
\beq{
&\braket{1_+,2_-}=-\frac{i }{2 A_2 \sin \pp_2}\comma \\
&\braket{1_-,2_+}=-\frac{iA_2}{\sin \pp_1}\left( \cos (\pp _1-\pp _2)-\cos \pp _3\right)\comma \\
&\braket{2_+,3_-}=-\frac{iA_2}{2A_3\sin \pp _3}\left( e^{i(\pp _1-\pp _2)}-e^{-i\pp _3}\right)\comma \\
&\braket{2_-,3_+}=-\frac{iA_3}{2A_2 \sin \pp _2}\left( e^{i(\pp _3-\pp _1)}-e^{i\pp _2}\right)\comma \\
&\braket{3_+,1_-}=-\frac{iA_3e^{-i \pp _1}}{\sin \pp _1}\left( \cos (\pp _1-\pp _3)-\cos \pp _2\right)\comma \\
&\braket{3_-,1_+}=-\frac{ie^{i\pp_1}}{2A_3\sin \pp _3} \period
}

Several remarks are in order. 
\begin{itemize}
	\item First,  the constants $c\comma A_2$ and $A_3$, which 
 are not yet fixed, may depend on the spectral parameter $\xi$ in  unknown ways. 
 This circumstance  would give us trouble when we later make a WKB expansion in 
$\xi$ to extract the information of the contour integrals from the  $SL(2,C)$ 
invariant products. Fortunately, this problem can be  circumvented  if we focus on 
certain combination of the products which do not depend on these constants. 
 For instance  the quantity 
\beq{
\braket{1_+,2_+}\braket{1_-,2_-}=-(2\sin \pp_1 \sin \pp_2 )^{-1}(\cos (\pp_1+\pp_2)-\cos \pp_3) \label{Aindcomb}
}
is such a combination. 
\item Second, in computing the invariant products $\braket{i,j}$,   which reflect the 
global structure of the solution,   from the local monodromy data, we have not made explicit use of the  $Z_2$ symmetry mentioned in subsection 2.1. As this symmetry stems from  the Virasoro conditions, 
our  method of analysis should be useful  even when such conditions are 
 not satisfied, for example for strings rotating in $S^5$ as well as in $AdS_5$. 
\end{itemize}
\subsection{WKB analysis of the auxiliary linear problem}
In the previous subsection, we have computed the $SL(2,C)$ invariant 
 products between the eigenvectors of the monodromy matrices
 in terms of the local monodromy data $\pp_i(\xi)$. 
These eigenvectors, called  $i_\pm$,  are classified by the 
 $\pm$ signs in the exponent of these solutions. In order to make use 
 of these  invariant products in the computation of the regularized area, 
 we must relate them to the contour integrals via WKB expansions 
 of these quantities. 

However, since the classification of the solutions 
as $i_\pm$ is not directly related to  the WKB expansion,  it will be more 
 convenient to first introduce a different classification which is more closely 
 connected to such an expansion. This is the notion of the ``small" solution 
 $s_i$ and the ``big"  solution $b_i$,  associated to each singularity  $z_i$. 
In the case of LSGKP strings,  due to the prescribed 
 behavior at the  singularities, namely $e^{2\alpha}\sim\sqrt{p\pbar}$, 
one of the two independent solutions of the \alp  must decrease, while the
 other increases,  as one approaches a singularity at $z_i$. 
 The former is called the small solution $s_i$ and 
 the latter the big solution $b_i$. Although  such  definitions  of small and big 
 solutions do not refer to the WKB expansion,  they are  actually closely tied 
 to it since the leading term of the WKB expansion near  $z_i$ 
 tells us whether it is $s_i$ or $b_i$.  The small solution is particularly important
 as it is uniquely determined. Contrarily, the big solution is ambiguous 
 as it can contain a multiple of small solution. 

In this subsection we shall 
develop  the WKB expansion of  the  $SL(2,C)$ invariant products of  small solutions $\braket{s_i, s_j}$ and observe that the same contour integrals that appear 
 in the expression of the regularized area show up  in that expansion. 
Later in subsection 3.6, we will elucidate the relation between 
 the $i_\pm$ eigenvectors and the small solutions $s_i$ and thereby 
 connect the local monodromy data to these  contour integrals. 
\subsubsection{WKB expansions }
Before we begin the WKB expansion of  the solutions of the \alp\!\!, 
 we should make a remark on the validity  of such an expansion. 
As it is a perturbative expansion in powers of $\xi$ (or $\xi^{-1}$), 
 it is merely an asymptotic expansion and hence is known to be reliable 
 only  on a set of curves called the WKB curves\cite{GMN:0907}. The WKB curves are the ones 
on the complex plane which satisfy the equation 
\beq{
{\rm Im}\, \left[ \frac{1}{\xi} \sqrt{p}\frac{dz}{dt}\right] =0 \comma}
where $t$  parametrizes the curve.  Along a WKB 
 curve, the phase of the leading part of the WKB solution is constant 
 and hence one can follow the change of the magnitude of the solution reliably. 
In what follows, we will always assume that WKB expansions are made 
along such WKB curves.  For a generic value of $\xi$, a WKB curve
 starts from a singularity and ends at another singularity. 

The auxiliary linear problem we need to solve was given in (\ref{eqalpxi}), 
namely 
\beq{
(\del + B_z (\xi ))\psi (\xi )=0\comma \hspace{11pt} (\delbar + B_{\barz} (\xi ))\psi (\xi )=0 \comma \label{wkbaux}
}
where the $\xi$-dependent flat connections $B_z(\xi), B_\zbar(\xi)$ take 
 the forms shown  in (\ref{Bxicon1})$\sim$ (\ref{Bxicon3}). 
As in the treatment of the LSGKP solution, one can simplify the equations
 by the gauge transformation discussed in (\ref{gaugetrA}), 
viz., 
\beq{
\psi = \pmatrix{cc}{p^{-1/4} e^{\alpha /2}&0\\0&p^{1/4} e^{-\alpha /2}}\tilde{\psi} \comma \qquad \psitil = \pmatrix{c}{\psitil _1 \\ \psitil _2} 
\period \label{gaugetrpsi}
}
If we further use the variable $\alhat = \al = {1\over 4} \ln p\pbar$ 
and switch to the worldsheet coordinate  $w$ defined by 
 $dw = \sqrt{p(z)} dz$, the equations for $\psitil$ simplify to 
\beq{
&\left[\del _w + \frac{1}{\xi}\pmatrix{cc}{0&-1\\-1&0}+\pmatrix{cc}{\del _w\hat{\alpha} &0\\0&-\del _w \hat{\alpha}} \right] \psitil =0 \comma &
\nonumber \\
&\label{app2}\\
&\left[\delbar _w + \xi\pmatrix{cc}{0&-e^{-2\hat{\alpha}}\\- e^{2\hat{\alpha}}&0} \right] \psitil =0 \period &   \nonumber 
}
To solve these equations in powers of $\xi$,  it is convenient to first analyze 
the  second order equations for $\psitil_1$,  obtained by 
 combining the above first-order equations,  and then compute   $\psitil_2$ 
 in terms of $\psitil_1$. The equations for $\psitil_1$ become 
\beq{
&\del _w ^2 \tilde{\psi} _1-\frac{1}{\xi ^2} \left( 1+\xi^2 ((\del _w\hat{\alpha} )^2-\del _w^2 \hat{\alpha} )\right)\tilde{\psi}_1 =0\comma\label{psitil11}\\
&\delbar _w^2 \tilde{\psi}_1 +2\delbar_w \hat{\alpha}\,\delbar_w\tilde{\psi} _1 -\xi ^2 \tilde{\psi} _1 =0\comma\label{psitil12}
}
and $\psitil_2$ are expressed in terms of $\psitil_1$ in two ways as 
\beq{
&\psitil_2=\xi \left( \del_w\psitil_1 +\del_w\hat{\alpha} \psitil_1\right)\comma\label{psitil21}\\
&\psitil_2=\frac{1}{\xi}e^{2\hat{\alpha}}\delbar _w\psitil_1\period\label{psitil22}
}

We now make the WKB expansion of  $\psitil_1$ in powers of $\xi$   in the form 
\beq{
\psitil _1 = \exp \left[ \frac{S_{-1}}{\xi} + S_0 +\xi S_1 + \xi ^2 S_2+\cdots\right] \comma \label{exppsitil1}
}
and substitute this into the equations (\ref{psitil11}) and (\ref{psitil12}). 
Then at each  order of $\xi$  we obtain equations involving $\del_w$ and 
 $\delbar_w$ derivatives and making use of the result of the previous order
 one can solve these equations successively.  As the procedure is more or less 
 straightforward, we will only sketch the computation at first two orders. 

For the function  $S_{-1}$ of the leading order,  one obtains the following two equations
\begin{align}
(\del _w S_{-1})^2 =1\comma \qquad (\delbar _w S_{-1})^2 =0 
\comma 
\end{align}
with the solutions 
\begin{align}
S_{-1} &= \pm w \period \label{Sminus1}
\end{align}
At the next order, we get  the equations of the form
\begin{align}
& \del^2_w S_{-1} + 2\del_w S_{-1} \del_w S_0 =0 \comma \\
&\delbar _w ^2 S_{-1} +2\delbar _w S_{-1} \delbar _w S_{0} +2\delbar \hat{\alpha} \delbar S_{-1} =0 \period 
\end{align}
With the result (\ref{Sminus1}) the first equation gives $\del_w S_0 =0$, 
while the second equation is automatically satisfied because $\delbar_w S_{-1} =0$. To fix $S_0$ completely, one needs to look at the next order of the
 expansion. One can then deduce an additional equation $\delbar_w S_0=0$ 
and hence the solution is $S_0 =$ constant,  which can be absorbed into the 
 normalization constant. 

Continuing in this fashion, we arrive at the following result for the WKB 
 expansion of $\psitil_i$, up to $\calO(\xi)$:
\beq{
&\psitil _1 = \exp \left[\pm \left( \frac{1}{\xi} \int dw + \xi \left(\frac{1}{2}\int dw\{ (\del _w\hat{\alpha} )^2 -\del_w^2 \hat{\alpha} \} +\int d\bar{w} e^{-2\hat{\alpha}}\right) +\cdots\right) \right] \comma
\label{wkbpsitil1} \\
&\psitil_2 =\pm\exp \left[ \pm \left( \frac{1}{\xi} \int dw + \xi \left( \del _w \hat{\alpha} (w_i)+\frac{1}{2}\int dw\{ (\del _w\hat{\alpha} )^2 +\del_w^2 \hat{\alpha} \} +\int d\bar{w} e^{2\hat{\alpha}}\right) +\cdots\right)\right] \period \label{wkbpsitil2}
}
It turns out that it is more convenient to reorganize this result by 
 grouping the terms in the exponent of $\psitil_1$ into two parts which are even or odd  under $\xi \rightarrow -\xi$. Namely we write (\ref{exppsitil1}) as 
\beq{
\psitil_1 = \exp \left[ S_{odd}+S_{even}\right] \period 
} 
Substituting this into the equations for $\psitil_1$, namely (\ref{psitil11}) and (\ref{psitil12}),  we can deduce four separate equations relating $S_{even}$ and 
$S_{odd}$. One can actually obtain two more equations from (\ref{psitil21}) 
and  (\ref{psitil22}),  expressing  the relations between 
$\psitil_1$ and $\psitil_2$. Combining these set of equations, one can deduce, 
 in particular, the following simple equation expressing  $S_{even}$ in 
 terms of $S_{odd}$:
\beq{
S_{even} = - \frac{1}{2}\ln \del _w S_{odd} \period  \label{even}
}
Using this relation, we can express $\psitil_i$ compactly as 
\beq{
&\psitil _1 = \frac{1}{\sqrt{\xi\del _w S_{odd}}}\exp \left[ \pm S_{odd}^{+}\right] \comma \nn\\
& \hspace{8cm} \label{WKBsol} \\
&\psitil _2 = \pm \sqrt{\xi\del _w S_{odd}^{+}}\left( 1\pm \frac{\delbar _w S_{even}}{\delbar _w S^{+}_{odd}}\right)\exp \left[ \pm S_{odd}^{+}\right]\period 
}
Here the symbol $S^{+}_{odd}$ signifies the value of $S_{odd}$ when 
the sign in  the solution for $S_{-1}$ is plus, namely for 
$S_{-1}=+w$ case. More explicitly,  switching back to the coordinate 
 $z$ in the expression (\ref{wkbpsitil1}), $S^{+}_{odd}$ is given by 
\beq{
S^{+}_{odd}=\frac{1}{\xi} \int _{z^{\ast}} dz\sqrt{p} + \xi \left(\frac{1}{2}\int _{z^{\ast}} dz\{ \frac{1}{\sqrt{p}}(\del \hat{\alpha} )^2 -\del(\frac{1}{\sqrt{p}}\del \hat{\alpha}) \} +\int _{z^{\ast}} d\barz \sqrt{\bar{p}}e^{-2\hat{\alpha}}\right) +\cdots
}
The initial point of the integration  $z^{\ast}$  can be  chosen
 arbitrarily. Since a different  choice only gives a  shift by a constant, 
 derivatives of $S^+_{odd}$ are independent of $z^{\ast}$. 
\subsubsection{Small and big solutions and SL(2,C) invariant products of  small solutions}
We now focus on the behavior the WKB solutions near 
 the singularities at $z_i$. At a generic value of $\xi$, one of the solutions in (\ref{WKBsol}) increases  while the other decreases as one approaches $z_i$. 
The former is called the big solution $b_i$ and the latter the small solution $s_i$. 
This is controlled by the behavior of the leading term in the exponent 
and hence the sign in front of $S^+_{odd}$. 

Which sign corresponds to the small solution depends on two factors. 
One is the position of the singularity $z_i$. A solution 
 with a particular sign which behaves as $s_i$ at $z_i$ may behave as a big solution $b_j$  at $z_j$.  In fact if $z_i$ and $z_j$ are connected by a WKB curve, 
such a situation must occur. In that case, the  sign of the exponent 
of the solution which is identified as $s_j$ at $z_j$ is opposite to the 
 one for $s_i$. 
Another factor is  the phase  of $\xi$. For example, 
the big and small solutions are interchanged  under  $\xi \rightarrow 
 e^{i\pi }\xi$. 

Let us clarify the relation between the solutions $i_\pm$, introduced 
 in the previous subsection, and the small and big solutions $s_i$ and $b_i$ discussed here. 
$s_i$ must coincide  (up to normalization) with  one of $i_\pm$ since 
such a decreasing solution is uniquely specified with a definite asymptotic 
 behavior. On the other hand $b_i$ is ambiguous since it can contain 
a multiple of $s_i$. Thus the correspondence is 
$\{ i_\pm, i_\mp\} \Leftrightarrow \{ s_i, b_i +\al_i s_i \}$. 
It will be of utmost importance to understand which of $i_\pm$ is 
 $s_i$ at each $z_i$. This is dictated by the analyticity of the solutions 
 on the double cover $D$ of the worldsheet.  Also it is intertwined with 
 the analyticity in the $\xi$-plane as it depends on the phase of $\xi$. 

Now for the problem at hand, reconstructing  the original solution 
from   (\ref{WKBsol}) by the gauge transformation 
(\ref{gaugetrpsi}), we can define  the small solution $s_i$, with 
normalization fixed, by 
\beq{
s_i= \frac{e^{-i\pi /4}}{\sqrt{2}} \pmatrix{c}{e^{\alpha /2}\left(\pm\xi\del  S_{(i),odd}^{+}\right)^{-1/2}\\  e^{-\alpha /2}\left(\pm\xi\del  S_{(i),odd}^{+}\right)^{1/2}\left( 1\pm \frac{\delbar  S_{(i),even}}{\delbar  S^{+}_{(i),odd}}\right)}\exp \left[ \pm S_{(i),odd}^{+}\right] \period 
}
Here,  $S_{(i),odd}^{+}$ denotes  $S_{odd}^{+}$ in which 
the initial point $z^\ast$  of the integrals is chosen 
 at the special point $z_{i}^{\ast}$ defined in subsection 3.3. 
Of course it is understood that the choice of the sign of the exponent  is made such that 
$s_i$ decrease as $z$ approaches $z_i$. 
Once we define  $s_i$ this way, the big solution $b_i$, which satisfies 
 the normalization condition $\braket{s_i\comma b_i}=1$, 
can be  given,  with the opposite sign choice,  by 
\beq{
b_i= \frac{\pm e^{-i\pi /4}}{\sqrt{2}} \pmatrix{c}{ e^{\alpha /2}\left(\mp\xi\del  S_{(i),odd}^{+}\right)^{-1/2}\\  e^{-\alpha /2}\left(\mp\xi\del  S_{(i),odd}^{+}\right)^{1/2}\left( 1\mp \frac{\delbar  S_{(i),even}}{\delbar  S^{+}_{(i),odd}}\right)}\exp \left[ \mp S_{(i),odd}^{+}\right]\period 
}
Ambiguity of $b_i$ of the form of $\al_i s_i$ does not affect the normalization 
 condition since  $\braket{s_i\comma s_i}=0$.

Let us now derive the form of the WKB expansion of $\braket{s_i,s_j}$, where $z_i$ and $z_j$ 
 are connected by a WKB curve. Such  products  will play crucial roles 
 in the computation of the regularized area in subsection 3.6. 

What is important here is that the signs of the exponent of such WKB-connected small solutions $s_i$ and $s_j$  are opposite.  Suppose  the small solution $s_i$ 
 corresponds to the plus sign, \ie   $s_i = i_+$ in the notation of  subsection  3.4.  Then that solution when followed to the point $z_j$ along the 
 WKB curve must be identified as the big solution $b_j$ there. This means that 
 the small solution $s_j$, which is  the other independent solution 
 at $z_j$,  must have the sign opposite to that of $b_j$ and is  
 identified as $s_j = j_-$. Similarly, if  $s_i = i_-$ then  $s_j = j_+$. 
Therefore we can write the WKB expansion of $s_i$ and $s_j$ as 
\beq{
&s_i= \frac{e^{-i\pi /4}}{\sqrt{2}} \pmatrix{c}{e^{\alpha /2}\left(\pm\xi\del  S_{(i),odd}^{+}\right)^{-1/2}\\  e^{-\alpha /2}\left(\pm\xi\del  S_{(i),odd}^{+}\right)^{1/2}\left( 1\pm \frac{\delbar  S_{(i),even}}{\delbar  S^{+}_{(i),odd}}\right)}\exp \left[ \pm S_{(i),odd}^{+}\right]\comma\nn\\
&s_j= \frac{e^{-i\pi /4}}{\sqrt{2}} \pmatrix{c}{e^{\alpha /2}\left(\mp\xi\del  S_{(j),odd}^{+}\right)^{-1/2}\\  e^{-\alpha /2}\left(\mp\xi\del  S_{(j),odd}^{+}\right)^{1/2}\left( 1\mp \frac{\delbar  S_{(j),even}}{\delbar  S^{+}_{(j),odd}}\right)}\exp \left[ \mp S_{(j),odd}^{+}\right]\period \label{WKBsol2}
}
Note that the prefactors in front of the exponential parts in these 
 expressions depend only on the derivatives of $S^+_{odd}$, which 
 as remarked previously do not depend on the initial points  $z^\ast_i$ 
 of the line integrals and hence on the subscripts $(i)$ and $(j)$. 
For this reason, in the product $\braket{s_i, s_j}$ these prefactors 
 cancel exactly and the result takes the following simple form:
\beq{
\braket{s_i\comma s_j}=\pm \exp\left[ \pm S_{i\to j}\right]\period \label{WKBpro}
}
Here the symbol $S_{i \to j}$ is defined by 
\beq{
S_{i\to j}&=S^{+}_{(i),odd}-S^{+}_{(j),odd}\nn\\
&=\frac{1}{\xi} \int _{z_i^{\ast}}^{z_j^{\ast}} dz\sqrt{p} + \xi \left(\frac{1}{2}\int _{z_i^{\ast}}^{z_j^{\ast}} dz\{ \frac{1}{\sqrt{p}}(\del \hat{\alpha} )^2 -\del(\frac{1}{\sqrt{p}}\del \hat{\alpha}) \} +\int _{z_i^{\ast}}^{z_j^{\ast}} d\barz \sqrt{\bar{p}}e^{-2\hat{\alpha}}\right) +\cdots 
\comma \label{Sij}
}
where  the integration is performed along  the WKB curve. 
 The signs in  (\ref{WKBpro}) are defined to be
 in unison with  those of the exponent of $s_i$. That is, if  the exponential part of $s_i$ is $\exp[+ S^+_{(i),odd}]$, 
then $\braket{s_i\comma s_j}=+ \exp\left[ + S_{i\to j}\right]$ and 
 so on.

It is actually more convenient to express the formula (\ref{Sij}) 
in a  manner independent of the  explicit positions  $i, j$ of $s_i$ and $s_j$. 
For this purpose, we introduce a notation  $s_\oplus$
for a  small solution for which the exponent of its WKB expansion is of the form $\exp[+ S^+_{odd}]$, 
without  referring to the position around which it is a small solution. 
The symbol  $s_\ominus$ should be understood in the same obvious manner. 
In other words, 
\beq{
s_\oplus \sim \exp \left[ \frac{1}{\xi}\int_{z_{\oplus}^{\ast}} dz\sqrt{p} +\cdots\right]\comma \hspace{11pt}s_{\ominus} \sim \exp \left[- \frac{1}{\xi}\int_{z_{\ominus}^{\ast}} dz\sqrt{p} +\cdots\right]\period\nn
}
 Then, the formula (\ref{Sij}) can be expressed,  for any pair of points 
connected by a WKB curve,  as 
 \beq{
&\braket{s_\oplus, s_\ominus}=\exp\left[ S_{\oplus\to\ominus}\right]\comma \nn\\
&S_{\oplus\to\ominus}=\frac{1}{\xi} \int _{z_{\oplus}^{\ast}}^{z_{\ominus}^{\ast}} dz\sqrt{p} + \xi \left(\frac{1}{2}\int _{z_{\oplus}^{\ast}}^{z_{\ominus}^{\ast}} dz\{ \frac{1}{\sqrt{p}}(\del \hat{\alpha} )^2 -\del(\frac{1}{\sqrt{p}}\del \hat{\alpha}) \} +\int _{z_{\oplus}^{\ast}}^{z_{\ominus}^{\ast}} d\barz \sqrt{\bar{p}}e^{-2\hat{\alpha}}\right) +\cdots \period\label{oplusominus}
} 

Now an important observation is that a part  of  $S_{\oplus\to\ominus}$ 
above can be expressed in terms of the contour integrals  $\int \lambda dz$ and $\int ud\barz+vdz$, which appear in the expression of the regularized area. 
The integrals $\int \lambda dz$ and $\int v dz$ are visible already. 
 To recognize the structure $\int ud\zbar$, we must rewrite the last 
 term of the  order $\xi$ part using the equation of motion for $\alhat$, 
which reads
\begin{align}
{\del \delbar \alhat \over \sqrt{p\pbar}} - e^{2\alhat} 
 + e^{-2\alhat} =0 \period
\end{align}
Then it is easy to rewrite $S_{\oplus\to\ominus}$ into the form
\beq{
S_{\oplus\to\ominus}=&\frac{1}{\xi} \int _{z_{\oplus}^{\ast}}^{z_{\ominus}^{\ast}} \lambda dz + \xi \int_{z_{\oplus}^{\ast}}^{z_{\ominus}^{\ast}}\sqrt{\bar{p}}d\barz \nn\\
&+\frac{\xi}{2} \left(\int _{z_{\oplus}^{\ast}}^{z_{\ominus}^{\ast}}( ud\barz +v dz )  -\int _{z_{\oplus}^{\ast}}^{z_{\ominus}^{\ast}} \del (\frac{1}{\sqrt{p}}\del \hat{\alpha} )dz + \delbar (\frac{1}{\sqrt{p}}\del  \hat{\alpha} )d\barz\right) +\cdots \nn
}
Notice that the last term is an  integral over a total derivative 
of $\del \alhat/\sqrt{p}$.  In the case of LSGKP strings, this contribution 
 vanishes since $\del \alhat\rightarrow 0$ at each singularity. 
Therefore for our problem $S_{\oplus\to\ominus}$ simplifies to 
\beq{
S_{\oplus\to\ominus}=\frac{1}{\xi} \int _{z_{\oplus}^{\ast}}^{z_{\ominus}^{\ast}} \lambda dz + \xi \int_{z_{\oplus}^{\ast}}^{z_{\ominus}^{\ast}}\sqrt{\bar{p}}d\barz +\frac{\xi}{2} \int _{z_{\oplus}^{\ast}}^{z_{\ominus}^{\ast}}( ud\barz +v dz ) +\cdots 
\period \label{exps}
}

So far we have been  considering  the WKB expansion around $\xi=0$. The expansion around $\xi=\infty$ can be carried out in a similar way. In particular, 
for that expansion the leading term around $\xi=\infty$ is given by
\beq{
S_{\oplus\to\ominus}=\xi \int_{z_{\oplus}^{\ast}}^{z_{\ominus}^{\ast}}\sqrt{\bar{p}}d\barz+\cdots\period
} 

These results will be used to determine the analyticity of $SL(2,C)$ invariant products in section 3.6, which in turn leads  to their evaluation. 
\subsection{Regularized area from SL(2,C)-invariant products}
We are now finally in a position to  combine all the results of the preceding
 subsections 
 and derive the formula for the  regularized area in terms of the parameters 
$\kappa_i$ which characterize GKP strings. The remaining steps are as follows. 
First, we express the contour integrals given in (\ref{riemann4}) in terms of the $SL(2,C)$ invariant products between small solutions, duly distinguishing the 
 different domains of the value of $\xi$. 
We then clarify the precise relations  between the small solutions and  the plus-minus solutions. Together with the result obtained in subsection 3.4, this will 
 allow us to  express the contour integrals expressing the regularized 
 area in terms of $\kappa_i$ in a completely explicit manner.  
The final result will be  the formula  (\ref{regarea2}). 
\subsubsection{WKB curves for  a specific  parameter regime }
As we saw in subsections 3.3 and  3.5, both the regularized area of our interest 
and the WKB expansion of the invariant products $\braket{s_i, s_j}$
 of small solutions can be expressed in terms of the same 
type of integrals, namely  $\int \lambda dz$ and $\int (ud\barz +vdz)$. 
To make use of this result, however,  the  integration contours 
 in both cases must be related.  Since the WKB expansion can only be 
 performed reliably along WKB curves, this means that we must 
obtain  the precise knowledge of the WKB curves and 
 express the contours relevant to  the regularized area in terms of 
such curves. 

\begin{figure}
\begin{center}
\picture{clip,height=5cm}{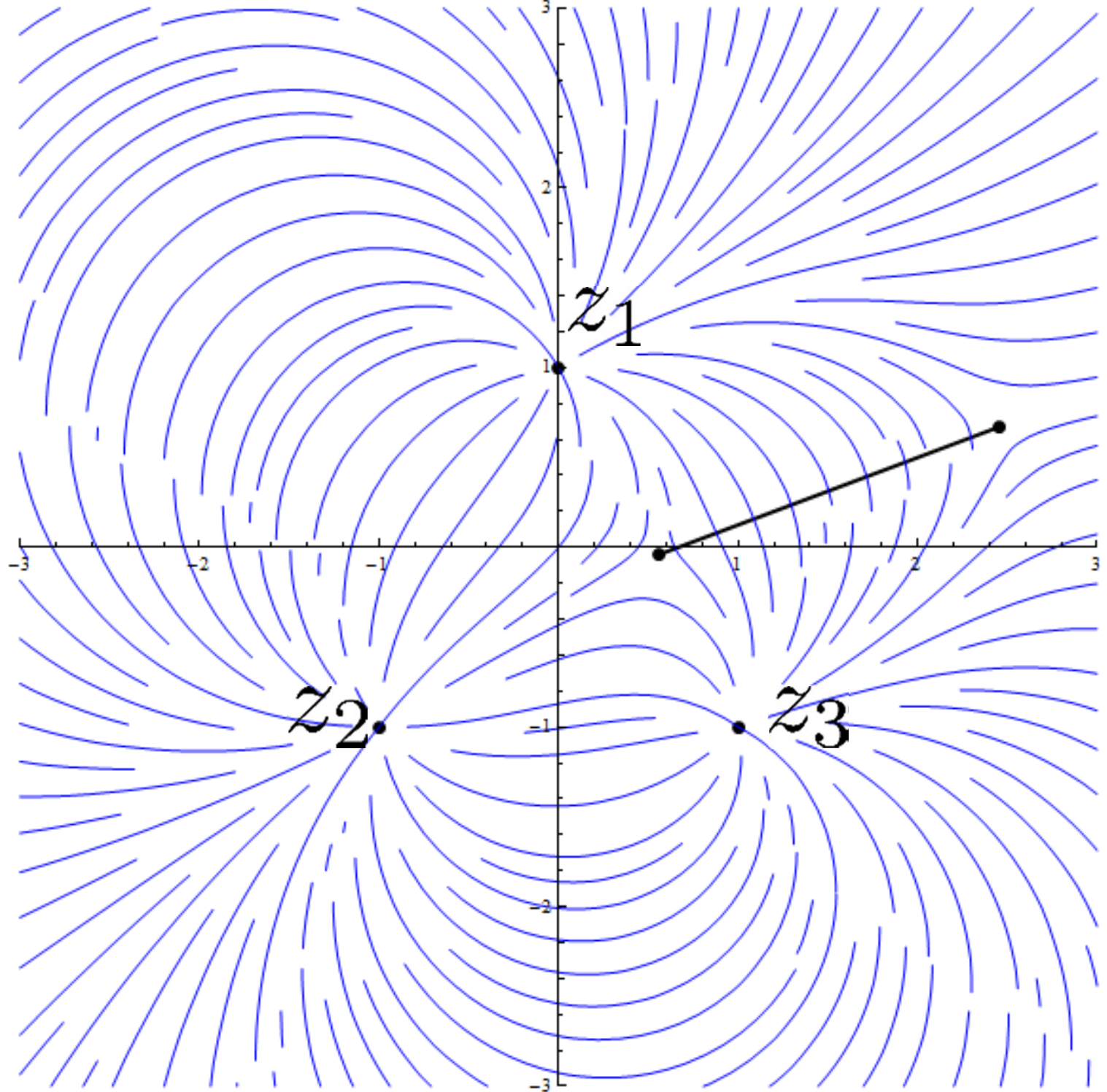}
\end{center}
\caption{Typical configuration of WKB curves for the parameter regime where the triangle inequality holds.  The parameters used are  $\kappa_1=\kappa_3=1, \kappa_2=\sqrt{3}$ and ${\rm Im}\, \xi <0$. }
\label{poswkb}
\end{figure}

As the structure of the WKB curves strongly depend on the phase of $\xi$ and 
on the residues $\delta_i$ of the poles,  it is not useful  to try to discuss the general situation from the beginning. In the case of the GKP strings, $\delta_i$'s 
 are pure imaginary and related to $\kappa_i$ as 
\begin{align}
\delta_i = {i\kappa \over 2} \period 
\end{align}
Therefore we shall first focus on a specific case where the parameters are in the range
\beq{
&\rm{Im}\,\xi >0 \comma \label{imxiplus}\\
&\kappa _2>\kappa _1\comma \hspace{11pt}\kappa _2>\kappa _3 \comma\hspace{11pt} \kappa _1+\kappa _3>\kappa _2 \period 
\label{kapparange1}
}  
The case for  $\rm{Im}\,\xi<0$ will be considered  later in subsection 3.6.4 and 
the cases with more general configurations  of  $\kappa _i$ will be treated  in subsection 3.6.5  and the appendix A. 
Typical configurations of  WKB curves are shown in figure \ref{poswkb}, which are obtained by using the formula (\ref{pz3pt}). The important feature of the WKB curves in this parameter regime is that 
every pair of  poles are connected directly by a WKB curve
 and hence the  contours  $d_j$,  which were defined at the end of  subsection 3.3 and  will play crucial roles,  can easily be expressed in terms of  three 
such WKB curves. (As we shall  see later, this is not the case for  generic $\kappa _i$.) 

\subsubsection{Relation between contour integrals and the  products $\boldsymbol{\braket{s_i, s_j}}$ }
Having seen the structure of the WKB curves in the parameter regime 
defined in  (\ref{imxiplus}) and (\ref{kapparange1}), we now relate 
 the contour integrals along $d_i$'s and the products $\braket{s_i, s_j}$ explicitly.  To do this, we must first specify the branch structure of the 
 function $\sqrt{p(z)}$.  We will take the square root branch cut to 
be along a line which connects the two zero's of $p(z)$. Then 
the behavior of  $\sqrt{p(z)}$ on  the double cover of the worldsheet
 is automatically dictated once we specify the behavior around
 one of its poles. In what follows, we do this by 
setting the behavior around $z_1$ on the first sheet to be  $\sqrt{p(z)}\overset{z\sim z_1}{\sim}\delta_1/(z-z_1) $. This determines the behaviors 
on the first sheet around  the poles to be 
\beq{
\sqrt{p(z)}& \overset{z\sim z_1}{\sim}\frac{\delta_1}{z-z_1}=\frac{i\kappa_1}{2(z-z_1)}\comma\nn\\
& \overset{z\sim z_2}{\sim}\frac{-\delta_2}{z-z_2}=\frac{-i\kappa_2}{2(z-z_2)}\comma\nn\\
& \overset{z\sim z_3}{\sim}\frac{\delta_3}{z-z_3}=\frac{i\kappa_3}{2(z-z_3)}\period\nn
} 
The corresponding behaviors  on the second sheet are obtained 
 by flipping the overall sign. 
 With  this definition,  the eigenvalues of the monodromy matrices are given by 
\beq{
\pp _1= 2\pi \delta _1 \left(\frac{1}{\xi}+\xi \right)\comma \hspace{11pt}\pp _2= -2\pi \delta _2 \left(\frac{1}{\xi}+\xi \right)\comma \hspace{11pt}\pp _3= 2\pi \delta _3 \left(\frac{1}{\xi}+\xi \right)\comma \label{rhos}
}
and the small solutions and the plus-minus solutions are related as
\beq{
s_1\sim 1_{+} \comma \hspace{11pt} s_2 \sim 2_{-} \comma \hspace{11pt} s_3 \sim 3_{+} \period\label{smallpm1}
}  
Here $\sim$ symbol  means  that the relation is yet up to normalization. 

With this preparation  we can now apply (\ref{oplusominus}) to relate the products $\braket{s_i\comma s_j}$ with  the contour integrals on $d_i$.  We will illustrate this by taking the contour $d_2$ 
as an example. From the definition of $d_2$, it can be decomposed into 
 three WKB curves as 
\beq{
d_2 = (\widehat{z}_2\to \widehat{z}_3)+(\widehat{z}_3\to z_1)+(z_1\to z_2)\comma \label{pathd2}
}  
where 
$(z_i\to z_j)$ denotes a WKB curve connecting $z_i$ and $z_j$ 
  and  the symbol $\widehat{z}_i$ with a hat  reminds us that  the point 
is on the second sheet. 

We first compute the contribution from the last part, $(z_1 \to z_2)$ 
 by using the formula (\ref{exps}) developed in the previous subsection. 
Since $s_1$ and $s_2$ are related to plus-minus solutions 
 as in  (\ref{smallpm1}),  the symbols $z_\oplus^\ast$ and $z_\ominus^\ast$
 in the formula correspond to $z_1$ and $z_2$ respectively. 
Therefore, the logarithm of $\braket{s_1 \comma s_2}$ can be evaluated as
\beq{
&\log \braket{s_1\comma s_2} = \left(\xi ^{-1}\int ^{z_2} _{z_1}\sqrt{p}dz  +\xi \int ^{z_2} _{z_1}\sqrt{\bar{p}}d\barz \right) +\frac{\xi}{2} \left(\int _{z_1} ^{z_2} ud\barz +v dz \right) +\cdots \period
}
Similarly, for the contribution coming from  the WKB curve $(\widehat{z}_3\to z_1)$,  $z_\oplus^\ast$ corresponds to $z_1$, while  $z_\ominus^\ast $ 
 is identified as $\widehat{z}_3$ since the sign of $\sqrt{p(z)}$ is flipped on the second sheet. We therefore obtain 
\beq{
&\log \braket{s_1\comma s_3} = \left(\xi^{-1}\int^{\widehat{z}_3}_{z_1}\sqrt{p}dz  +\xi \int^{\widehat{z}_3}_{z_1}\sqrt{\bar{p}}d\barz \right) +\frac{\xi}{2} \left(\int ^{\widehat{z}_3}_{z_1} ud\barz +v dz \right) +\cdots \period
}
Finally, in the same way, WKB expansion of $\braket{s_2, s_3}$ along   $(\widehat{z}_2\to \widehat{z}_3)$  gives the formula
 \beq{
&\log \braket{s_2\comma s_3} = \left(\xi^{-1}\int_{\widehat{z}_2}^{\widehat{z}_3}\sqrt{p}dz  +\xi \int_{\widehat{z}_2}^{\widehat{z}_3}\sqrt{\bar{p}}d\barz \right) +\frac{\xi}{2} \left(\int _{\widehat{z}_2}^{\widehat{z}_3} ud\barz +v dz \right) +\cdots \period
}
Putting the three contributions together, we get the expression for  the 
 integrals along $d_2$,  which is displayed  below  together with the 
ones along $d_1$ and $d_3$ obtained by similar analysis:
\beq{
&\frac{\braket{s_2\comma s_3}}{\braket{s_2\comma s_1}\braket{s_1\comma s_3}}=\exp\left[\frac{1}{\xi} \int _{d_1} \lambda dz + \xi \int_{d_1}\sqrt{\bar{p}}d\barz +\frac{\xi}{2} \left(\int _{d_1} ud\barz +v dz \right) +\cdots\right]\comma\\
&\frac{\braket{s_2\comma s_3}\braket{s_1\comma s_2}}{\braket{s_1\comma s_3}}=\exp\left[\frac{1}{\xi} \int _{d_2} \lambda dz + \xi \int_{d_2}\sqrt{\bar{p}}d\barz +\frac{\xi}{2} \left(\int _{d_2} ud\barz +v dz \right) +\cdots\right]\comma\label{ssd1} \\
&\frac{\braket{s_1\comma s_2}}{\braket{s_1\comma s_3}\braket{s_3\comma s_2}}=\exp\left[\frac{1}{\xi} \int _{d_3} \lambda dz + \xi \int_{d_3}\sqrt{\bar{p}}d\barz +\frac{\xi}{2} \left(\int _{d_3} ud\barz +v dz \right) +\cdots\right]\period
}

As we already calculated, in subsection 3.3,  the $SL(2,C)$ products between plus-minus solutions in terms of the local monodromy data (\ref{rhos}), 
if we can identify $s_i$ with $i_{\pm}$  precisely,  the contour integrals
 can be expressed in terms of the parameters $\delta_i$.  This 
however is not straightforward. The reason is that since the plus-minus 
 solutions are defined only up to $\xi$-dependent scaling factor, 
 the relations such as (\ref{smallpm1}) suffer from that ambiguity. 
Such uncontrolled  $\xi$ dependence ruins the  identification 
 of the coefficient of each power of $\xi$.  Therefore it is extremely important 
 to understand the conditions under which we can identify  $s_i=i_+$ (or $i_-$)
 precisely, without $\xi$-dependent relative factors. 

This turned out to be intimately connected with the analyticity in $\xi$. 
To show this, as a first step, we study how the computation of the contour integrals 
 described above changes as we vary $\xi$ to the region where ${\rm Im} \,\xi$
 is  negative. 
 The crucial change is  that the relation between the  small solutions and 
the plus-minus solutions gets  altered like 
\beq{
&{\rm Im} \, \xi >0 :\quad  s_1\sim 1_{+} \comma \hspace{5pt} s_2 \sim 2_{-} \comma \hspace{5pt} s_3 \sim 3_{+} \comma \nn\\
&{\rm Im} \, \xi <0 :\quad  s_1\sim 1_{-} \comma \hspace{5pt} s_2 \sim 2_{+} \comma \hspace{5pt} s_3 \sim 3_{-} \period \label{imxichange}
} 
This change inevitably  affects the WKB expansion of the $SL(2,C)$ invariant products:  The poles which were previously of the type $z_{\oplus}$ are now 
changed to $z_{\ominus}$ and vice versa. This modifies 
 the relation between the contour integrals on $d_i$ and the 
 $SL(2,C)$ invariant products  to 
\beq{
&\frac{\braket{s_1\comma s_2}\braket{s_3\comma s_1}}{\braket{s_3\comma s_2}}=\exp\left[\frac{1}{\xi} \int _{d_1} \lambda dz + \xi \int_{d_1}\sqrt{\bar{p}}d\barz +\frac{\xi}{2} \left(\int _{d_1} ud\barz +v dz \right) +\cdots\right]\comma\\
&\frac{\braket{s_3\comma s_1}}{\braket{s_3\comma s_2}\braket{s_2\comma s_1}}=\exp\left[\frac{1}{\xi} \int _{d_2} \lambda dz + \xi \int_{d_2}\sqrt{\bar{p}}d\barz +\frac{\xi}{2} \left(\int _{d_2} ud\barz +v dz \right) +\cdots\right]\comma \label{ssd2}\\
&\frac{\braket{s_3\comma s_1}\braket{s_2\comma s_3}}{\braket{s_2\comma s_1}}=\exp\left[\frac{1}{\xi} \int _{d_3} \lambda dz + \xi \int_{d_3}\sqrt{\bar{p}}d\barz +\frac{\xi}{2} \left(\int _{d_3} ud\barz +v dz \right) +\cdots\right]\period
}
Compared to  (\ref{ssd1}), the expressions on the left hand side are  inverted and acquired an overall minus sign, because $\braket{s_i,s_j}$ gets changed into $\braket{s_j, s_i} = -\braket{s_i, s_j}$. 
\subsubsection{Connection between normalization and  analyticity}
Now we shall show that convenient normalization of the eigenvectors 
 $i_\pm$ can be specified by imposing   certain analyticity property 
on the $SL(2,C)$ invariant products  $\braket{i_+, j_-}$. 

Let us  illustrate the idea by using a concrete example involving 
 $\braket{1_+,2_-}$ and  $\braket{1_-,2_+}$. 
In subsection 3.4, we showed that the products  $\braket{i_+, j_-}$
can be expressed in terms of the local monodromy data $\pp_i$ and 
 that their judicious combinations are free of rescaling ambiguities. 
We gave such an example in (\ref{Aindcomb}), which we display  here again:
\beq{
\braket{1_+,2_-}\braket{2_+,1_-}= \frac{1}{2\sin \pp_1\sin \pp_2}\left( \cos (\pp _1-\pp _2) -\cos \pp _3 \right)\period \label{T-like}
}
We wish to solve this equation for the individual product, $\braket{1_+,2_-}$
 and $\braket{2_+,1_-}$. We know from the preceding discussion that if  ${\rm Im} \,\xi >0$,  the eigenvectors $1_+$ and  $2_-$  are  proportional to  $s_1$ and $s_2$ respectively and hence $\braket{1_+,2_-}$ should be proportional to 
 $\braket{s_1, s_2}$, for which we can make  WKB expansion.  
Similarly, for ${\rm Im} \,\xi <0$, the product  $\braket{2_+,1_-}$ 
should be proportional  to $\braket{s_2, s_1}$. Now we show that 
such solutions to the equation \ref{T-like} are unique up to {\it $\xi$-independent }
 constant if we impose the following analyticity conditions,  inspired by 
 the WKB expansion of $\braket{s_1, s_2}$,  in respective domain of 
$\xi$:
\beq{
&\log \braket{1_+,2_-} = \left(\frac{1}{\xi}\int_{z_1}^{z_2}\sqrt{p}dz  +\xi \int_{z_1}^{z_2}\sqrt{\bar{p}}d\barz \right) + \text{functions regular on }{\rm Im}\,\xi \geq 0 \comma\label{analytic3}\\
&\log \braket{2_+,1_-}=  \left(\frac{1}{\xi}\int_{z_2}^{z_1}\sqrt{p}dz  +\xi \int_{z_2}^{z_1}\sqrt{\bar{p}}d\barz \right) + \text{functions regular on }{\rm Im}\,\xi \leq 0 \period\label{analytic4}
}
To prove it, assume that there are two different solutions to (\ref{T-like}) 
 like 
\begin{align}
&(i)\quad \braket{1_{-},2_{+}} = f \comma \quad \braket{2_{-},1_{+}}= g
\comma 
\nn\\
&(ii) \quad \braket{1_{-},2_{+}} = f' \comma \quad \braket{2_{-},1_{+}}= g'
\period \nn
\end{align}

Then, since $fg=f'g'$,  we have 
\beq{
\log f -\log f^{\prime} =\log g^{\prime} -\log g\period
}
From the analyticity we imposed, the left hand side of this  equation is regular on ${\rm Im}\,\xi \leq 0$, while the right hand side is regular on ${\rm Im}\,\xi\geq0$. Therefore, both sides  must be regular on the whole complex plane (including
 the infinity) and hence is a $\xi$-independent constant. This proves 
 the assertion. 

On the other hand, as should already be clear from the presentation above, 
the analyticity properties we imposed  are precisely  those of the WKB expansions 
 of  $\braket{s_1\comma s_2}$ if we identify small solutions and plus-minus solutions as\footnote{Strictly speaking, the notion of small solution does not 
exist for  ${\rm Im}\,  \xi =0$. However, since the limit ${\rm Im}\,  \xi \rightarrow 0$
  produces no singularity, we can extend the condition 
 of the analyticity to include ${\rm Im}\,  \xi=0$ in (\ref{analytic3}) 
 and (\ref{analytic4}). } 
\beq{&
{\rm Im}\,\xi > 0\Rightarrow s_1=1_+\comma \hspace{5pt}
 s_2=2_- \comma
\label{norm1} \\
&{\rm Im}\,\xi < 0\Rightarrow s_1=1_-\comma \hspace{5pt} s_2=2_+ 
\period
\label{norm2}
}
Therefore, the imposition of the analyticity   is totally equivalent to fixing  the normalization of $1_{\pm}$ and $2_{\pm}$ as 
 in (\ref{norm1}) and (\ref{norm2}). 

Summarizing, we have found that  to evaluate the contour integrals 
 we should (i) find the unique solution to (\ref{T-like}) which satisfy 
the analyticity properties (\ref{analytic3}) and (\ref{analytic4}), and 
(ii) WKB-expand $\braket{1_+,2_-}$  for     ${\rm Im}\,\xi \geq 0$,   or 
equivalently, $\braket{1_-,2_+}$  for  ${\rm Im}\,\xi \le  0$.
In what follows, we will show how these procedures can be carried out in practice to obtain the expression of the regularized area.
\subsubsection{Derivation of the formula for SL(2,C) products 
and computation of the regularized area}
As the first step, we wish to  solve the equation of the type (\ref{T-like}) 
under the prescribed analyticity. For this purpose, it should be noted 
 that although individually $\braket{1_-,2_+}$  and $\braket{2_-,1_+}$
 contain log-divergent terms 
of the form  $\pm\left(\xi^{-1} \int_{z_1}^{z_2}\sqrt{p}dz  +\xi \int_{z_1}^{z_2}\sqrt{\bar{p}}d\barz \right)$, they cancel each other in the 
 product. Therefore, we can only focus on the non-singular part
 of these $SL(2,C)$ invariant products. 

Now to solve  the equation (\ref{T-like}) under the analyticity requirement, 
the following  identity is useful:
\begin{align}
\frac{1}{2\pi i}\int _{-\infty}^{\infty}d\xi^{\prime}
 \frac{1}{\xi^{\prime}-\xi} \left(F(\xi ^{\prime})+G(\xi ^{\prime})\right) &= 
\bracetwo{
F(\xi ) \comma  \hspace{14pt}  ({\rm Im}\,\xi > 0)}{-G(\xi ) \comma \hspace{5pt} ({\rm Im}\,\xi < 0)}
\end{align}
Here,   $F(\xi)$ is a function regular on the upper half plane and exponentially suppressed at large $\left| \xi\right|$, while  $G(\xi)$ is function 
regular on the lower half plane and exponentially suppressed at large $\left| \xi\right|$.
 This  formula can be proved using the residue theorem by closing the contour on the upper half or lower half plane.
With the use of this identity,  one can show that the following expressions give 
the nonsingular part of $\log \braket{2_{-},1_{+}}$ and $\log \braket{1_{-},2_{+}}$:
\beq{
&\log \braket{2_{-},1_{+}}_{n.s.}= \frac{1}{2\pi i}\int _{-\infty}^{\infty}d\xi^{\prime} \frac{1}{\xi^{\prime}-\xi} \log\left(\frac{\cos (\pp _1-\pp _2)-\cos \pp _3}{2\sin \pp _1 \sin \pp _2}(\xi^{\prime}) \right)\comma\hspace{5pt} ({\rm Im}\,\xi > 0)\comma\nn\\
&\log \braket{1_{-},2_{+}}_{n.s.}=  -\frac{1}{2\pi i}\int _{-\infty}^{\infty}d\xi^{\prime} \frac{1}{\xi^{\prime}-\xi} \log \left(\frac{\cos (\pp _1-\pp _2)-\cos \pp _3}{2\sin \pp _1 \sin \pp _2}(\xi^{\prime}) \right)\comma\hspace{5pt} ({\rm Im}\,\xi < 0)\comma\nn
}
where the subscript $n.s.$ denotes the non-singular part. 
The expressions of  these quantities on the other half of the  $\xi$-plane 
are defined by the analytic continuation of the above results.
Now rewrite the  factor $\half (\cos (\pp _1-\pp _2)-\cos \pp _3)$ 
as 
\beq{
\frac{\cos (\pp _1-\pp _2)-\cos \pp _3}{2}=\sin \frac{\pp _1-\pp _2+\pp _3}{2}\sin \frac{-\pp _1+\pp _2+\pp _3}{2}  \period\nn
}
and substitute the forms of $\pp_i$   given in (\ref{rhos}). 
Then, switching to the notation in terms of small solutions, 
 the above expressions can be reorganized as 
\beq{
&{\rm Im}\,\xi > 0 \,:\nn\\
&\log \braket{s_1,s_2}_{n.s.}= h(\kappa _1)+h(\kappa _2)-h(\frac{\kappa _1+\kappa _2+\kappa _3}{2})-h(\frac{\kappa _1+\kappa _2-\kappa _3}{2})\comma\label{final}\\
&{\rm Im}\,\xi <0 \,:\nn\\
&\log \braket{s_1,s_2}_{n.s.}=  -h(\kappa _1)-h(\kappa _2)+h(\frac{\kappa _1+\kappa _2+\kappa _3}{2})+h(\frac{\kappa _1+\kappa _2-\kappa _3}{2})
\comma \label{final2}
}
where the function $h(x)$ is defined by 
\beq{
h(x) \equiv -\frac{1}{\pi i}\int _{0}^{+\infty}d\xi^{\prime} \frac{\xi}{\xi^{\prime ^2}-\xi ^2} \log \left(1-\exp \left( -2\pi x (\xi^{\prime ^{-1}}+\xi^{\prime} )\right)\right) \period 
}

In the same manner, $\braket{s_2,s_3}$ and $\braket{s_3,s_1}$ 
can be expressed as 
\beq{
&{\rm Im}\,\xi > 0 \,:\nn\\
&\log \braket{s_2,s_3}_{n.s.}=h(\kappa _2)+h(\kappa _3)-h(\frac{\kappa _1+\kappa _2+\kappa _3}{2})-h(\frac{-\kappa _1+\kappa _2+\kappa _3}{2})
\comma \label{final3}\\
&\log \braket{s_3,s_1}_{n.s.}=h(\kappa _3)+h(\kappa _1)-h(\frac{\kappa _1+\kappa _2+\kappa _3}{2})-h(\frac{\kappa _1-\kappa _2+\kappa _3}{2})
\comma \label{final4}\\
&{\rm Im}\,\xi <0 \,:\nn\\
&\log \braket{s_2,s_3}_{n.s.}=-h(\kappa _2)-h(\kappa _3)+h(\frac{\kappa _1+\kappa _2+\kappa _3}{2})+h(\frac{-\kappa _1+\kappa _2+\kappa _3}{2})
\comma \label{final5}\\
&\log \braket{s_3,s_1}_{n.s.}=-h(\kappa _3)-h(\kappa _1)+h(\frac{\kappa _1+\kappa _2+\kappa _3}{2})+h(\frac{\kappa _1-\kappa _2+\kappa _3}{2}
\period\label{final6}
} 

We can now finally evaluate the contour integrals of our interest. 
First  the integrals  $\int _{d_i} (ud\barz +vdz)$ can be extracted 
 as the coefficients  of $\xi$ in the non-singular part of the expansions given in  (\ref{ssd1}) and (\ref{ssd2}). Hence, we need the order $\xi$ part of the 
function $h(x)$.  It is given by 
\begin{align}
h(x)&= i \xi  K(x) + \calO(\xi^2) \comma \\
K(x ) &= \frac{1}{\pi} \int _{-\infty}^{\infty} d\theta \, e^{-\theta} \log \left( 1- e^{-4\pi x\cosh \theta }\right) \period 
\end{align}
Using this function $K(x)$, we obtain 
\beq{
\int _{d_1} (ud\barz +vdz) &= 2i\left( K(\frac{\kappa_1+\kappa_2+\kappa_3}{2})-2K(\kappa_1)\right.\\
&\left.  -K(\frac{-\kappa_1+\kappa_2+\kappa_3}{2})+K(\frac{\kappa_1-\kappa_2+\kappa_3}{2})+K(\frac{\kappa_1+\kappa_2-\kappa_3}{2})\right)\comma\nn\\
\int _{d_2} (ud\barz +vdz) &= -2i\left(  K(\frac{\kappa_1+\kappa_2+\kappa_3}{2}) -2K( \kappa _2)\right.\\
&\left. +K(\frac{-\kappa _1+\kappa _2+\kappa _3}{2})-K(\frac{\kappa_1-\kappa_2+\kappa_3}{2})+K(\frac{\kappa _1+\kappa _2-\kappa _3}{2})\right)\comma\nn\\
\int _{d_3} (ud\barz +vdz) &= 2i\left(  K(\frac{\kappa _1 +\kappa _2 +\kappa _3}{2})-2K(\kappa _3)\right.\\
&\left.  +K(\frac{-\kappa _1+\kappa _2+\kappa _3}{2})+K(\frac{\kappa _1-\kappa _2+\kappa _3}{2})-K(\frac{\kappa _1+\kappa _2-\kappa _3}{2})\right)\period\nn
}
On the other hand, the contour integrals $\int_{C_i} \lam dz$
 around the poles $z_i$ are easily evaluated as 
\beq{
&\int_{C_1}\lambda dz = -\pi  \kappa _1\comma \qquad 
 \int_{C_2}\lambda dz = +\pi  \kappa _2\comma \qquad 
 \int_{C_3}\lambda dz = -\pi  \kappa _3 \period 
}
Combining the above results, together with the constant contribution 
 discussed in subsection 3.3, the regularized area in the 
parameter regime (\ref{kapparange1}) is expressed in terms of $\kappa_i$
 as 
\beq{
\int d^2 z \lambda\, u&={\pi \over 12} +\pi \left[ -\kappa _1 K(\kappa _1)-\kappa _2 K(\kappa _2)-\kappa _3 K(\kappa _3)\right.\nn\\
&\left. +\frac{\kappa _1 +\kappa _2 +\kappa _3}{2}K(\frac{\kappa _1+\kappa _2+\kappa _3}{2})\right.
 \left. +\frac{-\kappa _1 +\kappa _2 +\kappa _3}{2}K(\frac{-\kappa _1+\kappa _2+\kappa _3}{2})\right.\nn\\
&\left. +\frac{\kappa _1 -\kappa _2 +\kappa _3}{2}K(\frac{\kappa _1-\kappa _2+\kappa _3}{2})\right. 
\left. +\frac{\kappa _1 +\kappa _2 -\kappa _3}{2}K(\frac{\kappa _1+\kappa _2-\kappa _3}{2})\right] \period \label{regarea1}
}

Since the result above is symmetric under the permutation of $\kappa_i$, 
 we can apply the same formula for such permuted regimes.  The characteristic 
feature of these regimes is that the ``triangle inequality"  
$\kappa _i < \kappa_j + \kappa_k\ {}_{(i \ne j \ne k)} $ is satisfied. 
Therefore, to cover the entire  range of parameters, we must discuss separately 
 the cases in which  the triangle inequality is not satisfied, namely when 
$\kappa _i > \kappa_j + \kappa_k\ {}_{(i \ne j \ne k)} $. 
The analysis for such regimes is  more involved because the configuration 
 of the WKB curves and the position of the branch cut change in such a way 
 that one cannot perform the WKB expansion straightforwardly. 
We will explain how such a situation should be handled 
in the appendix A.  We find, after careful analyses, that signs in 
 the formula (\ref{regarea1}) must be changed appropriately. 
The net result, however, turned out to be quite gratifying.  
The general formula which is valid for all the parameter regimes can be 
 written as 
\beq{
A_{reg}=\int d^2 z \lambda \, u&={\pi \over 12} +\pi \left[ -\kappa _1 K(\kappa _1)-\kappa _2 K(\kappa _2)-\kappa _3 K(\kappa _3)\right.\nn\\
&\left. +\frac{\kappa _1 +\kappa _2 +\kappa _3}{2}K(\frac{\kappa _1+\kappa _2+\kappa _3}{2})\right.\nn\\
&\left. +\left|\frac{ -\kappa _1 +\kappa _2 +\kappa _3}{2}\right|K(\left|\frac{-\kappa _1+\kappa _2+\kappa _3}{2}\right|)\right.\nn\\
&\left. +\left|\frac{\kappa _1 -\kappa _2 +\kappa _3}{2}\right|K(\left|\frac{\kappa _1-\kappa _2+\kappa _3}{2}\right|)\right.\nn\\
&\left. +\left|\frac{\kappa _1 +\kappa _2 -\kappa _3}{2}\right|K(\left|\frac{\kappa _1+\kappa _2-\kappa _3}{2}\right|)\right] 
\period \label{regarea2}
}
This is the final result expressing the most important part of 
 the three point coupling of the  LSGKP strings,  in terms of
 the parameters $\kappa_i$  by which the AdS energy and 
 the spin of the string are expressed. Substituting this result into (\ref{area0}), we obtain the fully explicit expression of the finite part of the area $A_{fin}$. 

\section{Semi-classical  wave function and  vertex operator for three-point 
 functions}
In the previous section, we have computed the $SL(2,C)$ invariant finite part of the area $A_{fin}$, which constitutes  the main part of the three point function for the LSGKP 
strings.  In what follows, we will discuss the remaining part of the 
three point function made up of the divergent part of the area defined 
 in subsection 3.1 and the contribution of the vertex operators. We will 
 argue that these contributions together give finite result and explain 
how the appropriate vertex operator should be constructed systematically. 
Although such a program has not yet been carried out for the LSGKP
 strings, we will illustrate our arguments by using an  example of rotating 
 strings  in flat spacetime. 
\subsection{Structure of three-point functions}
As described in the introduction,  the remaining part of the area 
 \beq{
A_{div} = 4\int d^2z \sqrt{p\pbar}
}  
and the contribution from the vertex operators 
\beq{
\sum _i \log V_i\left[ X_{\ast};z_i,x_i,Q_i\right]
}
diverge  separately. Therefore, to compute these quantities properly 
 one must introduce appropriate regularization parameters for each 
 of the contributions. As will be illustrated later for the case of rotating string
 in flat space, if the regularization parameters are judiciously taken, the 
 dependence on them will cancel  each other and one obtains a finite result. 

To demonstrate such a mechanism for the LSGKP strings and obtain 
 complete  three point functions,  we must first 
construct the  semi-classical vertex operators for LSGKP. 
In string theory, the correct vertex operator for a closed string state 
 must satisfy two requirements:
\begin{itemize}
\item First it must transform correctly under the global 
 symmetry group so that it carries  the target space quantum numbers  of the state in question. This requirement is relatively easy to satisfy as string coordinates  which transform covariantly are usually available. 
\item The second requirement is that it must be a conformal primary 
of dimension $(1,1)$ on the worldsheet.  In the present case, where we are 
 interested in states with large quantum numbers, $(0,0)$ primarity  is sufficient within the saddle point approximation.
\end{itemize}

One way to check that a candidate vertex 
 operator  is  indeed such a primary  is to evaluate its  two point function, 
 as discussed in \cite{BuchTseytlin}, 
and see if the result does not  depend  on the insertion points 
on the worldsheet. This method is rather cumbersome as one needs to evaluate  the entire two point function. We will make a comment on it shortly.

Below we propose an alternative  simpler method, 
 apparently not discussed in the literature, which is a saddle point 
 version of the conformal Ward identity.

As is well-known, the 
correlation functions of the primary operators $V_i$ with dimension $\Delta_i$
satisfy the following  Ward identity:
\begin{align}
&\langle T(z) V_1(z_1) V_2(z_2) \cdots V_n(z_n) \rangle  \nn\\
&\qquad = \sum_{i=1}^n \left( {\Delta_i \over (z-z_i)^2 }
+ {1\over z-z_i} {\del \over \del z_i} \right)\langle  V_1(z_1) V_2(z_2) \cdots V_n(z_n) \rangle \period \label{wardid}
\end{align}
It should be emphasized  that the vertex operators, as well as the energy-momentum tensor, must be 
properly renormalized so that the correlation functions are finite. When the theory is not free, such as in the case at hand, this renormalization procedure is in general quite non-trivial.
Now when all the vertex operators 
 carry large quantum numbers, of the order of a large 
 number $\ga$, $V_i$  can be expressed as 
\beq{
V_i &= e^{\ga  v_i} \period 
}
In such a case, since the $T(z)$ itself 
 is an operator with small quantum number, the insertion of $T(z)$ does not affect the saddle. Therefore we can evaluate (\ref{wardid}) with the same saddle 
 configuration appropriate for the correlation function without the insertion of 
$T(z)$. Now since the correct saddle point configuration must satisfy the Virasoro condition, the left hand side vanishes.  In particular, for the two point function the identity becomes
\begin{align}
0 &=   {1\over z-z_1} \langle \del_1 V_1(z_1) V_2(z_2)  \rangle +  {1\over z-z_2} \langle  V_1(z_1) \del_2 V_2(z_2)  \rangle    \label{2ptwarda} \\
&= \left(\sum_i {\ga \over z-z_i}  {\del v_i (z_i)
\over \del z_i}\biggr|_{saddle}\right) \langle 
V_1(z_1) V_2(z_2)  \rangle \period \label{2ptwardb}
\end{align}
Since $z$ is still arbitrary, this is equivalent to the conditions
\begin{align}
 {\del v_i (z_i)
\over \del z_i}\biggr|_{saddle} &=0 \period \label{saddleward}
\end{align}
which must hold at each point $z_i$. From similar consideration with 
the insertion of $\bar{T}(\barz)$, we obtain the same relations 
 with $\del$ replaced by $\delbar$. 
This is the simple form of the conformal Ward identity valid  for 
the  saddle point approximation and one can use it to check if the vertex 
 operators are conformally correct.

Here we should make a clarifying remark on  the passage  from (\ref{wardid}) to the expression  (\ref{2ptwarda}).  In the path integral formulation  the symbol  $\langle \cdots \rangle$ contains 
 the weight factor $e^{-S}$, where $S$ is the action. Since $S$ is independent of the coordinates $z_i$ of the vertex operators, it is legitimate to bring  the derivatives $\del_i$  inside the bracket $\langle \cdots  \rangle$ as in (\ref{2ptwarda}).  However, if we choose to compute the two-point function using the the saddle point method {\it before} applying the derivatives $\del_i$, 
the action $S$ as well as the vertex operators themselves acquire extra dependence on $z_i$ {\it through}  the saddle point cofiguration  $X_\ast(z; z_1, z_2)$, which certainly depends on the positions of the vertex operators. The derivatives $\del_i$ must then  act on such $z_i$ dependence as well. 
Fortunately, this apparent complication is actually absent for a simple reason. 
Because such extra contributions are produced  only through the functional dependence on $X_\ast(z; z_1, z_2)$,  the derivative  $\del_i$ acts as 
$\int d^2z  \del_i X_\ast(z; z_1, z_2) \delta /\delta X_\ast(z;z_1, z_2)$. 
When $\delta /\delta X_\ast(z;z_1, z_2)$ acts  on the expression $\left. V_1 V_2 e^{-S}\right|_{saddle}$, it generates  the equation  of motion with $-\sum_i \ln V_i$  as the source terms. Since $X_\ast$ is precisely a solution to such an equation, this vanishes identically and again we obtain 
(\ref{2ptwarda}).

 Now  if we apply (\ref{saddleward})  to simple (properly renormalized) 
vertex operators in flat spacetime, one can check that it is 
 equivalent, within the precision of the approximation, to the on-shell 
 condition. On the other hand, applying   it to the vertex operator of the form %
\begin{align}
(\del X \delbar X)^{S/2}(X_{-1}+X_{4})^{-\Delta} \comma \label{gkpvert}
\end{align}
which is 
used in \cite{BuchTseytlin} for the LSGKP string,  one finds that 
it is not satisfied due to the dependence on the $\sigma$ coordinate
 of the worldsheet. 
Also, it should be mentioned that in the direct computation of the two-point function using the above vertex operator,  certain  $\sigma$-dependence remains, 
 which was apparently subtracted in  \cite{BuchTseytlin} as a part of the 
 renormalization process.  As was already remarked,  the proper  renormalization procedure for interacting theory is  non-trivial and would  require further careful understanding.

The analyses above indicate that, although the vertex operator (\ref{gkpvert})  has the correct 
 transformation property with respect to  the target symmetry,  it does not
appear to  have the correct worldsheet property  even within the saddle point  approximation.

Therefore, it is fair to say that for the LSGKP strings  construction of the {\it precise} semi-classical vertex operators remains as an important future problem. 
In the next subsection, however, we will give a general discussion of 
how such  vertex operator  should be systematically constructed 
from the solution of the Hamilton-Jacobi equation.

\subsection{Semi-classical vertex operator from WKB wave function}
Let us begin with the general relation between the vertex operator and 
 the wave function. It is given by the so-called ``state-operator 
correspondence", which is expressed by 
\beq{
\int \mathcal{D}X \big| _{X=X_0(\sigma ) \text{ at }\tau = \tau _0} V[X(z=0)] \exp (-S\big| _{\tau < \tau _0}) = \Psi [X_{0}(\sigma )] 
\period \label{stateop} 
}
Here and hereafter, we use the worldsheet coordinate $z=e^{\tau+i\sig}$. 
On the left hand side, $V[X(z=0)] $ is the vertex operator inserted at $z=0$, 
$\exp (-S\big| _{\tau < \tau _0}) $ is the contribution of the action 
within a disk of radius $e^{\tau_0}$ around $z=0$, and the path integral 
 is performed over the string configuration with the prescribed behavior
 on the circumference at  $\tau=\tau_0$ of the disk.  This quantity 
 equals the right hand side 
$\Psi [X_{0}(\sigma )]$, which represents the wave function 
for the degrees of freedom living on the circumference. 

In the cases where the saddle point approximation is adequate, the path integral 
 can be replaced  by  the value at the saddle and the formula  simplifies to 
\beq{
\ln V[X_{cl}(z=0)] -S_{cl}\big| _{\tau < \tau _0} = \ln \Psi [X_{cl}(\tau = \tau _0, \sigma )]\period \label{stateop2}} 
At the same time, the wave function can be expressed in terms of an appropriate 
solution $\mathcal{W}$ of the Hamilton-Jacobi equation as 
\beq{
\ln \Psi [X (\tau = \tau _0, \sigma )]=-\mathcal{W}[X (\tau = \tau _0, \sigma )]
\period 
}
However, except for simple cases such as those in flat spacetime, there are 
difficulties in constructing  the vertex operator using the relation (\ref{stateop2}) 
directly. First, non-linearity of the equations of motion for a string in curved 
 spacetime makes it a highly non-trivial problem to obtain the appropriate solution 
 of  the Hamilton-Jacobi equation and express it in terms of the set of 
target space variables  $X[\sig]$. Second, even if one succeeds in obtaining 
 the wave function,  the contribution of the action $S_{cl}\big| _{\tau < \tau _0}$
is in general a complicated function of $\tau_0$ and hence separating  the 
vertex operator part properly is not an easy process. 
 
These difficulties may  be  greatly alleviated if one can find a judicious 
canonical transformation  of variables. Suppose we make  such a 
transformation from $X$ to a new set of variables $\phi_i$. Then, the 
relation  (\ref{stateop2}) becomes 
\beq{
\ln V[\phi_{cl}(z=0)] -\tilde{S}_{\phi\comma cl}\big| _{\tau < \tau _0} = \ln \Psi [\phi_{i\comma cl}(\tau = \tau _0)] \period 
} 
Here $\tilde{S}_{\phi_i}$ is the action for  $\phi_i$,  given by 
\beq{
\tilde{S}_{\phi}=\int d\tau \,  \left(\sum_{i} P _{\phi_i} \del _{\tau}\phi_i -H\right)  \comma 
}
where $P_{\phi_i}$  are the momenta conjugate to $\phi_i$ and  $H$ 
denotes the worldsheet energy.  In particular, if we can choose $(P_{\phi_i}, 
\phi_i)$ to be  action-angle variables, then $P _{\phi _i} \del _{\tau}\phi _i$
 becomes  a number and the evaluation of the action part 
simplifies enormously. At the same time, the Hamilton-Jacobi equations 
reduce to 
\beq{
\frac{\del \mathcal{W}}{\del \phi _i}= P _{\phi_i} =\text{constant} 
}
 and hence the wave function can be constructed simply as 
\beq{
\Psi [\phi_{i}]=\exp \left[ -\sum _i P _{\phi _i} \phi _i \right]\period 
}
Note that in this description, even for a system like a string which 
 has infinite degrees of freedom, it is possible that for certain configurations 
 only a finite number of $P_{\phi_i}$ are non-vanishing and hence effectively  we only need to deal with a system with  finite degrees of freedom. 

In making use of such canonical transformations, it is important to keep in mind
 that $V[X_{cl}(\sigma)]$ and $V[\phi _{i\comma cl}]$ are not in general equal. 
Indeed while $V[X(\sigma)]$ is expressed as 
\beq{
V[X(\sigma)] = \bra{X(\sigma )}V\ket{0} \comma \label{VX} 
}
where $\ket{0}$ denotes the $SL(2,R)$ invariant vacuum, 
the quantity $V[\phi _{i}]$ is given by 
\beq{
V[\phi _{i}]=\bra{\phi_i} V \ket{0} = \int \mathcal{D} X \,\braket{\phi _i | X}\bra{X}V\ket{0}\comma 
}
which is obviously different from (\ref{VX}).  In cases where we may use the 
 saddle point approximation,  $V[\phi _{i}]$ reduces to 
$V[\phi _{i}] =\braket{\phi _i | X}\bra{X}V\ket{0}$, where $X$ 
 is the saddle point configuration. Even in such a situation the factor 
$\braket{\phi _i | X}$ may produce a finite difference\footnote{In the case 
 of a string in flat spacetime this factor turns out 
 to be unity.}. 

Although we have not yet been able to  construct 
 the proper vertex operator for the LSGKP string through above procedure, 
 it is quite instructive to demonstrate the validity of the argument by
 using  a specific solvable example. It is a rotating folded string solution 
 in flat space \cite{Tseytlin} given by 
\beq{
&X^{\mu}=-i\, k^{\mu}\ln |z|  \comma\hspace{5pt}\mu\neq 1,2
\comma \\
&X\equiv X_1+iX_2=\frac{\omega}{2i}(z-\barz )\comma \hspace{5pt}\bar{X}\equiv X _1 -iX_2 = \frac{\omega}{2i}(1/z-1/\barz ) \period 
}
(For simplicity we have set $\al'=1$.)  We will treat the zero mode and 
the non-zero mode separately, as they are completely decoupled in this case. 

For the zero mode, we can identify the angle variable as $x^{\mu}=\frac{1}{2\pi}\int d\sigma X^{\mu}$ and the solution of the Hamilton-Jacobi equation 
 is given by 
\beq{
\mathcal{W}_{\text{zero mode}} =-ik_{\mu}x^{\mu} +\frac{k^2}{2} \tau
\period
}
Then  the relation (\ref{stateop2}) for the zero mode part reads  
\beq{
\log V_{\text{zero mode}}+\frac{k^2}{2}(\tau _0 -\log \epsilon )=ik_\mu x^\mu(\tau _0)-\frac{k^2}{2}\tau _0 \period 
}
Note that to regularize the divergence, we have cut off 
  the integration region for  the action by $\tau = \log   \ep$. Now  if we use 
 the relation $x^\mu (\tau=\tau _0) + ik^\mu (\tau _0-\log \epsilon)=ik_\mu x^\mu (\tau=\log\epsilon)=ik_\mu X^\mu (\tau = \log \epsilon)+\mathcal{O}(\epsilon)$, 
the vertex operator can be written as\beq{
\log V_{\text{zero mode}}= ik_\mu X^\mu (\tau = \log \epsilon )-\frac{k^2}{2}\log \epsilon 
\period 
}
This vertex, in addition to the usual expression $ik_\mu X^\mu$, 
  contains a  divergent term of the form $-\frac{k^2}{2}\log \epsilon$. 
But this is  the correct form because this precisely  cancels 
 the divergence produced by cutting off the area at $\tau=\ep$. 
Therefore, as we have pointed out already, the total amplitude is finite. 

Let us now discuss the non-zero mode part.  From the mode expansion 
\beq{
X^{\mu} = x^{\mu} + i \sum _{n \neq 0} \frac{1}{n} \left( \alpha _{n}^{\mu} e^{-in\sigma} + \tilde{\alpha} _{n} ^{\mu} e^{in\sigma} \right) \nn\\
\del _{\tau} X^{\mu} = -i p^{\mu} -\frac{i}{2}\sum _{n\neq 0} \left( \alpha _{n}^{\mu} e^{-in\sigma} - \tilde{\alpha} _{n} ^{\mu} e^{in\sigma} \right) \nn
}
we can identify the relevant angle  variables $\theta, \tilde{\theta}$
 and the corresponding action variables $J, \tilde{J}$ as 
\beq{
\theta \equiv \log \left( \alpha _{-1}^{1}+i\alpha _{-1}^{2}\right)\comma  & \qquad  J\equiv -\frac{1}{2}(\alpha _{-1}^{1}+i\alpha _{-1}^{2})(\alpha _{-1}^{1}-i\alpha _{-1}^{2})\comma \\
\tilde{\theta} \equiv \log \left( \tilde{\alpha} _{-1}^{1}+i\tilde{\alpha} _{-1}^{2}\right)\comma  & \qquad \tilde{J}\equiv -\frac{1}{2}(\tilde{\alpha} _{-1}^{1}+i\tilde{\alpha} _{-1}^{2})(\tilde{\alpha} _{-1}^{1}-i\tilde{\alpha} _{-1}^{2})\period 
}
For the rotating string solution above,  action variables for all the other modes 
 can be set to zero. The solution of the Hamilton-Jacobi equation is given by 
\beq{
\mathcal{W}_{\text{oscillator}}=-\omega ^2\tau -\frac{\omega ^2}{2} \theta-\frac{\omega ^2}{2} \tilde{\theta}  \period 
}
In this case there is no
 contribution from the action $\tilde{S}$ and hence the vertex 
 operator for this part is obtained through (\ref{stateop2}) as 
\beq{
\log V_{\text{oscillator}} &=\omega ^2\tau +\frac{\omega ^2}{2} \theta + \frac{\omega ^2}{2} \tilde{\theta}\nn\\
&=\frac{\omega ^2}{2} \log \left( \int \frac{dz}{2\pi i z} \del X \int \frac{d\barz}{2\pi i \barz} \delbar X \right) \period 
}
Since the spin of this string is given by $S=\omega^2$, we finally obtain 
the expression 
\beq{
V_{\text{oscillator}} = (\del X \delbar X)^{S/2}\comma 
}
which is quite reasonable. 

As it should be clear from the above demonstration, in order to construct 
the correct vertex operator, one must properly take into account the contribution 
 of  the action in the application of the state-operator correspondence. 
This applies to the GKP string as well:  One should find  appropriate 
 action-angle variables and then apply the formula (\ref{stateop2}), 
which includes the effect of the action. 
\section{Summary and future directions}
In this work, we have discussed how to evaluate  the three point functions 
 of heavy string states, carrying large quantum numbers of order $\sqrt{\lam}$,  by the use of the saddle point method.  By adapting the powerful 
integrability-based method invented previously for the gluon scattering 
 problem and making several non-trivial modifications, we have developed
 a way to compute the regularized area, which is   the essential part of the $SL(2,C)$ invariant three point coupling,   without the knowledge of the explicit form of the saddle point 
configuration. In particular, we have applied the method  to the case of 
 large spin limit of the GKP strings and  were able to express the 
 regularized area explicitly in terms of the parameters $\kappa_i$  that  
 characterize the external GKP states.  Reliable computation 
of the remaining  $SL(2,C)$ variant part of  the 
three point function,  which contains  the contributions of the vertex 
operators,    must await future investigations, since at present time the  precise  form of the semi-classical vertex operator for the GKP string is  not known. Nevertheless, we have presented a general discussion of 
 how such  vertex operators should be constructed  systematically 
using the semi-classical  wave functions obeying the Hamilton-Jacobi equations 
and gave an explicit  example to illustrate the method. 

There are some possible directions for future work. 

One  important task, of course,  is to analyze the  Hamilton-Jacobi equation 
 relevant to  the LSGKP string by developing a judicious integrability-based 
 method and find the precise form of the semi-classical vertex operator for 
LSGKP. This will allow us to complete the computation of the three point 
 functions of these states, which we set out to perform in this work. 
Also it would be desirable to extend the computation to the case of 
 general GKP strings, without taking the large spin limit. 
Such investigations are now in progress.

Another obvious extension of our work is the computation of   
higher point functions, in particular that of four point functions of LSGKP 
 strings. As already remarked in the main text, the generalized Riemann 
 bilinear identity that we derived in this work should be indispensable 
 for that purpose. It would be quite interesting if one can understand 
such  characteristic  property of the conformal field theory  as the crossing symmetry etc. from the string theory side at strong 't Hooft coupling. 

Clearly, the method that we developed in this article should be applicable 
 to various other types of heavy strings, which for example belong to the 
class of so-called  finite gap solutions. Successful applications in that direction 
 would deepen our understanding of  the integrability  structure of string theory
 in curved spacetime. 

Such study of the finite gap solutions might also  lead to the discovery of 
 some exact three-pronged string solution, which will undoubtedly 
 be useful in performing and checking the calculation of three point functions. 

Finally, it should be important to study how one can 
compute the next order in the $1/\sqrt{\lam}$ expansion, making 
 use of the integrability structure of the system. 

We hope to return to discuss progress in some of these problems in near 
 future. 
\par\bigskip\noindent
{\large\bf Acknowledgment}\par\smallskip\noindent
We are grateful to R.~Janik for useful comments on the manuscript and E.~Buchbinder and A.~Tseytlin for raising  clarifying questions.
The research of  Y.K. is supported in part by the 
 Grant-in-Aid for Scientific Research (B) 
No.~20340048, while that of S.K. is supported in part 
 by JSPS Research Fellowship for Young Scientists,   from the Japan 
 Ministry of Education, Culture, Sports,  Science and Technology. 
\appendix
\setcounter{equation}{0}
\renewcommand{\theequation}{A.\arabic{equation}}
\renewcommand{\thefigure}{A.\arabic{figure}}
\section*{Appendix A:\ Regularized area  for general configuration of  $\boldsymbol{\kappa_i}$'s}
\addcontentsline{toc}{section}{Appendix A:\ Regularized area  for general configuration of  $\boldsymbol{\kappa_i}$'s}
Below we give some details of the analysis needed for the computation 
 of  the regularized area for more general configuration of the parameters $\kappa_i$ than discussed in the main text, namely the cases 
 where the triangle inequality is not satisfied. 
Specifically we consider, without loss of generality,  the case where 
the parameters satisfy 
\begin{align}
\kappa _2 >\kappa _1+\kappa _3  \period \label{configkappa}
\end{align}
 Results for other cases are  obtained by permutations of $\kappa _i$'s.

\begin{figure}[htb]
\begin{minipage}{0.5\hsize}  
\begin{center}
   \picture{clip, height=5cm}{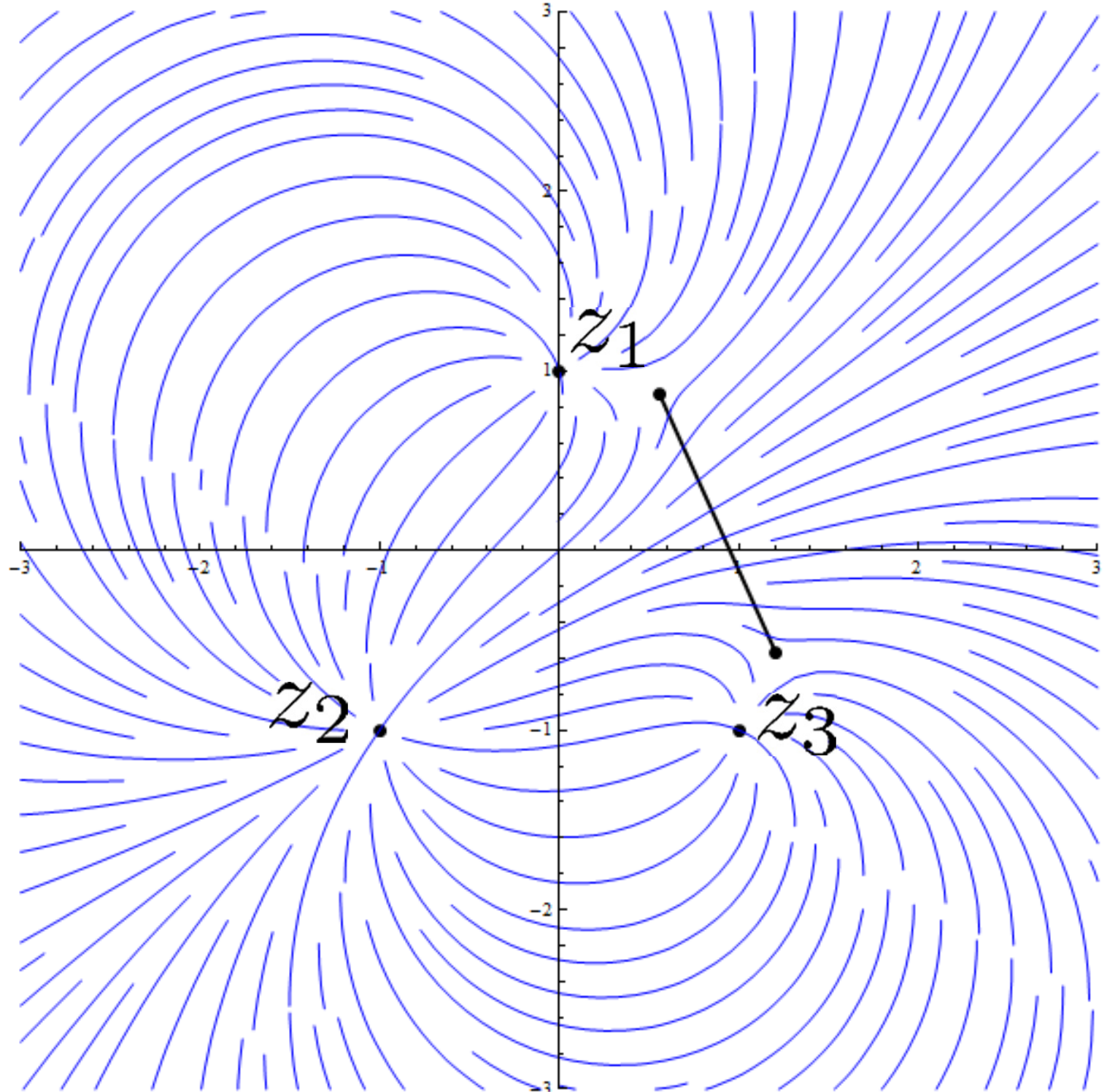}\\
   $(a)$
  \end{center}
 \end{minipage}
\begin{minipage}{0.5\hsize}
\begin{center}
\picture{height=5cm, clip}{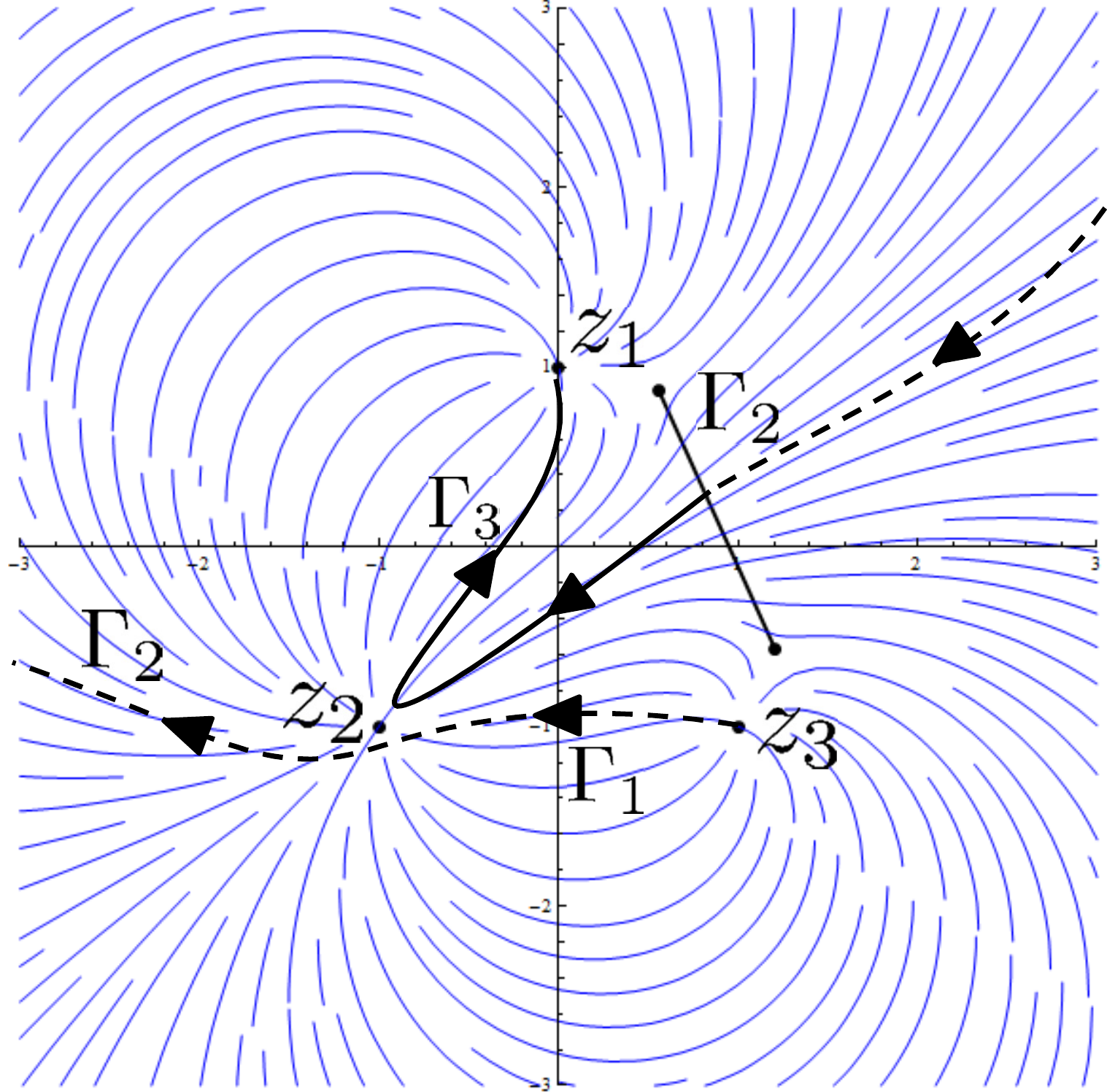}\\
$(b)$
\end{center}
\end{minipage}
\caption{$(a)$  Behavior of WKB curves for the parameter regime where the 
triangle inequality is not satisfied. The parameters used are 
$\kappa_1=\kappa_3=1, \kappa_2=\sqrt{6}$ and ${\rm Im}\, \xi >0$. 
$(b)$ Deformation of the contour connecting $\widehat{z}_3$ to $z_1$. 
The deformed contour consists of  three parts  $\Ga_1, \Ga_2$ and $\Ga_3$.}
\label{negwkb}
\end{figure}

Compared to the case where the triangle inequality is satisfied, there 
 are substantial  changes in the configuration of the WKB curves and 
 the position of the branch cuts as exhibited in figure \ref{negwkb}. 
The most crucial difference is that  two poles, $z_1$ and $z_3$ in figure \ref{negwkb}, 
 are no longer connected directly by a single WKB curve. Since  reliable 
 expansion must be made along WKB curves, we must deform the path 
 connecting  $z_3$ and $z_1$ appropriately as in figure \ref{negwkb} 
 so that the resultant  path consists of the following three WKB curves 
$\Gamma_1 \sim \Gamma_3$:
\begin{align}
&\Gamma_1:\quad \mbox{connecting $\widehat{z}_3$ and $\widehat{z}_2$} \nn\\
&\Gamma_2:\quad \mbox{connecting $\widehat{z}_2$ and $z_2$}\nn\\
& \Gamma_3:\quad \mbox{connecting $z_2$ and $z_1$ }\nn
\end{align}

Contributions form $\Ga_1$ and $\Ga_2$, which lie  respectively in 
 the second and the first sheet, are easy to compute  since 
 we can show that, for ${\rm Im}\,\xi>0$, they are related to $\braket{s_2,s_3}^{-1}$ and $\braket{s_1,s_2}^{-1}$ respectively. 
On the other hand, evaluation of the  contribution  from $\Ga_2$, 
which connects the points on top of  each other on different sheets, 
requires care. Because $\widehat{z}_2$ is on the second sheet, 
 the relevant quantity is 
\beq{
\braket{s_2,s_2^{\prime}} \label{propri}
}  
where $s_2^{\prime}$ is the solution which can be obtained from $s_2$ 
by the analytic continuation  along a path encircling $z_1$ clockwise.
Now since $s_2$ is related to the plus-minus solutions in the following way,
\beq{
&{\rm Im}\,\xi >0\Rightarrow s_2 \sim 2_{-} \comma\nn\\
&{\rm Im}\,\xi <0\Rightarrow s_2 \sim 2_{+} \comma\nn
}
we need to evaluate $\braket{2_{-}\comma 2^{\prime} _{-}}$ and $\braket{2_{+}\comma 2^{\prime} _{+}}$, where  $2_{\pm}^{\prime}$ 
are obtained from $2_{\pm}$ by the same analytic continuation 
as for $s_2'$.  
Since $2_-$ can be expressed in terms of the solutions $1_\pm$ as 
\beq{
2_{-} = -\braket{1_{-},2_{-}}1_{+} +\braket{1_{+},2_{-}}1_{-}\comma
} 
$2_{-}^{\prime}$ can be calculated as
\beq{
2_{-}^{\prime} = -\braket{1_{-},2_{-}}e^{-i\pp _1(\xi )} 1_{+} +\braket{1_{+},2_{-}}e^{+i\pp(\xi )}1_{-}\period
}
Therefore, the product $\braket{2_{-},2_{-}^{\prime}}$ can be evaluated as
\beq{
\braket{2_{-},2_{-}^{\prime}}&=-2i\sin \pp _1 (\xi)\braket{1_{+},2_{-}} \braket{1_{-},2_{-}}\nn\\
&= -(cA_2^2  \sin ^2 (-\pp _2)) ^{-1} \left(\sin (\frac{-\pp _1+(-\pp _2)-\pp _3}{2})\sin (\frac{-\pp _1+(-\pp _2)+\pp _3}{2})\right)
\period \nn
}
Similarly, $\braket{2_{+},2_{+}^{\prime}}$ can be evaluated as
\beq{
&\braket{2_{+},2_{+}^{\prime}}=4c A_2^2\left(\sin (\frac{\pp _1+(-\pp _2)+\pp _3}{2})\sin (\frac{\pp _1+(-\pp _2)-\pp _3}{2})\right)
\period \nn
}
Combining, we obtain the following normalization independent quantity:
\beq{
\braket{2_{+},2_{+}^{\prime}}\braket{2_{-},2_{-}^{\prime}}&=-4 (\sin ^2 (-\pp _2)) ^{-1}\times\nn\\
&\left[\sin (\frac{-\pp _1+(-\pp _2)-\pp _3}{2})\sin (\frac{\pp _1+(-\pp _2)+\pp _3}{2})\right.\nn\\
&\left. \sin (\frac{-\pp _1+(-\pp _2)+\pp _3}{2})\sin (\frac{\pp _1+(-\pp _2)-\pp _3}{2})\right] \period \label{proan}
}
From this result,  one can determine the relevant part of $\braket{s_2,s_2^{\prime}}$  as follows:
\beq{
{\rm Im}\,\xi> 0\hspace{44pt}&\nn\\
\log \braket{s_2,s_2^{\prime}}_{n.s.}&=  2h(-\pp _2) -h(\frac{\pp _1+(-\pp _2)+\pp _3}{2})-h(\frac{-\pp _1+(-\pp _2)-\pp _3}{2})\nn\\
&-h(\frac{-\pp _1+(-\pp _2)+\pp _3}{2})-h(\frac{\pp _1+(-\pp _2)-\pp _3}{2})\comma \\
{\rm Im}\,\xi< 0\hspace{44pt}&\nn\\
\log \braket{s_2,s_2^{\prime}}_{n.s.}&=  -2h(-\pp _2) +h(\frac{\pp _1+(-\pp _2)+\pp _3}{2})+h(\frac{-\pp _1+(-\pp _2)-\pp _3}{2})\nn\\
&+h(\frac{-\pp _1+(-\pp _2)+\pp _3}{2})+h(\frac{\pp _1+(-\pp _2)-\pp _3}{2})\period 
}

Finally, summing up the contributions from all three curves
  $\Gamma_1 \sim \Gamma_3$, we find that the net result  is simply 
given by making the following substitution to the expression 
 for the $\kappa _2 <\kappa _1+\kappa _3$ case:
\beq{
K(\frac{\kappa _1 -\kappa _2 +\kappa _3}{2})\longmapsto -K(\frac{-\kappa _1 +\kappa _2 -\kappa _3}{2}).
}
From this result and the corresponding ones with $\kappa_i$ permuted, 
one can easily deduce the general  formula given in (\ref{regarea2}). 


\end{document}